\documentclass[journal,twoside,web]{ieeecolor}
\pdfoutput=1
\usepackage{generic}
\usepackage[noadjust]{cite}

\usepackage{amsmath,amssymb,amsfonts,amsthm}
\usepackage{algorithm}
\usepackage[algo2e]{algorithm2e}
\usepackage{algorithmic}
\usepackage{wasysym,xcolor}
\usepackage{hyperref}
\usepackage{mathrsfs}
\usepackage{graphicx,float}
\usepackage{capt-of}
\usepackage{textcomp,soul,comment}
\usepackage{pgfplots}

\usepackage{caption}
\makeatletter
\let\NAT@parse\undefined
\makeatother
\usepackage{hyperref}

\pgfplotsset{compat=1.10}
 \def\BibTeX{{\rm B\kern-.05em{\sc i\kern-.025em b}\kern-.08em
    T\kern-.1667em\lower.7ex\hbox{E}\kern-.125emX}}
\usepackage{fancyhdr}
\usepackage{xcolor}

\def\doi{10.1109/TAC.2024.3365569}
\fancyfootoffset{+0.0em}

\fancypagestyle{copyright}{
    \vspace{-2.0em}
	\cfoot{\vspace{0.0em}\footnotesize\textcopyright{} 2024 IEEE. Personal use of this material is permitted.  Permission from IEEE must be obtained for all other uses, in any current or future media, including reprinting/republishing this material for advertising or promotional purposes, creating new collective works, for resale or redistribution to servers or lists, or reuse of any copyrighted component of this work in other works.}
	\chead{\footnotesize This article has been accepted for publication in IEEE Transactions on Automatic Control. This is the author's version which has not been fully edited and content may change prior to final publication. Citation information: \href{http://dx.doi.org/\doi}{DOI \doi}}
}

\newcommand{\rahel}[1]{{\color{black} #1}}
\newcommand{\rrahel}[1]{{\color{black} #1}}

\renewcommand{\baselinestretch}{1.05}
\parskip = \medskipamount
                 
\usepackage{amssymb,url,amsmath,mathrsfs}
\usepackage{epsfig,adjustbox} 
\usepackage{times} 
\usepackage{euscript}
\usepackage{xcolor}
\usepackage{xr-hyper}
\usepackage{hhline,multirow,makecell}
\usepackage{color}
\usepackage{comment,soul}
\usepackage{caption}

\newcommand{\notn}[2]{& {#1} \quad \quad && \text{#2}\\}

\newtheorem{remark}{Remark}

\newtheorem{theorem}{Theorem}
\newtheorem{proposition}{Proposition}
\newtheorem{assumption}{Assumption}
\newtheorem{lemma}{Lemma}

\allowdisplaybreaks

\begin{document}
\title{Active Learning-based \\
Model Predictive Coverage Control}
\author{Rahel Rickenbach, Johannes K\"ohler, Anna Scampicchio, Melanie N. Zeilinger, Andrea Carron
\thanks{
An accompanying video detailing the contributions of the paper is available at the following link: \url{https://youtu.be/3u-JZxx3L3M}. \\
All authors are members within the Institute for Dynamic Systems and Control (IDSC), ETH Z{\"u}rich.
{\tt\footnotesize [rrahel|jkoehle|ascampicc|\\mzeilinger|carrona]@ethz.ch}}}

\maketitle
\thispagestyle{copyright}
\pagestyle{empty}

\begin{abstract}
The problem of coverage control, i.e., of coordinating multiple agents to optimally cover an area, arises in various applications. 
However, coverage applications face two major challenges: (1) dealing with nonlinear dynamics while respecting system and safety critical constraints, and (2) performing the task in an initially unknown environment. We solve the coverage problem by using a hierarchical framework, in which references are calculated at a central server and passed to the agents' local model predictive control (MPC) tracking schemes.
Furthermore, to ensure that the environment is actively explored by the agents a probabilistic exploration-exploitation trade-off is deployed. In addition, we derive a control framework that avoids the hierarchical structure by integrating the reference optimization in the MPC formulation.  
Active learning is then performed drawing inspiration from Upper Confidence Bound (UCB) approaches. For all developed control architectures, we guarantee \rahel{closed-loop constraint satisfaction} and convergence to an optimal configuration. Furthermore, all methods are tested and compared on hardware using a miniature car platform.
\end{abstract}

\begin{IEEEkeywords}
Coverage Control, NL Predictive Control, Cooperative Control, Machine Learning, Agents and Autonomous Systems
\end{IEEEkeywords}

\section{Introduction}
\label{sec:introduction}
 
\begin{figure} [t]
\centering
\includegraphics[width=0.45\textwidth]{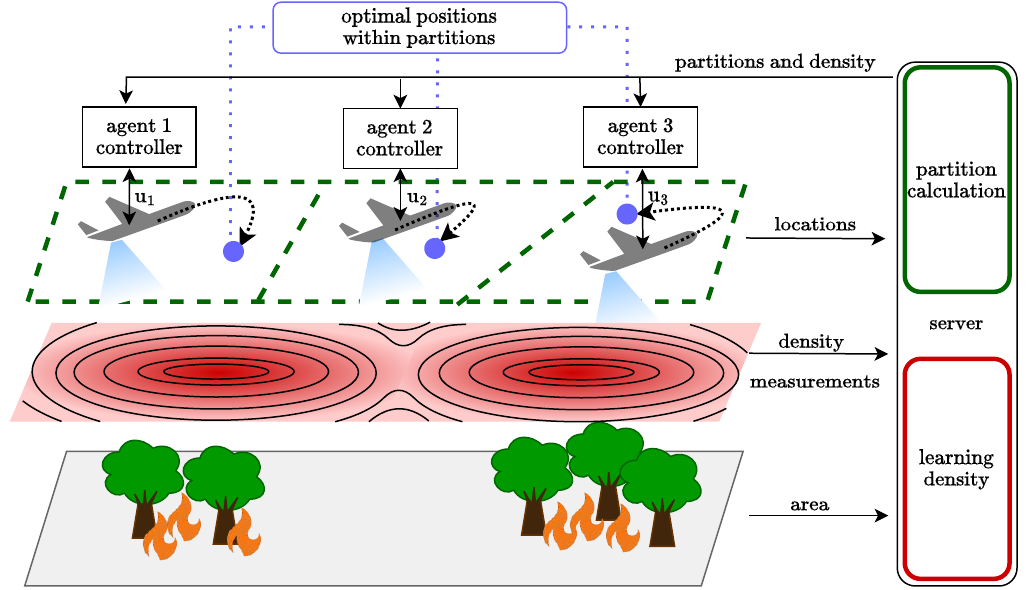}
\caption{Illustration of coverage control problem and partitioning of environment at the example of firefighting planes and environmental demands defined by the resulting heat map.}
\label{fig:explanatoryfigure}
\vspace{-1.2em}
\end{figure}

A key problem in multi-agent systems consists in optimally covering a finite area with respect to environmental demands, which  
are reflected by a density function of measurable values of interest. 
This task goes under the name of \textit{coverage control}, and it can be rephrased as placing the agents at the centroids of their Voronoi partition induced by the density function~\cite{Du1999}. Applications are versatile. One is, e.g.,  
the autonomous re-positioning of self-driving taxis according to the population density, providing a faster and more environmentally friendly service. Another example, illustrated in Figure~\ref{fig:explanatoryfigure}, are firefighting planes that are autonomously and optimally distributing themselves with respect to the heat map of a certain area.
From these examples, the challenges that emerge are twofold and intertwined. The first is designing a coverage control architecture for agents with nonlinear dynamics while ensuring collision avoidance and respecting safety constraints. The second consists in dealing with initially unknown environments, i.e., unknown density functions characterizing the optimal coverage problem. In this case, the considered area needs to be explored during the process, relying on the agents' sensing capabilities. 
The goal of this paper is to design a coverage control framework that addresses both of the aforementioned challenges. 

\subsubsection*{Related Work}
\label{subsubsec:relatedwork}
In classical works such as~\cite{Cortes2004,Bullo2012}, the coverage control problem is rigorously solved under the assumptions that the environment is perfectly known and that dynamics are single integrators. Nonlinear dynamics and state/input constraints can be taken into account by using a  model predictive control (MPC) scheme~\cite{Grune2011,rawlings2017model}, as pursued in~\cite{Carron2017,Kohler2018,Farina2015}. In all the aforementioned references, the implemented coverage algorithm is hierarchical, i.e., a reference is calculated before being passed to a tracking MPC~\cite{Limon2018}. Moreover, they all require some non-trivial offline design for the terminal ingredients in MPC to prove convergence and recursive feasibility. Furthermore, none of them addressed coverage control in an unknown environment.\\
To cope with this second challenge, ``exploitation" targeted to the coverage control task needs to be combined with ``exploration" given by data collection performed via active learning~\cite{Li2006,Campbell1990} to improve the estimate of the initially unknown density function. The problem of coverage control in an unknown environment is investigated in~\rahel{\cite{Schwager2009,Todescato2017,McDonald2021,prajapat2022,le2012}}, where different strategies to balance exploration and exploitation are proposed. 
However, in these works the agents' dynamics are once again assumed to be single integrators.
The general coverage control problem encompassing an unknown environment and nonlinear constrained dynamics, \rahel{while ensuring persistent collision avoidance,} has not yet been addressed in the literature. 

\begin{figure} [t]
\centering
\includegraphics[trim={0cm 0cm 0.8cm 0cm},clip,width=0.45\textwidth]{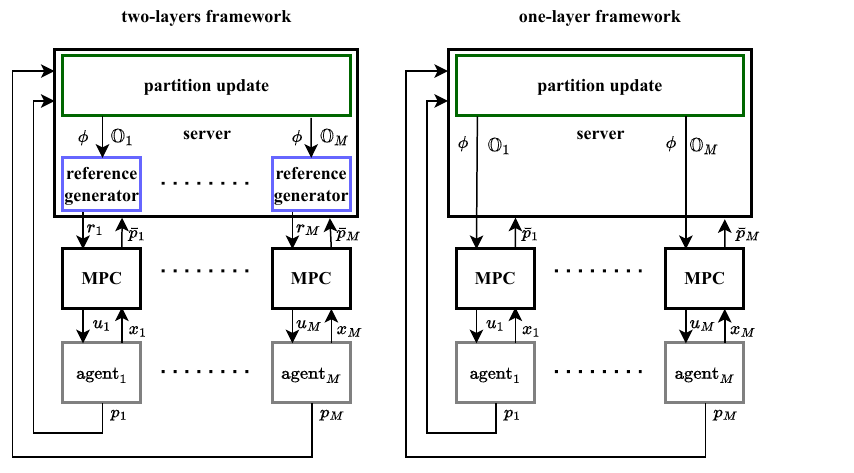}
\caption{Simplified illustration of the developed two- and one-layers algorithm. Considering density $\phi$, partitions $\mathbb{O}$, references $r$, state $x$, inputs $u$, positions $p$, as well as setpoint positions $\bar{p}$.} 
\label{fig:coverageclarification}
\vspace{-1.5em}
\end{figure}

\subsubsection*{Contributions}
\label{subsubsec:contributions}

In \rahel{presence of complex agents' dynamics and in need of environment exploration, ensuring collision avoidance when agents change their configuration becomes a key challenge in coverage control. To tackle this problem, we propose a tracking MPC-based coverage framework, which (a) includes safety-ensuring constraints, and (b) avoids the difficult design of terminal constraints needed, e.g., in \cite{Carron2017}, by leveraging the tools proposed in ~\cite{Grune2011,Soloperto2021}.
Specifically, we propose and analyze two strategies. The first, visualized on the left of Figure~\ref{fig:coverageclarification}, extends \cite{Carron2017} by including collision avoidance constraints. Its rationale consists in passing references, calculated by the server, to each agent's individual tracking MPC: accordingly, we will refer to this as the ``two-layers approach". The second architecture, summarized on the right of Figure ~\ref{fig:coverageclarification}, overcomes the hierarchical structure by directly integrating the calculation of the next optimal configuration into the MPC cost of each agent; for this reason, it will be referred to as the ``one-layer approach". To the best of the authors' knowledge, this strategy is novel in the literature. Furthermore, we complement both methods with active learning strategies: for the two-layers approach, we leverage the probabilistic exploration-exploitation decision
proposed in~\cite{Todescato2017}, while for the one-layer we draw inspiration from Bayesian optimization methods~\cite{Auer2002,Srinivas2010} and use an Upper Confidence Bound (UCB) approach. \\ Both control architectures are based on an MPC problem, which is challenged by the time-varying nature of its safety-ensuring constraints. In fact, as the latter change with respect to the agents' configuration, recursive feasibility, closed-loop constraint satisfaction and convergence to a locally optimal configuration require special investigation. Rigorous proofs of those two key properties are the main theoretical contributions of this paper, carried out for the two proposed control strategies and in both cases of known and initially unknown environment.
Lastly, we test and compare all the proposed approaches on hardware using the miniature racing cars Chronos in combination with CRS, an open-source software framework for control and robotics~\cite{carron2022}.}

\subsubsection*{Outline}
\label{subsubsec:roadmap}
The remainder of the paper is structured as follows. Section~\ref{sec:problemFormulation} states the coverage control problem, followed by a presentation of preliminaries regarding nonlinear tracking MPC and Bayesian learning in Section~\ref{sec:controlandlearningframework}. \rahel{Section \ref{sec:colavoidanceandrecursivefeasibility} introduces a strategy to ensure safety and recursive feasibility under changing Voronoi partitions.} Sections~\ref{sec:twolayers} and~\ref{sec:onelayer} detail the proposed two-layers and one-layer methods, respectively. In addition, we state the main theoretical results on convergence, satisfaction of safety-critical constraints and recursive feasibility for both set-ups with known and unknown environment. The proofs are deferred to the Appendix. Section~\ref{sec:experiments} gathers all the experimental results, and Section~\ref{sec:conclusions} ends the paper by discussing the benefits of the proposed approaches and drawing conclusions.
 
\subsubsection*{Notation}
\label{subsubsec:notation}
Throughout the paper, $\Vert \! \cdot \! \Vert$ indicates the Euclidean norm and $\mathcal{N}$ the Gaussian distribution. The set of all non-negative real numbers is given by $\mathbb{R}_{+}$ and the set of natural numbers by $\mathbb{N}$. We define the ball $\mathbb{B}_{\rahel{o}}^{b} = \{x \in \mathbb{R}^b\vert \,\Vert x \Vert\leq \rahel{o} \}$ and indicate with $\ominus$ the Pontryagin set difference \cite[Sec.~3.1]{kouvaritakis2016}. Considering some arbitrary small but fixed constant $\epsilon > 0$, for each set $\mathbb{V}\subset\mathbb{R}^b$ we denote $\mathbb{V}^{\mathrm{int}}:= \mathbb{V} \ominus \mathbb{B}_{\epsilon}^{b}$. Given a compact set $\mathbb{D}\subset\mathbb{R}^b$ and a  continuous function $f: \mathbb{D} \rightarrow \mathbb{R}$, we define
\begin{equation}
    f(\rahel{y})_{\mathbb{D}} := \min_{\tilde{\rahel{y}} \in \mathbb{D}}f(\rahel{y} - \tilde{\rahel{y}}).
    \label{eq:minsetdistance}
\end{equation}

\section{Problem Formulation}
\label{sec:problemFormulation} 
This section states the coverage control problem, together with the specification of the nonlinear constrained dynamics that are considered in the paper.

\subsection{Dynamics and Constraints}
We consider a \rrahel{polytopic} area \mbox{$\mathbb{A}\in \mathbb{R}^{D}$}, on which $M$ agents move according to time-invariant, discrete-time, and nonlinear dynamics. For agent $i$, these are
\begin{equation}
\begin{split}
    &x_{i,k+1} = f_{i}(x_{i,k}, u_{i,k}) \\
    &p_{i,k} = C_{i}x_{i,k},
\end{split}
\label{eq:nonlineardynamics}
\end{equation}
where $x_{i,k} \in \mathbb{R}^{n_{i}}$ and $u_{i,k} \in \mathbb{R}^{m_{i}}$ denote the state and input at time $k$, respectively. The vector $p_{i,k} \in \mathbb{R}^D$ indicates the position of agent $i$ at time $k$, which is usually chosen as Cartesian coordinates in $D$=2 or $D$=3. Additionally, all agents are subject to state and input constraints, i.e., $x_{i,k} \in \mathbb{X}_{i}$ and $u_{i,k} \in \mathbb{U}_{i}$ for $k \in \mathbb{N}$. We consider relatively general heterogeneous dynamics for the different agents and make the subsequent assumption.
\begin{assumption} The agent's dynamics $f_{i}$ are assumed to be known and Lipschitz continuous with Lipschitz constant $\mathcal{L}_{i}$. Moreover, each agent's state can be measured perfectly and its state and input constraints are compact.
\label{assumption:dynamics}
\end{assumption}

\subsection{Coverage Control}
\label{subsec:coverage}
We indicate with \mbox{$p = [p_{1}, \hdots, p_{M}]^{T} \in \mathbb{R}^{M \times D}$} the agents' position, and denote with \mbox{$\mathbb{O} = \{\mathbb{O}_{1}, \hdots, \mathbb{O}_{M}\}$} an arbitrary collection of polytopes partitioning $\mathbb{A}$. Accordingly, the task of coverage control can be mathematically stated as the following optimization problem:
\begin{equation}
\begin{split}
    \min_{p,\,\mathbb{O}} \:\underbrace{\sum_{i=1}^{M} \int_{\mathbb{O}_{i}} \Vert q - p _{i} \Vert^{2}\phi(q)dq}_{H(p,\mathbb{O})},\, \ p_{i} \in \mathbb{O}_{i},
\end{split}
\label{eq:cortescost}
\end{equation}
where $\phi: \mathbb{A} \rightarrow \mathbb{R}_+$ is the so-called \textit{density function}, which acts as a measure of information on the environment~$\mathbb{A}$, \rahel{the squared Euclidean norm represents} the agents' \textit{sensing capability}, and $H(p, \mathbb{O})$ is known as the \textit{locational optimization cost}. It was shown in~\cite{Du1999} that the optimal set  $\mathbb{O}$ in~\eqref{eq:cortescost} is given by the \textit{Voronoi tessellation}~\cite{Senechal1995} \rahel{with respect to a generating position configuration $p$, denoted by \mbox{$\mathbb{W}_{p} = \{\mathbb{W}_{p,1}, \hdots, \mathbb{W}_{p,M}\}$}}, with
\begin{equation}
\begin{split}
    \mathbb{W}_{p,i} = \{ q \in \mathbb{A} \; \vert\; \Vert q - p_{i} \Vert \leq \Vert q - p_{j} \Vert, \, \forall j \neq i \}.
\end{split}
\label{eq:voronoidef}
\end{equation}
\rahel{Note that all the sets in $\mathbb{W}_{p}$ are convex, since $\mathbb{A}$ is convex~\cite{Du1999}.}
Each agent's optimal position is given by the \textit{centroids} $c(\mathbb{W}_{p}, \phi) = [c_1(\mathbb{W}_{p,1}, \phi), \hdots, c_M(\mathbb{W}_{p,M}, \phi)]$, with
\begin{equation}
\begin{split}
     c_i(\mathbb{W}_{p,i}, \phi) = \left(\int_{\mathbb{W}_{p,i}}\phi(q)dq\right)^{-1} \left(\int_{\mathbb{W}_{p,i}}q\phi(q)dq\right).
\end{split}
\label{eq:voronoicenterdef}
\end{equation}
\rahel{This locally} optimal solution in terms of both partition and positions is called \rrahel{a} \textit{centroidal Voronoi configuration} and the coverage control problem can hence be reduced to ensuring $p$ converges to $c(\mathbb{W}_{p}, \phi)$ with $\mathbb{W}_{p}$ according to~\eqref{eq:voronoidef}. Classically, for integrator dynamics this is achieved using the Lloyd algorithm~\cite{Lloyd1982}, which consists in an iterative update of the form $p_{i, k+1} = c(\mathbb{W}_{p_{k},i}, \phi)$. We study \rahel{this coverage problem while accounting for} more general nonlinear dynamics, collision avoidance constraints, and both for known environments, i.e., with known density~$\phi$, and in the case in which $\phi$ is unknown and needs to be learned online.

\begin{remark}\label{rmk:g}
    \rahel{The following exposition can be extended to consider a coverage cost \eqref{eq:cortescost} with a more general sensing capability \mbox{$g(\Vert q-p_i \Vert)$: $\mathbb{R}_{+} \rightarrow \mathbb{R}_{+}$} instead of the squared distance~\cite{Cortes2004}. In this case, the optimal positions are minimizers of $H(p, \mathbb{O})$ and have no closed-form solution.}
\end{remark}

\section{Control and Learning Framework}
\label{sec:controlandlearningframework}
In this section, we introduce the core control and learning tools of the developed methods that allow for a safe coverage movement. 
Therefore, we recall a nonlinear tracking MPC formulation and complement such a set-up with \rahel{position constraints }(Section~\ref{subsec:nonlineartrackingmpc}). Further, we define the Bayesian linear regression strategy that will be deployed in all learning-based approaches dealing with an unknown environment (Section~\ref{subsec:measurementcol}).

\subsection{Nonlinear Tracking MPC}
\label{subsec:nonlineartrackingmpc}

\rahel{For each agent, consider the problem of reaching an external reference $r_i \in \mathbb{P}_i$ while satisfying state, input, and position constraints, indicated with $x_i \in \mathbb{X}_i$, $u_i \in \mathbb{U}_i$, and $p_i \in \mathbb{P}_i$ respectively. A way of accomplishing this task consists in resorting to a nonlinear tracking MPC~\cite{Limon2018,Soloperto2021}, which we now review.}\\  
\rrahel{We introduce the concept of agent-specific \textit{artificial setpoints}, $s_i = (\bar{x}_{i},\bar{u}_{i})$, which should lie in the following set that also depends on the position constraint $\mathbb{P}_i$}
\begin{equation}
\begin{split}
    & \mathbb{S}_{\mathbb{P},i} \! = \! \{(x,u) \! \in \! \mathbb{R}^{n_{i}+m_{i}}\,\vert\, \\  & C_{i}x \in \mathbb{P}_{i}^{\mathrm{int}}, x \! \in \mathbb{X}^{\mathrm{int}}_{i}, u \! \in \mathbb{U}^{\mathrm{int}}_{i}, x \! = \! f_{i}(x,u) \}.
\end{split}
\label{eq:steadystatesetwithcollavoidance}
\end{equation}
\rahel{The steady-state position corresponding to $s_i$ is obtained by \mbox{$\bar{p}_{i} = C_{i}\bar{x}_{i}$}, which lies in the projected position space
\begin{equation}
\begin{split}
    \mathbb{S}_{\mathbb{P},i}^{\mathrm{p}} = & \{p \in \mathbb{P}_{i}^{\mathrm{int}} \;\vert \; p = C_{i}x: \exists (x,u) \text{ s.t. } (x,u) \in \mathbb{S}_{\mathbb{P}_{i}}\} \notag.
\end{split}
\label{eq:steadystatesetwithcollavoidancepositionspace}
\end{equation} Following the the MPC notation in~\cite{kouvaritakis2016}, we let $x_{i,l\vert k}$ indicate the predicted state of agent $i$ at prediction step $l$ for a controller applied at time step $k$, and use the symbol ``$\cdot$" to denote all values of the index $l=\{0,\dots,N-1\}$. The nonlinear tracking MPC cost then reads as}
\begin{equation}
\begin{split}
&J_{i}(x_{i,\cdot \vert k}, u_{i,\cdot \vert k}, s_{i,k}, r_{i,k}) \\ & = V_{N,i}(x_{i,\cdot \vert k},u_{i,\cdot \vert k},s_{i,k}) + \ell_{T,i}(\bar{p}_{i,k} - r_{i,k} )\\
& = \sum_{l=0}^{N-1}\ell_{i}(x_{i,l\vert k}, u_{i,l\vert k}, s_{i,k}) + \ell_{T,i}(\bar{p}_{i,k} - r_{i,k} ).
\end{split}
\label{eq:generalnonlineartrackingmpccost}
\end{equation}
Similar to standard MPC~\cite{kouvaritakis2016}, the \textit{tracking cost}~$V_{N,i}$ sums a continuous \textit{stage cost} $\ell_i$ over a finite horizon~$N$, and steers the system to the \rahel{artificial setpoint~$s_{i,k}$. It enables the tracking of piece-wise constant references~$r_{i,k}$, whose difference to their steady-state position $\bar{p}_{i,k}$ is penalized in the continuous \textit{target cost} $\ell_{T,i}$.}
For the $i$-th agent at time $k$, current state~$x_{i,k}$, \rahel{reference~$r_{i,k}$, and position constraint $\mathbb{P}_i$ the resulting MPC problem 
reads as}
\begin{subequations}
\begin{align}
\min_{x_{i,\cdot \vert k}, u_{i,\cdot \vert k},s_{i,k}}&J_{i}(x_{i,\cdot \vert k}, u_{i,\cdot \vert k}, s_{i,k}, r_{i,k}) \label{eq:generalcostwithcolavoidance} \\ &\qquad x_{i,0\vert k} = x_{i,k} \label{eq:generalinit}\\
&\qquad x_{i, l+1\vert k} = f_{i}(x_{i,l\vert k}, u_{i,l\vert k}) \label{eq:generaldyn} \\
&\qquad x_{i,\cdot\vert k} \in \mathbb{X}_{i},\; u_{i,\cdot\vert k} \in \mathbb{U}_{i}  \label{eq:generalstateconst}\\
&\qquad V_{N,i}(x_{i,\cdot \vert k},u_{i,\cdot \vert k}, s_{i,k}) \leq V_{\max,i} \label{eq:generalvarbound}\\
&\qquad p_{i,l\vert k} = C_{i}x_{i,l\vert k} \in \mathbb{P}_{i} \label{eq:generalvoronoiconst}\\
&\qquad s_{i,k} \in \mathbb{S}_{\mathbb{P},i}. \label{eq:generalsteadystatesetconstwithcolavoidance} \\
& \qquad l = 1, \hdots, N-1 \label{eq:1toN-1withcolavoidance}.
\end{align}
\label{eq:nonlineartrackingmpcwithcolavoidance}
\end{subequations}
\noindent \rahel{Its construction follows closely the one of a nonlinear tracking MPC as in~\cite{Limon2018,Soloperto2021}. Therefore, constraint~\eqref{eq:generalvarbound} exploits a user-chosen constant~$V_{\max,i}>0$ to ensure that the agent stays in a region of attraction around the setpoint $s_i$ without the need for utilizing a terminal constraint; see \cite{Soloperto2021,Boccia2014} for further details.} \\
For each agent, a solution of~\eqref{eq:nonlineartrackingmpcwithcolavoidance} exists given continuity of $\ell_{i},\ell_{T,i}$, and compact constraints. It consists of the optimal state and input trajectories, as well as the optimal setpoint, indicated with ~$x_{i,\cdot \vert k}^{*}$, $u_{i,\cdot \vert k}^{*}$, and \mbox{$s_{i,k}^{*} = (\bar{x}_{i, k}^{*}$, $\bar{u}_{i,k}^{*})$}, respectively. We denote by \rahel{$V_{N,i}^{*}(x_{i,k},s_{i,k}, \rahel{\mathbb{P}_{i}}) = \min_{u_{i,\cdot \vert k}, \in \mathbb{U}_{i},x_{i,\cdot \vert k} \in \mathbb{X}_{i}, C_ix_{i,\cdot \vert k} \in \mathbb{P}_{i}}V_{N,i}(x_{i,\cdot \vert k},u_{i,\cdot \vert k}, s_{i,k})$} the optimal tracking cost for the considered \rahel{position constraint and an arbitrary setpoint}. Furthermore, we denote by $\ell^*_i(x_k,\rahel{s_k}) = \min_{u \in \mathbb{U}_{i}} \ell_{i}(x_{i,k},u,s_{i,k})$ the one-step optimal stage cost
whose minimizing input is assumed to be equal to $\bar{u}_{i,k}$. The actual control problem is then solved in a receding horizon fashion, i.e., at every time $k$ we solve~\eqref{eq:nonlineartrackingmpcwithcolavoidance} and apply the first input of the obtained input sequence, that is $u_{i,k}=u^*_{i,0|k}$. \\
In the following, we discuss the stability properties of the MPC scheme~\eqref{eq:nonlineartrackingmpcwithcolavoidance} using the theory in~\cite{Boccia2014}. To this end, we require a local stabilizability condition, also known as \textit{exponential cost controllability} (\cite[Assumption 1]{Boccia2014}) in the literature.
\begin{assumption} For every agent $i \in \{1,\hdots,M\}$ there exist constants $\rahel{\chi_{i}}, \gamma_{i} > 0$ \rahel{such that, for any 
feasible artificial setpoint $s \in \mathbb{S}_{\mathbb{P},i}$, any horizon length $N \in \mathbb{N}$ and any state $x \in \mathbb{R}^{n_{i}}$ with a stage cost $\ell_{i}^{*}(x,s) \leq \rahel{\chi_{i}}$}, we have
\begin{equation}
    V_{N,i}^{*}(x,s) \leq \gamma_{i} \ell_{i}^{*}(x,s). 
\label{eq:assumption2gammavmax}
\end{equation}
\label{assumption:expocostcontrollability}
\end{assumption}
\vspace{-1.4em}
\noindent In addition, we require that the stage cost $\ell_{i}$ is positive definite w.r.t. the setpoint $s_{i}$, which is often achieved using a quadratic stage cost $\ell_{i}$. However, in~\cite{Muller2017}, it was shown that Assumption~\ref{assumption:expocostcontrollability} cannot be satisfied for non-holonomic systems using a quadratic stage cost. To avoid this issue for experiments with non-holonomic robots (cf. Section~\ref{sec:experiments}), we use the results in~\cite{Rosenfelder2021, Coron2020, Worthmann2016} to devise a polynomial stage cost and introduce the following distance function with even exponents $\eta_{i,j} \in \mathbb{N}, j = 1,\hdots,n_i+m_i$:
\begin{equation}
d_i(\zeta) = \sqrt{\zeta_1^{\eta_{i,1}} + \zeta_2^{\eta_{i,2}} + \cdots + \zeta_{n_i+m_i}^{\eta_{i,n_i+m_i}}}.\label{eq:distance}
\end{equation}
\rahel{In accordance to it, we present Assumption~\ref{assumption:boundedbyd}, which is similar to Assumption 1 in~\cite{Soloperto2021}}.
\begin{assumption} For every agent $i \in \{1,\hdots,M\}$ there exist constants $\alpha_{1,i}, \alpha_{2,i} > 0$ such that, for all \rahel{states} $x \in \mathbb{X}_{i}$, \rahel{artifical setpoints} $s \in \mathbb{S}_{\mathbb{P},i}$, and their respective \rahel{stage cost minimizing input} $\bar{u} \in \mathbb{U}_{i}$
, the following holds: 
\begin{equation}
\begin{split}
    \alpha_{1,i}d_{i}((x,\bar{u}) \! - \! s)^{2} \! \leq  \! \ell_{i}^{*}(x,s) \! \leq \! \alpha_{2,i}d_{i}((x,\bar{u}) \! - \! s)^{2}.
\end{split}
\label{eq:boundstagecosttwolayer}
\end{equation}
Moreover, there exist constants $\xi_{1,i}, \xi_{2,i} \geq 0$ such that for any two admissible setpoints \mbox{$s_{1} \! = \! (\bar{x}_{1},\bar{u}_{1})$} and \mbox{$s_{2} \! = \! (\bar{x}_{2},\bar{u}_{2})$ $\in \mathbb{S}_{\mathbb{P},i}$,}
\begin{equation}
\begin{split}
    \ell_{i}(x,u,s_{1}) \leq  \xi_{1,i}\cdot \ell_{i}(x,u,s_{2}) + \xi_{2,i}\cdot d_{i}(s_{1}-s_{2})^{2}.
\end{split}
\label{eq:boundtwosteadystatestwolayer}
\end{equation}
\label{assumption:boundedbyd}
\end{assumption}
\vspace{-1.5em}
\begin{remark}\label{remark:commentAssumption23}
\rahel{Assumptions \ref{assumption:expocostcontrollability} and \ref{assumption:boundedbyd} hold, e.g., with a quadratic stage cost $\ell_i$ and a standard norm $d_{i}(\zeta) = \Vert \zeta \Vert$ if the linearized dynamics are stabilizable~\cite[Section A]{Soloperto2021}.} 
\label{remark:remarkgammamax}
\end{remark}
\vspace{-0.0em}
\noindent Given these requirements, it is proven in~\cite{Boccia2014,Soloperto2021} that, for a sufficiently long prediction horizon, the tracking cost is a valid Lyapunov function with respect to a fixed steady state.
\begin{theorem}\cite[Theorem 4]{Boccia2014} Let Assumptions~\ref{assumption:expocostcontrollability} and~\ref{assumption:boundedbyd} hold. Then, for any \rahel{position constraint set, $\mathbb{P}_{i} \subset \mathbb{A}$, with $\mathbb{P}^{\mathrm{int}}_{i} \neq \emptyset$,} any \rahel{choice of} $V_{\max,i} > 0$ and any $\bar{\alpha}_{N,i} \in (0,1)$, there exists a horizon \rahel{length} $N^{*} \in \mathbb{N}_{\geq 0}$ such that, for all $k \in \mathbb{N}$, $N \geq N^{*}$, $s_k \in \mathbb{S}_{\mathbb{P},i}$ and all $x_k$ with $V_{N,i}^{*}(x_{k},s_k) \leq V_{\max,i}$, 
\begin{equation}
\begin{split}
    &V_{N,i}^{*}(f_i(x_{k},u_{0\vert k}^{*}),s_k^*,\mathbb{P}_i) - V_{N,i}^{*}(x_{k},s_k^*,\mathbb{P}_i) \\ & \leq  - \bar{\alpha}_{N,i} \ell_{i}^{*}(x_{k},s_k^*).
\end{split}
\label{eq:theorem5twolayer}
\end{equation}
\label{theorem:theo4boccia}
\end{theorem}
\vspace{-1.8em}

\subsection{Data Collection and Bayesian Linear Regression}
\label{subsec:measurementcol}
In an unknown environment, the agents are assumed to be equipped with sensors and able to take a noisy measurement of the density $\phi$ at their current position. Note that measurements do not necessarily have to be taken at each time step, but their collection can be, e.g., conditioned on reaching a predefined location, or triggered after a fixed number of time steps. Indexing by $h$ the number of collected data \rahel{per agent}, and assuming that the noise is independent, identically distributed (i.i.d.) and zero mean Gaussian with variance $\sigma^2$, the measurements model for the $i-$th agent is
\begin{equation}
   m_{i,h} = \phi(p_{i,h}) + \nu_{i,h},\; \text{with } \nu_{i,h} \sim \mathcal{N} (0,\sigma^{2}). 
   \label{eq:measmod}
   \vspace{-0.3em}
\end{equation}
\rahel{The data locally collected by the agents are then communicated to a server, which then estimates $\phi$ in a centralized fashion. Further, we define the 
data-set collected by all $M$ agents after $t$ measurement steps as}
\begin{equation}
\begin{split}
    I_{t} = \{ (p_{i,h},m_{i,h})\: \vert \: i=1,\cdots, M,\; h=1, \cdots, t\}.
\end{split}
\label{eq:setofmeasurements}
\vspace{-0.3em}
\end{equation}
In this set-up, we make use of the following assumption\rahel{, also used similar in, e.g.,~\cite{Schwager2009}.}
\begin{assumption}
The density function $\phi$ is represented by a linear combination of $\upsilon$ known, Lipschitz continuous features collected in a vector $\Phi: \mathbb{A} \rightarrow \mathbb{R}^{1 \times \upsilon}$, i.e., there exists an unknown parameter vector  $\theta \in \mathbb{R}^{\upsilon}$, s.t., $\phi(p) = \Phi(p)\theta$, $\forall p\in\mathbb{A}$. Furthermore, the features are linearly independent on the partition $\mathbb{W}_{p,i}$ in the sense that $\max_{p\in\mathbb{W}_{p,i}}\Phi(p)\Sigma\Phi(p)^\top\geq c_\Phi\|\Sigma\|$ for any positive semi-definite matrix $\Sigma\in\mathbb{R}^{\upsilon\times \upsilon}$ and some constant $c_\Phi>0$.
\label{assumption:bayesianconvergence}
\end{assumption}
To estimate the unknown vector $\theta$ from collected data we solve the problem within the Bayesian linear regression framework~\cite{Sarkka2013}. To this aim, we endow the unknown vector with a Gaussian prior, such that \mbox{$\theta \sim \mathcal{N}(\mu_0,\Sigma_0)$}, and adopt a recursive scheme to update mean and covariance matrix in view of new data. Specifically, denoting by $\bar{\Phi}_{t+1} = [\Phi(p_{1,t+1})^{\top} \: \cdots \: \Phi(p_{M,t+1})^{\top}]^{\top} \in \mathbb{R}^{M \times \upsilon}$ the matrix of features measured at $t+1$, and by $\bar{m}_{t+1} = [m_{1,t+1} \: \cdots \: m_{M,t+1}]^{\top} \in \mathbb{R}^{M}$ the column vector of respective measurements, the updating rule reads as follows:
\begin{subequations}
\begin{align}
    \theta_{t+1} & \sim \mathcal{N}(\mu_{t+1}, \Sigma_{t+1})\\
    \mu_{t+1} & = \Sigma_{t+1}\Big(\Sigma_{t}^{-1}\mu_{t} + \frac{1}{\sigma^{2}}\bar{\Phi}^{\top}_{t+1}\bar{m}_{t+1}\Big)\\
    \Sigma_{t+1} & = \Big(\frac{1}{\sigma^{2}}\bar{\Phi}^{\top}_{t+1}\bar{\Phi}_{t+1} + \Sigma_{t}^{-1}\Big)^{-1}. \label{eq:bayessigmaupdate}
\end{align}
\label{eq:bayesupdate}
\vspace{-0.2em}
\end{subequations}
\noindent Accordingly, at any point $p \in \mathbb{A}$ we can define the estimated density $\hat{\phi}_{t+1}(p)$ and the variance $\text{Var}_{t+1}(p)$ as
\begin{subequations}
\begin{align}
    \hat{\phi}_{t+1}(p) & = \Phi(p)\mu_{t+1} \label{eq:thetaupdate}\\
    \text{Var}_{t+1}(p) & = \Phi(p)\Sigma_{t+1}\Phi(p)^{\top} \label{eq:variancephiblr}.
\end{align}
\label{eq:densityupdateblr}
\end{subequations}
\vspace{-1.5em}

\begin{remark}
Other coverage control approaches in an unknown environment rely on Gaussian Processes (GPs) for estimation of the unknown density function, i.e., in~\cite{Todescato2017} and~\cite{prajapat2022}, but they suffer from increasing complexity and memory requirements. The parametric set-up stated in Assumption~\ref{assumption:bayesianconvergence} thereby allows for a recursive update with fixed computational complexity. This is particularly relevant for the one-layer approach, where the GP would otherwise enter the MPC formulation.
\rahel{Note that the adopted parametric representation can be seen as an approximation of a Gaussian process when selecting the rows of $\Phi(\cdot)$ as a Fourier basis: see, e.g., \cite{lazaro-gredilla_sparse_2010}.}
\end{remark}

\vspace{-0.7em}
\section{Partition Update, Collision Avoidance, and Recursive Feasibility}
\label{sec:colavoidanceandrecursivefeasibility}
\rahel{\noindent Like other coverage control approaches, the proposed framework relies on Voronoi partition updates as a key step to further decrease the locational optimization cost.}
Hence, in the following, we introduce our \rahel{partition update strategy, as well as the} safety-guaranteeing position constraint selection. 
Since these are intertwined, the update might jeopardize the recursive feasibility of the tracking MPC: therefore, we present a strategy to overcome this issue.\\
\rahel{ \noindent 
Since the nonlinear tracking MPC provides us with optimal setpoints, we update the Voronoi partitions with respect to their positions $\bar{p}_{i,w}^* = [C_{i},0]s_{i,w}^{*}$ for all $i={1,...,M}$, and denoting with $w$ the time index at which the update occurs. 
To prevent agents from colliding with each other, it needs to be ensured that, for all time instances $k \in \mathbb{N}$, 
\begin{equation}
\Vert p_{i,k} - p_{j,k}\Vert \geq (r_{i,\max} + r_{j,\max}), \ \forall i\neq j, 
\label{eq:colavoidancerequirement}
\end{equation}
where $r_{i,\max}$ is the radius of the ball covering the $i-$th agent. 
Letting $r_{\max} = \max\{r_{1,\max},\hdots,r_{M,\max}\}$, we define \mbox{$\bar{\mathbb{W}}_{\rahel{\bar{p}_w^*},i} := \mathbb{W}_{\rahel{\bar{p}_w^*},i} \ominus \mathbb{B}_{r_{\max}}^D$}  for all $i = \{1,\hdots,M\}$. 
Consequently, $p_{i} \in \bar{\mathbb{W}}_{\rahel{\bar{p}_w^*},i}$ for all $i = \{1,\hdots,M\}$ ensures~\eqref{eq:colavoidancerequirement}. Thus, collision avoidance can be achieved 
using the position constraint $\mathbb{P}_i = \bar{\mathbb{W}}_{\rahel{\bar{p}_w^*},i}$ in the tracking MPC \eqref{eq:nonlineartrackingmpcwithcolavoidance}.}
\noindent However, feasibility of the nonlinear tracking MPC \rahel{with collision avoidance now} explicitly depends on the partition $ \bar{\mathbb{W}}_{\bar{p}^*,i}$ and any update might cause infeasibility of optimization problem~\eqref{eq:nonlineartrackingmpcwithcolavoidance} or invalidate the closed-loop properties in Theorem~\ref{theorem:theo4boccia}. \rahel{For this reason, the time instant $w$ will differ from the general time index $k$: i.e., the Voronoi partitions will not be updated at each time step. The proposed decision rule ensuring recursive feasibility given a candidate partition works as follows.}
Let \rahel{$\mathbb{W}_{\bar{p}_{k}^*}$ be the candidate Voronoi update at the current time-step $k>w$,} 
and consider the following input and state sequences
\begin{subequations}
    \begin{align}
    \hat{u}_{i, \cdot \vert k} = &[u^{*}_{i,1\vert k}, \hdots, u^{*}_{i,N-2\vert k}, \bar{u}_{i,k}^{*}, \bar{u}_{i,k}^{*}]^{\top}\label{eq:uhat}, \\
    \hat{x}_{i, \cdot \vert k} = &[x^{*}_{i,1\vert k}, \hdots, x^{*}_{i,N-1\vert k}, f(x^{*}_{i,N-1\vert k},\bar{u}_{i,k}^{*}), \label{eq:xhat} \\ &f(f(x^{*}_{i,N-1\vert k},\bar{u}_{i,k}^{*}),\bar{u}_{i,k}^{*})]^{\top}.\nonumber
    \end{align}
    \label{eq:lemma1proposal}
\end{subequations}
\noindent The idea is to only update the partitions if the candidate sequences remain feasible, i.e., if for all $l=\{1,...,N+1\}$ and all $i=\{1,...,M\}$, 
\begin{equation}
    \hat{x}_{i,l \vert k}\in\mathbb{X}_i, \ C_{i}\hat{x}_{i,l \vert k} \in \bar{\mathbb{W}}_{\bar{p}_{k}^*,i} \ \text{and} \  C_{i}\bar{x}_{i,k}^{*} \in \bar{\mathbb{W}}^{\mathrm{int}}_{\bar{p}_{k}^*,i},
    \label{eq:feasibilitycond}
\end{equation}
which will be later used to ensure closed-loop properties (cf. proof of Theorem~\ref{theorem:twolayers} and~\ref{theorem:onelayer} in the Appendix). \rahel{By construction, a partition updating setpoint $\bar{p}$ remains in $\bar{\mathbb{W}}^{\mathrm{int}}_{\bar{p}_w,i}$, the interior of its respective partition. Thus, if the agents' positions $p$ get close enough to such setpoints $\bar{p}$, then these positions lie in the updated Voronoi partition. This is formalized in the following lemma.  
\begin{lemma}
    For any $\bar{p} \in \mathbb{A}^{M}$ satisfying $\Vert \bar{p}_i - \bar{p}_j \Vert \geq 2(r_{\mathrm{max}} + \epsilon)$ for all $i \neq j$, it holds that $\bar{p} \in \bar{\mathbb{W}}^{\mathrm{int}}_{\bar{p}}$. Furthermore, $\forall p$ with $\Vert \bar{p} - p \Vert \leq \epsilon$, it holds $p \in \bar{\mathbb{W}}_{\bar{p}}$.
\label{lemma:inclusioninnewinterior}
\end{lemma}}
\vspace{-0.0em}

\rahel{\noindent Consequently, starting from a configuration $p$ for which the distance between all agents is bigger than or equal to $2(r_{\max} + \epsilon)$, which we further refer to as an \textit{initially feasible configuration}, the presented update and collision avoidance framework, in combination with Theorem~\ref{theorem:theo4boccia}, provides consistent safety guarantees and keeps the MPC recursively feasible.}

\noindent 

\section{Two-layers Coverage MPC Algorithm}
\label{sec:twolayers}
The two-layers approach presented in this section builds upon the MPC scheme proposed in~\cite{Carron2017}. However, leveraging the results of Sections \ref{subsec:nonlineartrackingmpc} and \ref{sec:colavoidanceandrecursivefeasibility}, the presented set-up does not leverage terminal ingredients, allowing for the consideration of more complex dynamical systems and the inclusion of collision avoidance constraints. In Section~\ref{subsec:twolayers} we study the solution with known density, while in~\ref{subsec:twolayerslearning} the set-up is adjusted to include learning.

\vspace{-0.3em}
\subsection{Known Environment}
\label{subsec:twolayers}
In this subsection, we consider the density $\phi$ as known. The overall presented strategy encompasses two main tasks: (i) \rahel{centralized computation of the Voronoi partitions and their respective centroids, according to the agents' current setpoint position;} (ii) \rahel{solution of the tracking MPC scheme \eqref{eq:nonlineartrackingmpcwithcolavoidance} with collision avoidance strategy from Section~\ref{sec:colavoidanceandrecursivefeasibility} 
and the centroids as references.} After solving task (ii), a new Voronoi partition can be computed \rahel{as in (i)}, and the procedure can be repeated. \\
We now discuss the conditions that ensure convergence of this iterative set-up to \rrahel{a centroidal} Voronoi partition solving problem~\eqref{eq:cortescost}. For the first task, we require an update rule on the partitions, which is built upon~\cite[Proposition 3.3]{Cortes2004}:  
\begin{proposition}\cite[Proposition 3.3]{Cortes2004} Assume that $p_{0} \in \mathbb{A}^{M}$ and the Voronoi partition is updated in finite time.
Further consider a continuous mapping $T: \mathbb{A}^{M} \rightarrow \mathbb{A}^{M}$, with $\rahel{\bar{p}_{w_{+}}^* = T(\bar{p}_w^*)}$, where $w_{+}$ indicates the next update step. If the mapping \rahel{$T(\bar{p}_w^*)$} fulfills the following properties at each update step $w$:
\begin{itemize}
    \item $ \Vert \rahel{\bar{p}_{i,w_{+}}^*} - c_i(\mathbb{W}_{\rahel{\bar{p}_{w}^*},i}, \phi) \Vert \leq \Vert \rahel{\bar{p}_{i,w}^*} - c_i(\mathbb{W}_{\rahel{\bar{p}_{w}^*},i}, \phi) \Vert,$ \newline  $\forall i \in \{1, ..., M\}$;
    \item As long as the positions do not describe a centroidal Voronoi partition,
    $\exists j \in \{1, ..., M\}$ such that$ \newline \Vert \rahel{\bar{p}_{j,w_{+}}^*} - c_j(\mathbb{W}_{\rahel{\bar{p}_{w}^*},j}, \phi) \Vert < \Vert \rahel{\bar{p}_{j,w}^*} - c_j(\mathbb{W}_{\rahel{\bar{p}_{w}^*},j}, \phi) \Vert$,
\end{itemize}
then, \rahel{$\bar{p}_{w}^*$} converges to a centroidal Voronoi partition.
\label{proposition:cortesconvergence}
\end{proposition}

As integrator dynamics are able to reduce their Euclidean distance to a given reference at each time-step $k$, their partitions can be updated accordingly. However, as arbitrary nonlinear dynamics do not generally decrease the distance to their corresponding centroid for all instances in time, 
a partition update rule has to be applied to meet the requirements of Proposition~\ref{proposition:cortesconvergence} and allow for the construction of a suitable mapping $T$. For this purpose, we rely on the conditions proposed in~\cite[Section~4]{Carron2017}: \rahel{i.e., defining $e_{i,w} \! = \! \Vert \bar{p}_{i,w}^* \! - \! c_i(\mathbb{W}_{\bar{p}_{w}^*,i}, \phi) \Vert$ and for later use also their combining vector $e_{w}=[e_{1,w}, \hdots, e_{M,w}]$}, the partition is updated only if
\begin{subequations}
\begin{align}
    & \bullet \; \Vert \rahel{\bar{p}_{i,k}^*} - c_i(\mathbb{W}_{\rahel{\bar{p}_{w}^*},i}, \phi) \Vert \leq e_{i,w} \;\: \forall i \in \{1,..., M\}, \label{eq:updatereq1} \\
    & \bullet \; \exists j \! \in \! \{1,..., M\} \text{ s.t. } \Vert \rahel{\bar{p}_{j,k}^*} \! - \! c_j(\mathbb{W}_{\rahel{\bar{p}_{w}^*},j}, \phi) \Vert \! < \! e_{j,w}. \label{eq:updatereq2} 
\end{align}
\label{eq:updatereq}%
\end{subequations}
\rahel{Note that, for recursive feasibility, we additionally always require the conditions from Section \ref{sec:colavoidanceandrecursivefeasibility} to hold.}\\
We now study the second task, i.e., the tracking MPC having the centroids as references, and discuss its convergence. The optimization problem at time $k$ is defined as in~\eqref{eq:nonlineartrackingmpcwithcolavoidance}, \rahel{with position constraints \mbox{$\mathbb{P}_i = \mathbb{W}_{\bar{p}_{w}^*}$}, and reference $r = c(\mathbb{W}_{\bar{p}_{w}^*}, \phi)$ with  $\bar{p}_{w}^*$ set to the most recent steady-state position configuration that fulfilled~\eqref{eq:updatereq} and~\eqref{eq:feasibilitycond}. The cost function $J_{i}(x_{i}, \bar{x}_{i}, u_{i}, \bar{u}_{i}, r_{i})$ follows the definition in~\eqref{eq:generalnonlineartrackingmpccost}. To ensure convergence, we leverage the following condition on the offset cost $\ell_T$, adapted from the literature~\cite{Soloperto2021}.}
\begin{assumption}  
For every agent $i \in \{1, ..., M\}$, given any \rahel{$\bar{p}_i \in \mathbb{S}^{\mathrm{P}}_{\mathbb{A},i}$}, corresponding $\mathbb{W}_{\bar{p}}$ and reference vector $r \in \mathbb{A}^{M}$, we define the set \begin{equation}
\begin{split}
    \rahel{\mathbb{T}_{\bar{p},r,i}} = \{s\in \mathbb{S}_{\mathbb{W}_{\bar{p},i}} \: \vert \: \ell_{T,i}(C_{i}\bar{x} - r_{i}) = l_{\mathbb{W}_{\bar{p}},r,i,\min}\},
\end{split}
\label{eq:Tid2layer}
\end{equation}
which is the union of all setpoints resulting in a target cost value equal to $l_{\mathbb{W}_{\bar{p}},r,i,\min} = \min_{s\in \mathbb{S}_{\mathbb{W}_{\bar{p},i}}} \ell_{T,i}(C_i\bar{x}_i-r_i)$. Then, for every agent $i \in \{1,\hdots,M\}$ there are constants  \mbox{$\beta_{1,i}, \beta_{2,i} > 0$} such that, for any $\bar{s} := (\bar{x},\bar{u}) \in \mathbb{S}_{\mathbb{W}_{\bar{p},i}}$ and any $\epsilon^{\prime} \in [0,1]$, there exists a setpoint $\hat{s} := (\hat{x},\hat{u}) \in \mathbb{S}_{\mathbb{W}_{\bar{p},i}}$ satisfying 
\vspace{-0.8em}
\begin{subequations}
\begin{align}
    d_{i}(\hat{s}-\bar{s}) &\leq \beta_{1,i}\epsilon^{\prime} d_{i}(\bar{s})_{\rahel{\mathbb{T}_{\bar{p},r,i}}} \label{eq:assumption31twolayer} \\
    \ell_{T,i}(C_{i}\hat{x} \! - \! r_{i}) \! - \! \ell_{T,i}(C_{i}\bar{x}-r_{i}) &\leq \! - \! \beta_{2,i}\epsilon^{\prime} d_{i}(\bar{s})_{\rahel{\mathbb{T}_{\bar{p},r,i}}}^{2}, \label{eq:assumption32twolayer}%
\end{align}
\label{eq:assumption3twolayer}
\end{subequations} 
where $d_i(\cdot)$ is the distance function defined in~\eqref{eq:distance} and $d_{i}(\cdot)_{\rahel{\mathbb{T}_{\bar{p},r,i}}}$ follows definition~\eqref{eq:minsetdistance}. Furthermore, there is a constant  \mbox{$\beta_{T,i} > 0$} such that, for any $s := (\bar{x},\bar{u}) \in \mathbb{S}_{\mathbb{W}_{\bar{p}_w^*,i}}$,
\begin{align}
    \ell_{T,i}(C\bar{x}-r_i) - l_{\mathbb{W}_{\bar{p}},r,i,\min} \leq \beta_{T,i}\epsilon^{\prime} d_{i}(s)_{\rahel{\mathbb{T}_{\bar{p},r,i}}}
    \label{eq:upperboundtargetcost}
\end{align}
\label{assumption:steadystatedecreasetwolayers}%
\end{assumption}
\vspace{-1.0em}
\noindent Assumption~\ref{assumption:steadystatedecreasetwolayers} states that, for any setpoint $\bar{s}$ not being a minimizer of the target cost (i.e., resulting in a target cost value not equal to its attainable minimal value $l_{\mathbb{W}_{\bar{p}_w^*},r,i,\min}$), there exists another setpoint in its neighborhood such that the target cost can be actually reduced. Additionally, it claims that the target cost of an arbitrary setpoint is upper bounded by a function of its distance to the target cost's minimizing setpoint. Assumption~\ref{assumption:steadystatedecreasetwolayers} is generally satisfied by choosing a target cost of the same powers as the stage cost, and by having $\mathbb{S}^{\mathrm{P}}_{\mathbb{W}_{\bar{p}_w^*,i}}$ and $\mathbb{A}$ as polytopes. The latter condition is not restrictive for most dynamical systems of interest in this work. \\
The overall procedure consists in constructing the Voronoi partitions with respect to the agents' steady-state position and setting the references equal to their centroids. Then, the MPC defined with according position constraints is applied recursively until conditions~\eqref{eq:updatereq1},~\eqref{eq:updatereq2} and~\eqref{eq:feasibilitycond} are fulfilled and the partitions, as well as the centroids, are updated. It is summarized in Algorithm~\ref{alg:twolayermpcalg} and we state our main result in Theorem~\ref{theorem:twolayers}, whose proof can be found in Appendix~\ref{subsec:twolayersproof}.

\begin{algorithm}[h!]
\SetAlgoLined
Set $k = 0$, $w=0$ and \rahel{ construct $\mathbb{W}_{\bar{p}_{w}^{*}}$ with $\bar{p}_0^*=p_0$} \par
Set \rahel{$\mathbb{P}_i = \mathbb{W}_{\rahel{\bar{p}_{w}^*}}$,} $r = c(\mathbb{W}_{\rahel{\bar{p}_{w}^*}},\phi)$ and calculate $e_{w}$.\par
 \For{k=0,1,\dots}{
 \ForAll{$i \in \{1, \hdots,M\}$}{
 Solve~\eqref{eq:nonlineartrackingmpcwithcolavoidance} and obtain $\hat{x}_{i,\cdot \vert k}$, $\hat{u}_{i,\cdot \vert k}$, \rahel{ $\bar{x}^{*}_{i,k}$,  $u^{*}_{i,0\vert k}$.} \par
 Apply $u^{*}_{i,0\vert k}$ to obtain  $x_{i,k+1}$. \par}
 Construct $\mathbb{W}_{\rahel{\bar{p}_{k}^{*}}}$ according to \rahel{$\bar{p}_{k}^{*}$}. \par
\If{\textup{conditions~\eqref{eq:updatereq1},~\eqref{eq:updatereq2} and~\eqref{eq:feasibilitycond} are fulfilled}}{
   Set $w = k$ and update $\mathbb{W}_{\rahel{\bar{p}_{w}^*}}$ = $\mathbb{W}_{\rahel{\bar{p}_{k}}}$.\par
   Update \rahel{$\mathbb{P}_i = \mathbb{W}_{\rahel{\bar{p}_{w}^*}}$,} $r = c(\mathbb{W}_{\rahel{\bar{p}_{w}^*}},\phi)$ and calculate $e_{w}$.}
   }
 \caption{Two-Layers Coverage MPC}
 \label{alg:twolayermpcalg}
\end{algorithm}

\begin{theorem} 
Let Assumptions~\ref{assumption:dynamics},~\ref{assumption:expocostcontrollability},~\ref{assumption:boundedbyd}, and~\ref{assumption:steadystatedecreasetwolayers} hold, and consider a horizon length $N \geq N^{*}$. Additionally, suppose that at time step $k = 0$ the MPC problem described in~\eqref{eq:nonlineartrackingmpcwithcolavoidance} is feasible for all agents. Then, the overall control problem according to Algorithm~\ref{alg:twolayermpcalg} is recursively feasible, the agents do not collide and satisfy $x_{i,k} \in \mathbb{X}_{i},\; u_{i,k} \in \mathbb{U}_{i}, \forall i \in \{1,\hdots,M\}$, $\forall k \in \mathbb{N}$. Furthermore, the partition update condition given by~\eqref{eq:updatereq1},~\eqref{eq:updatereq2} and~\eqref{eq:feasibilitycond} is fulfilled after a finite amount of time, and the agents' position configuration converges to a centroidal Voronoi partition.
\label{theorem:twolayers}
\end{theorem}

\vspace{-0.7em}
\subsection{Unknown Environment}
\label{subsec:twolayerslearning}
To deal with an initially unknown $\phi$, the motion planning scheme in the tracking MPC considered above must be adjusted to balance density learning and coverage. In particular, the reference of each agent during exploration is chosen as the point of maximal variance within the Voronoi region\footnote{Finding the point of maximal variance over the whole Voronoi region implies evaluating~\eqref{eq:densityupdateblr} on an infinite number of points. In practice, we perform a sufficiently dense gridding of each $\mathbb{W}_{\rahel{\bar{p}_w^*},i}$ and evaluate variances on those points.} $\bar{\mathbb{W}}_{\rahel{\bar{p}_w^*},i}^{\mathrm{int}}$, denoted by $v_i(\mathbb{W}_{\rahel{\bar{p}_w^*},i}, \text{Var}) = \text{arg}\!\max_{\tilde{p} \in \bar{\mathbb{W}}_{\rahel{\bar{p}_w^*},i}^{\mathrm{int}}} \text{Var}(\tilde{p})$ and belonging to the vector $v(\mathbb{W}_{\rahel{\bar{p}_w^*}}, \text{Var}) = [v_1(\mathbb{W}_{\rahel{\bar{p}_w^*},1}, \text{Var}), \hdots, v_M(\mathbb{W}_{\rahel{\bar{p}_w^*},M}, \text{Var})]$. In case of exploitation, the reference is set to the current partition's centroid with respect to the available density estimate $\hat{\phi}_t$, i.e., $r_{i,k} = c_i(\mathbb{W}_{\rahel{\bar{p}_w^*},i}, \hat{\phi}_t)$. As proposed in~\cite{Todescato2017}, denoting with $\text{Var}_{\max} = \max_{p \in \mathbb{A}} \text{Var}(p)$ the maximal variance value computed over the set $\mathbb{A}$, and with $F: [0,\infty] \rightarrow [0,1]$ an arbitrary strictly monotonically increasing function such that \mbox{$F(\epsilon)=0 \Leftrightarrow \epsilon=0$}, the decision between exploration and exploitation is performed according to a Bernoulli random variable $\mathcal{B}(F(\text{Var}_{\max})) \in \{0,1\}$. The rationale is the following: at each time step, a sample from $\mathcal{B}$ is drawn whose probability of success (i.e., of returning a value equal to 1) is $F(\text{Var}_{\max})$. In this case, exploration is selected, and both the Voronoi tessellation and the references are kept fixed for all agents. The exploration modality continues regardless of the new samples of $\mathcal{B}(F(\text{Var}_{\max}))$ and the data-set is expanded as $I_{t+1} = I_{t} \cup [(m_{1,k},p_{1,k}), \hdots, (m_{M,k},p_{M,k})]$ at each time-step until the agents' positions are sufficiently close to the point of maximum variance of their Voronoi region, i.e., when $e_{v,i,k} = \Vert v_i(\mathbb{W}_{\bar{p}_w^*,i}, \text{Var}_t) - p_{i,k} \Vert \leq \rho$ for an arbitrarily small $\rho >0$. At this point, exploration modality is left and a new sample from $\mathcal{B}$ is drawn. As for the exploitation phase, Voronoi partitions are updated only if conditions~\eqref{eq:updatereq1},~\eqref{eq:updatereq2} and~\eqref{eq:feasibilitycond} are met. The overall procedure is summarized in Algorithm~\ref{alg:twolayermpcalglearningimp}. The main result is given in the following theorem that is proven in Appendix~\ref{subsec:twolayerslearningproof}.
\begin{algorithm}[h!]
\SetAlgoLined
 Set $k = 0$, $w=0$ and \rahel{construct $\mathbb{W}_{\bar{p}_{w}^{*}}$ with $\bar{p}_0^*=p_0$}\par
 Set $t = 1$, $\rho$, $\mu_{0}, \Sigma_{0}$ and $I_{t} = [(m_{1,k},p_{1,k}),..., (m_{M,k},p_{M,k})]$.\par
 Compute $\hat{\phi}_{t}(p), \text{Var}_{t}(p) \ \ \forall p \in \mathbb{A}$. \par
 Compute $c(\mathbb{W}_{\rahel{\bar{p}_{w}^{*}}}, \hat{\phi}_{t}), v(\mathbb{W}_{\rahel{\bar{p}_{w}^{*}}}, \text{Var}_{t}), \text{Var}_{\max,t}$ and $e_{w}$.\par
 Set exploration flag = false. \par
  \For{k=0,1, \dots}{
  \eIf{\textup{exploration flag} $\Vert$ \textup{(}$\mathcal{B}(F(\textup{Var}_{\textup{max},t})) == 1$\textup{)}}{
   Set exploration flag = true. \par
   \ForAll{$i \in \{1, \hdots, M \}$}{
   Set \rahel{$\mathbb{P}_i = \mathbb{W}_{\rahel{\bar{p}_{w}^*}}$,} $r_{i,k} = v_i(\mathbb{W}_{\rahel{\bar{p}_{w}^{*}},i}, \text{Var}_{t})$.\par
   Solve~\eqref{eq:nonlineartrackingmpcwithcolavoidance} and obtain $u^{*}_{i,0\vert k}.$ \par
   Apply $u^{*}_{i,0\vert k}$ to obtain $x_{i,k+1}$ and $e_{v,i,k+1}$. \par} 
   $I_{t+1} \! = \! {\footnotesize I_{t} \! \cup \![(m_{1,k+1},p_{1,k+1}),..., (m_{M,k+1},p_{M,k+1})]}$ \par
   Update~\eqref{eq:thetaupdate}-\eqref{eq:variancephiblr} $\forall p \in \mathbb{A}$. \par
   Update $c(\mathbb{W}_{\rahel{\bar{p}_{w}^{*}}}, \hat{\phi}_{t+1})$,$v(\mathbb{W}_{\bar{p}_{w}^{*}}, \! \text{Var}_{t+1}),\text{Var}_{\max,t+1}$.\par
   $t=t+1$ \par
   \If{$\|e_{v,i,k+1}\|\leq\rho$, 
   $\forall i \in \{1, \hdots, M \}$}
   {
   Set exploration flag = false. \par}
  }
  {\ForAll{$i \in \{1, \hdots, M \}$}{
  Set \rahel{$\mathbb{P}_i = \mathbb{W}_{\rahel{\bar{p}_{w}^*}}$,} $r_{i,k} = c(\mathbb{W}_{\rahel{\bar{p}_{w}^{*}},i}, \hat{\phi}_{t})$. \par
  Solve~\eqref{eq:nonlineartrackingmpcwithcolavoidance} and obtain $\hat{x}_{i,\cdot \vert k}$, $\hat{u}_{i,\cdot \vert k}$, \rahel{  $\bar{x}^{*}_{i,k}$,  $u^{*}_{i,0\vert k}$.} \par
 Apply $u^{*}_{i,0\vert k}$ to obtain  $x_{i,k+1}$. \par}
 Construct $\mathbb{W}_{\rahel{\bar{p}_{k}^{*}}}$ according to $\rahel{\bar{p}_{k}^{*}}$. \par
\If{\textup{conditions~\eqref{eq:updatereq1},~\eqref{eq:updatereq2} and~\eqref{eq:feasibilitycond} are fulfilled}}{
   $w = k$, $\mathbb{W}_{\rahel{\bar{p}_{w}^{*}}}$ = $\mathbb{W}_{\rahel{\bar{p}_{k}}} \xrightarrow{update}$$c(\mathbb{W}_{\rahel{\bar{p}_{w}^{*}}},\hat{\phi}_t)$, $e_{w}$.}}
   }
  \caption{Two-Layers, Learning-Based Coverage MPC}
 \label{alg:twolayermpcalglearningimp}
\end{algorithm}

\begin{theorem} 
Let Assumptions~\ref{assumption:dynamics},~\ref{assumption:expocostcontrollability},~\ref{assumption:boundedbyd},~\ref{assumption:bayesianconvergence}, and~\ref{assumption:steadystatedecreasetwolayers} hold and consider a horizon length $N \geq N^{*}$ and $\rho>0$ sufficiently small. Additionally, suppose that at time step $k = 0$ the MPC problem described in~\eqref{eq:nonlineartrackingmpcwithcolavoidance} is feasible for all agents. Then, the overall control problem according to Algorithm~\ref{alg:twolayermpcalglearningimp} is recursively feasible, the agents do not collide, and satisfy $x_{i,k} \in \mathbb{X}_{i}, u_{i,k} \in \mathbb{U}_{i}, \forall i \in \{1,\hdots,M\}$, $\forall k \in \mathbb{N}$. Furthermore, the partition update condition given by~\eqref{eq:updatereq1},~\eqref{eq:updatereq2} and~\eqref{eq:feasibilitycond} is fulfilled after a finite amount of time, and the agents' position configuration converges to a centroidal Voronoi partition.
\label{theorem:twolayerslearning}
\end{theorem}
\section{One-Layer Coverage MPC Algorithm}
\label{sec:onelayer}
In the one-layer approach, the reference is not pre-calculated and passed to the MPC, but the locational optimization function is jointly optimized with the MPC cost, cf. Figure~\ref{fig:coverageclarification}. The combined optimization is then expected to reduce the time and energy required for exploration.
\vspace{-0.5em}
\subsection{Known Environment}
\label{subsec:onelayer}
For the case of a known density $\phi$, the MPC optimization cost of the one-layer framework will encompass the same tracking cost $V_{N,i}$ considered in the previous sections, but the target cost $\ell_{T,i}$ will be equal to the corresponding summand of the locational optimization cost defined in~\eqref{eq:cortescost}. In particular, the latter is to be optimized with respect to the steady-state position $\bar{p}_{i,k} = C_i\bar{x}_{i,k} \in \mathbb{S}^{\mathrm{p}}_{\mathbb{W}_{\bar{p}^{*}_{w},i}}$ (see~\eqref{eq:steadystatesetwithcollavoidance}), and the integral is evaluated over the current Voronoi partition $\mathbb{W}_{\bar{p}^{*}_{w},i}$.
Thus, the resulting continuous objective reads as follows:
\begin{align}
&J_{i}(x_{i,\cdot \vert k}, u_{i,\cdot \vert k}, s_{i,k},\mathbb{W}_{\bar{p}^{*}_{w},i}, \phi) = \label{eq:fullnonlineartrackingmpccostonelayer} \\ & V_{N,i}(x_{i,\cdot \vert k}, u_{i,\cdot \vert k}, s_{i,k}) +   \ell_{T,i}(\bar{p}_{i,k},\mathbb{W}_{\bar{p}^{*}_{w},i}, \phi) = \notag\\ 
& \sum_{l=0}^{N-1}\ell_{i}(x_{i,l\vert k}, u_{i,l \vert k}, s_{i,k}) + \lambda \int_{\mathbb{W}_{\bar{p}^{*}_{w},i}} \Vert q - \bar{p}_{i,k} \Vert^{2}\phi(q)dq,\notag
\end{align}
with \rahel{$\lambda > 0$} representing a scaling factor. The overall MPC program is then stated as
\begin{equation}
\begin{split}
\min_{x_{i,\cdot \vert k}, u_{i,\cdot \vert k},s_{i,k}}&J_{i}(x_{i,\cdot \vert k}, u_{i,\cdot \vert k}, s_{i,k},\mathbb{W}_{\bar{p}^{*}_{w},i}, \phi) \\
&\eqref{eq:generalinit}-\eqref{eq:1toN-1withcolavoidance} ,
\end{split}
\label{eq:fullnonlineartrackingmpconelayer}
\end{equation}
and is solved within the iterative coverage scheme presented in Algorithm~\ref{alg:onelayeralg}. To show convergence to a centroidal Voronoi configuration while dealing with non-convex target costs, Assumptions~\ref{assumption:ballwithminconvex} and~\ref{assumption:steadystatedecreaseonelayer} are imposed.

\begin{algorithm}[h!]
\caption{One-Layer Coverage MPC}
\SetAlgoLined
Set $k = 0$, $w=0$ and 
construct $\mathbb{W}_{\bar{p}_{w}^{*}}$ with $\bar{p}_0^*=p_0$\par
\For{k=0,1, \dots}{\ForAll{$i \in \{1, \hdots, M \}$}{
 \rahel{Set $\mathbb{P}_i = \mathbb{W}_{\bar{p}_{w}^{*}}$} \par
 Solve~\eqref{eq:fullnonlineartrackingmpconelayer} and obtain  $\hat{x}_{i,\cdot \vert k}$, $\hat{u}_{i,\cdot \vert k}$,  $\bar{x}^{*}_{i,k}$,  $u^{*}_{i,0\vert k}$. \par
 Apply $u^{*}_{i,0\vert k}$ to obtain $x_{i,k+1}$. \par}
 Construct $\mathbb{W}_{\bar{p}_{k}^{*}}$ according to $\bar{p}_{k}^{*}$. \par
  \If{\textup{condition~\eqref{eq:feasibilitycond} is fulfilled}}{
  $w = k$, $ \mathbb{W}_{\bar{p}_{w}^{*}} = \mathbb{W}_{\bar{p}_{k}^{*}}^{'}$.}
 }
\label{alg:onelayeralg}
\end{algorithm}

\begin{assumption} For every agent $i \in \{1, ..., M\}$, given any $\bar{p} \in \mathbb{S}^{p}_{\mathbb{A},i}$ and any $\bar{p}' \in \mathbb{S}^{\mathrm{p}}_{\mathbb{W}_{\bar{p},i}}$ there exist an $r > 0$ such that defining $S^{\mathrm{p}}_{\bar{p}',\mathbb{W}_{\bar{p},i}} := \mathbb{B}_{r}^{D}(\bar{p}') \cap \mathbb{S}^{\mathrm{p}}_{\mathbb{W}_{\bar{p},i}},$ the set $\mathrm{argmin}_{p \in S^{\mathrm{p}}_{\bar{p}',\mathbb{W}_{\bar{p},i}}} \ell_{T,i}(p,\mathbb{W}_{\bar{p},i}, \rahel{\phi})$ is convex.
\label{assumption:ballwithminconvex}
\end{assumption}
\begin{assumption} 
Let $l_{\bar{p}',\mathbb{W}_{\bar{p}},i,\min} \! = \! \min_{p \in S^{\mathrm{p}}_{\bar{p}',\mathbb{W}_{\bar{p},i}}}\ell_{T,i}(p,\mathbb{W}_{\bar{p},i},\rahel{\phi})$, and define the corresponding set of local minimizers as
\begin{equation}
\begin{split}
    \rahel{\mathbb{T}_{\bar{p}',\bar{p},i}} \! = \! \{p \in S^{\mathrm{p}}_{\bar{p}',\mathbb{W}_{\bar{p},i}}\,\vert\, \ell_{T,i}(p,\mathbb{W}_{\bar{p},i},\rahel{\phi}) \! = \! l_{\bar{p}',\mathbb{W}_{\bar{p}},i,\min}\}.
\end{split}
\label{eq:Tid}
\end{equation}
Then, for every agent $i \in \{1,\hdots,M\}$, any $\bar{p}^{\prime} \in \mathbb{S}^{p}_{\mathbb{A},i}$ and any $\bar{s}' := (\bar{x}',\bar{u}') \in \mathbb{S}_{\mathbb{W}_{\bar{p},i}}$ there are constants $\beta_{1,i}, \beta_{2,i} > 0$ as well as a $\mathcal{K}$-function $\kappa$ such that, for any $\epsilon^{\prime} \in [0,1]$, there exists a setpoint $\hat{s} := (\hat{x},\hat{u}) \in S_{\bar{p}',\mathbb{W}_{\bar{p},i}}$ satisfying
\begin{subequations}
\begin{align}
    d_{i}(\hat{s}-\bar{s}') &\leq \beta_{1,i}\epsilon^{\prime} \kappa(\Vert \bar{p}' \Vert)_{\rahel{\mathbb{T}_{\bar{p}',\bar{p},i}}}
    \label{eq:assumption31},\\
    \ell_{T,i}(\hat{p},\mathbb{W}_{\bar{p},i},\rahel{\phi}) \! - \! \ell_{T,i}(\bar{p}',\mathbb{W}_{\bar{p},i},\rahel{\phi}) &\leq \! - \! \beta_{2,i}\epsilon^{\prime} \kappa(\Vert \bar{p}' \Vert)_{\rahel{\mathbb{T}_{\bar{p}',\bar{p},i}}}^{2} \label{eq:assumption32},
\end{align}
\label{eq:assumption3}
\end{subequations}
with $\hat{p} = C_i\hat{x}$ and $\bar{p}' = C_i\bar{x}'$. Furthermore, there is a constant  \mbox{$\beta_{T,i} > 0$} such that, for any $s := (\bar{x},\bar{u}) \in \mathbb{S}_{\mathbb{W}_{\bar{p},i}}$ it holds:
\begin{align}
    \ell_{T,i}(C\bar{x}-r_i,\mathbb{W}_{\bar{p},i},\rahel{\phi}) - l_{\bar{p}',\mathbb{W}_{\bar{p}},i,\min} \leq \beta_{T,i}\epsilon^{\prime} d_{i}(s)_{\rahel{\mathbb{T}_{\bar{p}',\bar{p},i}}}
    \label{eq:assumption33}
\end{align}
\label{assumption:steadystatedecreaseonelayer}
\end{assumption}
\vspace{-1.0em}
\noindent Assumption~\ref{assumption:ballwithminconvex} builds upon the fact that the constructed Voronoi partitions are never empty (see Section~\ref{sec:colavoidanceandrecursivefeasibility}) \rahel{and ensures convexity of $\rahel{\mathbb{T}_{\bar{p}',\bar{p},i}}$. Assumption~\ref{assumption:steadystatedecreaseonelayer} is a generalization of Assumption~\ref{assumption:steadystatedecreasetwolayers} to the case of non-convex target costs. This becomes crucial in the next section, when we consider non-convex target costs $\ell_{T,i}$ for the learning. Note that, in the chosen set-up, $\ell_{T,i}(\bar{p},\mathbb{W}_{\bar{p},i})$ is convex in $p$~\cite[Lemma 6.1]{Bullo2012} and the set of minimizers is convex, i.e., Assumption~\ref{assumption:ballwithminconvex} holds with $ S^{\mathrm{p}}_{\bar{p}',\mathbb{W}_{\bar{p},i}} =\mathbb{S}^{\mathrm{p}}_{\mathbb{W}_{\bar{p},i}}$. } In accordance to~\cite[Remark 1]{Soloperto2021}, Assumption~\ref{assumption:steadystatedecreasetwolayers} holds taking functions $d$ and $\kappa$ quadratic if $\ell_{T,i}$ is strongly convex quadratic and sets are polytopic. \\
\noindent Next, we state the theorem collecting all the theoretical guarantees of the proposed scheme. The proof is presented in Appendix~\ref{subsubsec:onelayerknownenvironmenttheory}.
\begin{theorem} Let Assumptions~\ref{assumption:dynamics},~\ref{assumption:expocostcontrollability},~\ref{assumption:boundedbyd},~\ref{assumption:ballwithminconvex} and~\ref{assumption:steadystatedecreaseonelayer} hold, and consider a horizon length $N \geq N^{*}$. Additionally, suppose that at time step $k = 0$ the MPC problem described in~\eqref{eq:fullnonlineartrackingmpconelayer} is feasible for all agents. Then, in accordance to Algorithm~\ref{alg:onelayeralg} it remains recursively feasible, the agents do not collide and satisfy $x_{i,k} \in \mathbb{X}_{i},\; u_{i,k} \in \mathbb{U}_{i}, \forall i \in \{1,\hdots,M\}$, $\forall k \in \mathbb{N}$. Furthermore, the partition update condition given by~\eqref{eq:feasibilitycond} is fulfilled after a finite amount of time, and the agents' position configuration converges to a centroidal Voronoi partition.
\label{theorem:onelayer}
\end{theorem}

\vspace{-0.8em}
\subsection{Unknown Environment}
\label{subsec:onelayerlearning}
The adaptation of the one-layer approach to an unknown environment involves the following modification of the target cost $\ell_{T,i}$ entering~\eqref{eq:fullnonlineartrackingmpconelayer}, inspired by the Upper Confidence Bound method in Bayesian Optimization~\cite{Auer2002,Srinivas2010}. Specifically, the uncertainty on a potential setpoint $\bar{p}_i$, quantified by the variance $\text{Var}(\bar{p}_i)$ and scaled by a parameter $S > 0$, is subtracted from~\eqref{eq:fullnonlineartrackingmpconelayer}. Hence, the target cost now reads as
\begin{equation}
\begin{split}
&\ell_{T,i}(\bar{p}_{i}, \mathbb{W}_{\bar{p}^{*}_{w},i}, \hat{\phi}, \text{Var}) = \\ & \lambda \left(\int_{\mathbb{W}_{\bar{p}^{*}_{w},i}} \Vert q - \bar{p}_{i} \Vert^{2}\hat{\phi}(q)dq - S\cdot \text{Var}(\bar{p}_{i}) \right).
\end{split}
\label{eq:fullnonlineartrackingmpccostonelayerlearning}
\end{equation}
Apart from this modification of the cost and the data collection, the overall procedure follows the one proposed in Section~\ref{subsec:onelayer} and is summarized in Algorithm~\ref{alg:onelayerlearningalg}. Note that~\eqref{eq:fullnonlineartrackingmpccostonelayerlearning} is non-convex, so Assumptions~\ref{assumption:ballwithminconvex} and~\ref{assumption:steadystatedecreaseonelayer} become crucial. However, they hold with respect to the modified target cost $\ell_{T,i}$ defined in~\eqref{eq:fullnonlineartrackingmpccostonelayerlearning}, given its continuity with respect to $\bar{p}_i$\footnote{Given arbitrary positions $p_1$, $p_2$ and $q \in \mathbb{A}$, by continuity of norms we have that $\lim_{p_1 \rightarrow p_2} \|q-p_1\|^2-\|q-p_2\|^2 =0$. By considering the locational cost difference $\lim_{p_1\rightarrow p_2} H_{i}(p_{1}, \mathbb{W}_{\bar{p},i}) - H_{i}(p_{2}, \mathbb{W}_{\bar{p},i})$, and noting that the limit of the integrand exists and is integrable, conclusion follows by swapping limit and integration operations. }
Also, Assumption~\ref{assumption:ballwithminconvex} seems natural in such a scenario, ensuring the existence of a direction in which $\ell_{T,i}(\bar{p}',\mathbb{W}_{\bar{p},i})$ is decreasing and fulfilling equation~\eqref{eq:assumption32} as long as $\bar{p}' \notin \mathbb{T}_{\bar{p}',\bar{p},i}$. 

\begin{algorithm}[h!]
\caption{One-Layer, Learning-Based Coverage MPC}
\SetAlgoLined
Set $k = 0$, $w = 0$ and construct $\mathbb{W}_{\bar{p}^{*}_{w}}$ with $\bar{p}_0^*=p_0$\par
Set $t = 1$, $\mu_{0}$, $\Sigma_{0}$ and $I_{t} = [(m_{1,k},p_{1,k}), \hdots, (m_{M,k},p_{M,k})]$. \par 
Update $\hat{\phi}_{t}(p), \text{Var}_{t}(p) \ \ \forall p \in \mathbb{A}$. \par
 \For{k = 0,1,\dots}{\ForAll{$i \in \{1, \hdots, M \}$}{
  Set \rahel{$\mathbb{P}_i = \mathbb{W}_{\bar{p}_{w}^{*}}$ and}  $\ell_{T,i}$ = $\ell_{T,i}(\bar{p}, \mathbb{W}_{\bar{p}^{*}_{w},i}, \hat{\phi}_{t}, \text{Var}_{t})$ \par
  Solve \eqref{eq:fullnonlineartrackingmpconelayer} and obtain  $\bar{x}^{*}_{i,k}, u^{*}_{i,0\vert k}, \hat{u}_{i,\cdot \vert k}$, $\hat{x}_{i,\cdot \vert k}$. \par
  Apply $u^{*}_{i,0\vert k}$ to obtain $x_{i,k+1}$. \par}
 $I_{t+1} = I_{t} \cup [(m_{1,k},p_{1,k}), \hdots, (m_{M,k},p_{M,k})]$ \par
 Update~\eqref{eq:thetaupdate}-\eqref{eq:variancephiblr} $\forall p \in \mathbb{A}$. \par
 Construct $\mathbb{W}^{'}_{\bar{p}_{k}^{*}}$ according to $\bar{p}_{k}^{*}$. \par
  \If{\textup{condition~\eqref{eq:feasibilitycond} is fulfilled}}{
  $w = k$, $ \mathbb{W}_{\bar{p}_{w}^{*}} = \mathbb{W}_{\bar{p}_{k}^{*}}^{'}$.} $t=t+1$
 }
\label{alg:onelayerlearningalg}
\end{algorithm}

\noindent Define the set of feasible positions in the MPC problem of agent $i$ at time $k$ by $\mathbb{F}_{i,k}=\{p\in\mathbb{A}|~\exists (x_{\cdot|k},u_{\cdot|k},s_k)$ s.t.~\eqref{eq:fullnonlineartrackingmpconelayer} is feasible with $[C_i,0]s_k=p\}$ and accordingly $\mathbb{F}_{k}=\bigcup_{i=1}^{M}\mathbb{F}_{i,k}$. 
The theoretical results are summarized in the following theorem, which is proven in Appendix~\ref{subsec:onelayerunknownenvironmenttheory}. 
\begin{theorem} Let  Assumptions~\ref{assumption:dynamics},~\ref{assumption:expocostcontrollability},~\ref{assumption:boundedbyd},~\ref{assumption:bayesianconvergence},~\ref{assumption:ballwithminconvex} and~\ref{assumption:steadystatedecreaseonelayer} hold regarding the target cost defined in~\eqref{eq:fullnonlineartrackingmpccostonelayerlearning}, and consider a horizon length $N \geq N^{*}$. Additionally, suppose that at time step $k = 0$, the MPC problem described in~\eqref{eq:fullnonlineartrackingmpconelayer} is feasible for all agents. Then, the overall control problem according to Algorithm~\ref{alg:onelayerlearningalg} is recursively feasible, the agents do not collide and satisfy $x_{i,k} \in \mathbb{X}_{i},\; u_{i,k} \in \mathbb{U}_{i}, \forall i \in \{1,\hdots,M\}$, $\forall k \in \mathbb{N}$. The partition update condition given by~\eqref{eq:feasibilitycond} is fulfilled after a finite amount of time, the density estimate converges in probability, and the agents' position configuration converges to a centroidal Voronoi partition with respect to the converged density estimate $\hat{\phi}_{\infty}$. Furthermore, there exists a uniform constant $\Delta H\geq 0$, such that
$\sup_{p\in\mathbb{F}_k}\lim_{k\rightarrow\infty} \mathrm{Var}_k(p)\leq \frac{ \Delta H}{S}$, in probability. 
\label{theorem:onelayerlearning}
\end{theorem}
\vspace{-0.1em}
\noindent In accordance to Theorem~\ref{theorem:onelayerlearning}, the remaining uncertainty by the time of convergence can be tuned by altering the variance scaling factor $S$ introduced in equation~\eqref{eq:fullnonlineartrackingmpccostonelayerlearning}, as well as the controller horizon length $N$, and hence the set of feasible positions $\mathbb{F}_{i,k}$.
\vspace{-0.3em}
\section{Experimental Results}
\label{sec:experiments}
In the following section, we introduce the mathematical set-up of our experiments (Section~\ref{subsec:mathematicalsetup} and~\ref{subsec:expermpccost}) and provide details on the used software and hardware framework (Section~\ref{subsec:hardwaredetails}). Furthermore, the obtained experimental hardware results for both control architectures (one-layer and two-layers) and for both known and unknown environments are presented in Sections~\ref{subsec:experimenttwolayer}-~\ref{subsec:experimentonelayerlearning}.
\vspace{-0.6em}
\subsection{Mathematical Set-up}
\label{subsec:mathematicalsetup}
We consider a fleet of $M$=4 cars that covers an area $\mathbb{A}= [-1.55,1.55] \times [\rahel{-1.55,1.55}]$ meters. The nonlinear continuous dynamics for the $i-$th car are given by the kinematic bicycle model~\cite{Rajamani2012}, 
\begin{equation}
\begin{bmatrix}
    \dot{x}_{p,i} \\
    \dot{y}_{p,i} \\
    \dot{\psi_{i}}\\
    \dot{\delta_{i}}
    \end{bmatrix} = 
\begin{bmatrix}
    \cos{(\psi_{i})} \\
    \sin{(\psi_{i})} \\
    \frac{1}{L_i} \cdot \tan{(\delta_{i})}\\
    0
\end{bmatrix} u_{v,i} + 
\begin{bmatrix}
    0 \\
    0 \\
    0\\
    1
\end{bmatrix} u_{d,i} = g_{1,i} u_{v,i} + g_{2,i} u_{d,i}
\label{eq:dynamics}
\end{equation}
where $p_{i} =[x_{p,i},y_{p,i}]^{\top} \in \mathbb{R}^{2}$ is the position in Cartesian coordinates, and~$\psi_{i}$ its orientation with respect to the $x$-axis. The orientation of the wheels with respect to their neutral position is given by~$\delta_{i}$. Steering~$u_{d,i}$ and velocity~$u_{v,i}$ are available as inputs. Lastly, we denote with $L_i$ the wheelbase length. The discrete model is obtained using the forward Euler method, with a sampling time $T_{s}$. It is set to 0.05 seconds for all experiments, except for the one-layer, learning-based approach, where it has been set to 0.1 seconds due to the increased complexity of its target cost. The coverage is performed in accordance to density $\phi_{1}$ in case of a known environment, and $\phi_{2}$ for the experiments in which the environmental information is initially unknown, where
\begin{align}
    &\phi_{1}(x_p,y_p) = 5.0e^{-0.5((x_{p}-1.4)^2 + (y_{p}-1.7)^2)} \\
    &\phi_{2}(x_p,y_p) = -0.5x^{2} - 0.5y^{2} - 0.5x - 0.5y + 12. 
\end{align}
The cars are able to collect noisy measurements of the density at their current location according to~\eqref{eq:measmod}, with $\nu_{i,h} \sim \mathcal{N} (0,0.1) \ \forall i \in \{1,...,4\}$. For the application of the Bayesian linear regression, the feature vector is chosen as $\Phi = [x_p^2  \: y_p^2  \: x_p  \: y_p  \: 1]^{\top}$ and the Gaussian prior is set to $\mu_0 = [0  \: 0  \: 0  \: 0  \: 0]^{\top}$ and $\Sigma_0 = \mathbb{I}$.

\subsection{MPC Cost Function}
\label{subsec:expermpccost}
In consideration of the non-holonomic system dynamics presented in Section~\ref{subsec:mathematicalsetup}, the stage cost of the MPC is designed according to~\cite{Rosenfelder2021}. Therefore, the Lie brackets of the vectors $g_{1,i}$ and $g_{2,i}$, introduced in \eqref{eq:dynamics}, given by
\begin{align}
    & g_{3,i} := [g_{1,i},g_{2,i}] = \begin{bmatrix} 0 & 0 & \frac{1}{L_i}(\tan^{2}{(\delta_{i})} + 1) & 0 \end{bmatrix}^{\top} \notag \\
    & g_{4,i} := [g_{1,i},g_{3,i}] = \begin{bmatrix} \frac{\sin{(\psi_{i})}}{L_i(\sin^{2}{(\delta_{i})}-1)} & \frac{\cos{(\psi_{i})}}{L_i\cos^{2}{(\delta_{i})}} & 0 & 0 \end{bmatrix}^{\top}, \notag
\end{align}
are used to build a stage cost that allows for parallel parking manoeuvres. Concerning a full dimensional state and input reference, indicated by $x_{r,i}$ and $u_{r,i}$, the stage cost reads as
\begin{align}
    &\ell_{i} = \sum_{j = 1}^{4}Q_{j}(g_{j,i}^{\top}(x_{r})(x_{i}-\bar{x}_{i}))^{\rho_{j}} + \sum_{z = 1}^{2}R_{z}(u_{z,i}-\bar{u}_{z,i})^{\rho_{u}}. \notag 
\end{align}
By setting~$\rho_{1}= \rho_{2} = \rho_{u} = 12$,~$\rho_{3} = 6 $ and~$\rho_{4} = 4$, the stage cost satisfies the cost-controllability condition in Assumption~\ref{assumption:expocostcontrollability}, as shown in~\cite{Rosenfelder2021} based on~\cite{Coron2020}. Hence, the derived stability guarantees are applicable. The reference location is either set to the pre-calculated reference (two-layers approaches), or the previously calculated setpoint, (one-layer approaches). The reference in orientation,~$\psi_{r,i}$, is set to the angle enclosed by the x-axis and the error vector to the reference, and~$\delta_{r,i}$ is chosen equal to zero for all agents. While the stage cost is used for all of the presented approaches, the implemented target costs differ. In consideration of Assumption~\ref{assumption:steadystatedecreasetwolayers}, the target cost of the one-layer approaches is formed using the same structure,
\begin{align}
    &\ell_{T,i} \! = \! \sum_{j = 1}^{4}Q_{j}^{\prime}(g_{j,i}^{\top}(x_{r})(\bar{x}_{i}-x_{r,i}))^{\rho_{j}} \! + \! \sum_{z = 1}^{2}R_{z}^{\prime}(\bar{u}_{z,i}-u_{r,z,i})^{\rho_{u}}. \notag
\end{align}
For the one-layer approaches, the target cost follows the structures described in Sections~\ref{subsec:onelayer} and~\ref{subsec:onelayerlearning}. To speed up computation we approximate the included integral with a quadratic function whose coefficients are learned using Bayesian linear regression. The controller's horizon length is set to $N=30$ for all conducted experiments except for the one-layer learning-based approach where, due to the doubled sampling time, it is divided in half and set to $N=15$.

\subsection{Software and Hardware Details}
\label{subsec:hardwaredetails}
For the experiments, the miniature RC cars of Chronos in combination with the CRS software framework are used~\cite{carron2022}. The vehicles' position, velocity and orientation feedback is provided by a \rahel{Qualisys} motion capture system, and the control inputs are transmitted using WiFi-connected micro-controllers. \rahel{A HP NotebookPro with 32 GB RAM and a Dual-Core Intel Core i7 processor is used as a server. Its operating system is Ubuntu 22.04.} The closed-loop system is implemented using ROS (Robotics Operating System) embedded in a C++/ Python framework using ACADOS as a solver~\cite{Verschueren2019,Verschueren2018}. The approximate initial positions and orientations, as well as the car's wheelbase length, are given in Table \ref{tab:model_param}. The radius covering each agent $i$ is set to its length, i.e., $r_{i,\max} = L_i$.

\begin{table}[h]
\begin{center}
\vspace{1ex}
\begin{tabular}{l|cccccc}
\hline car & $x_{p,i,0}$ & $y_{p,i,0}$ & $\psi_{i,0}$ & $\delta_{i,0}$ & $L_{i}$ \\ \hline  \hline
1  & \rahel{-1.30m} & \rahel{-1.36m} & 0.785 rad & 0 rad & \rahel{0.099m} \\
2 & \rahel{-0.96m} & \rahel{-0.92m} & 0.785 rad & 0 rad & \rahel{0.099m} \\
3 & \rahel{-0.87m} & \rahel{-1.33m} & 0.000 rad & 0 rad & \rahel{0.115m} \\
4  & \rahel{-1.34m} & \rahel{-0.92m} & 1.570 rad & 0 rad & \rahel{0.115m}  \\
\hline
\end{tabular}
\caption{Values of initial configuration and model parameters of miniature RC cars for all conducted experiments.}
\label{tab:model_param}
\end{center}
\end{table}

\vspace{-2.6em}
\subsection{Results Two-Layers Coverage MPC}
\label{subsec:experimenttwolayer}
\begin{figure} [h]
\centering
\input{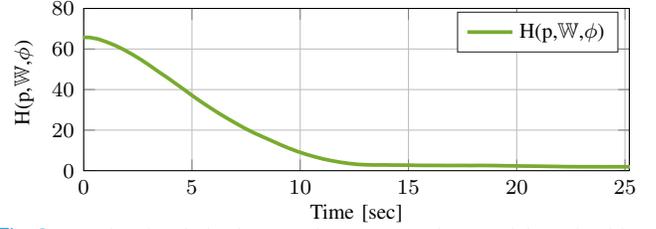}
\vspace{-0.7em}
\caption{\rahel{Locational optimization cost decrease over time, applying Algorithm~\ref{alg:twolayermpcalg} with respect to a known density $\phi_{1}$.}}
\label{fig:twolayercoveragecost}
\vspace{-0.6em}
\end{figure}

\begin{figure}[h!]
\begin{minipage}[t]{0.15\textwidth}
\centering
\includegraphics[trim={5.2cm 8.1cm 4.6cm 7.7cm},clip, width = 0.8\textwidth]{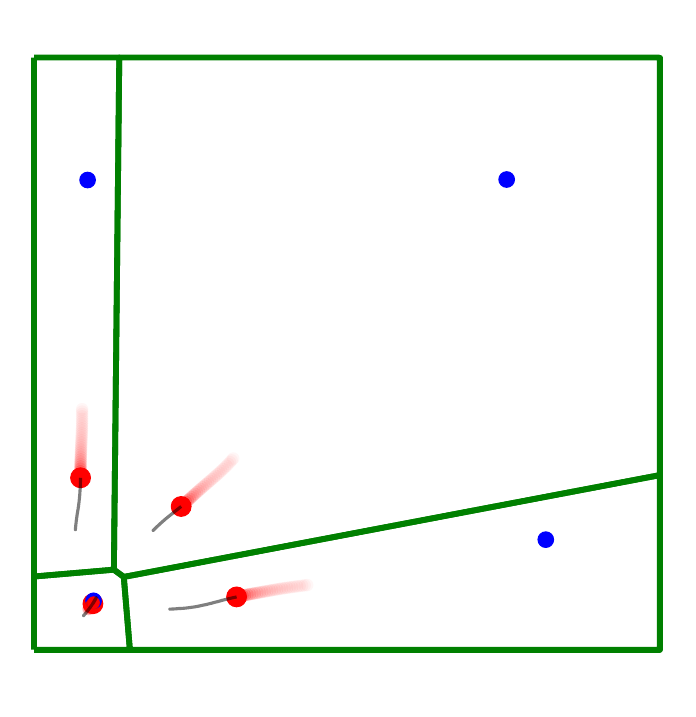}
\centering
\end{minipage}
\hfill
\begin{minipage}[t]{0.15\textwidth}
\centering
\includegraphics[trim={5.2cm 8.1cm 4.6cm 7.7cm},clip, width = 0.8\textwidth]{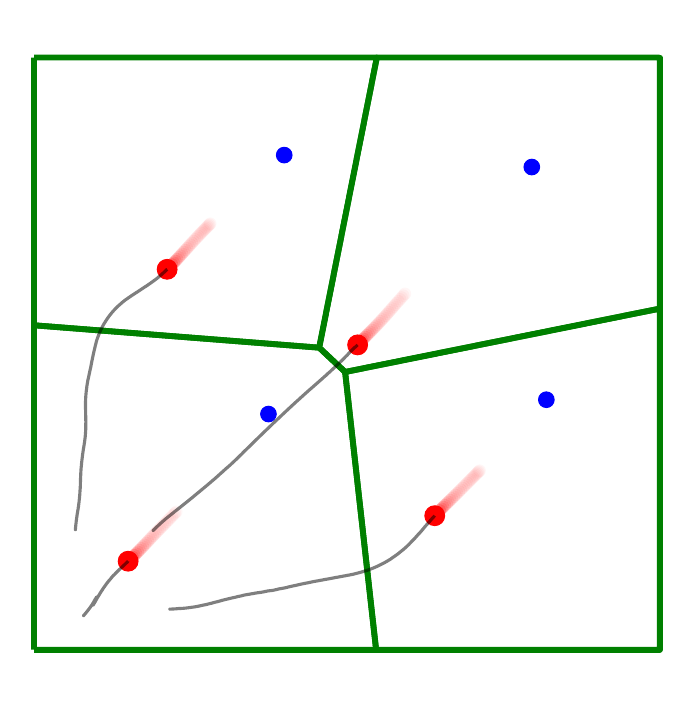}
\centering
\end{minipage}
\hfill
\begin{minipage}[t]{0.15\textwidth}
\centering
\includegraphics[trim={5.2cm 8.1cm 4.6cm 7.7cm},clip, width = 0.8\textwidth]{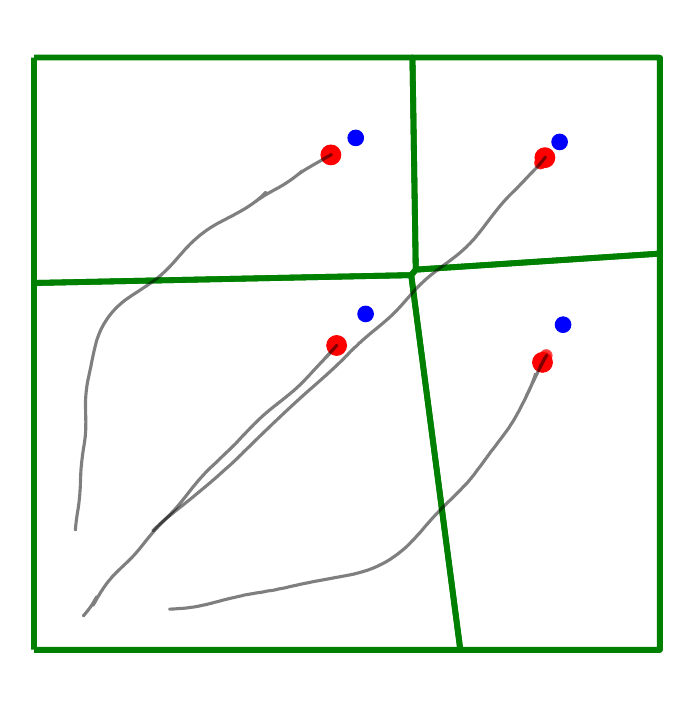}
\centering
\end{minipage}
\caption{\rahel{Configurations of cars at 2, 8, and 24 seconds applying Algorithm~\ref{alg:twolayermpcalg} in the described set-up. The agents' location and their predicted trajectory are given in red, the Voronoi partitions in green, their centroids in blue, and the traveled paths are visualized in light grey.}}
\label{pics:twolayersconfig}
\vspace{-0.8em}
\end{figure}

As shown in Figure~\ref{fig:twolayercoveragecost}, applying Algorithm~\ref{alg:twolayermpcalg} results in a decrease of the locational optimization cost $H(p,\mathbb{W})$ defined in~\eqref{eq:cortescost}. We further display in Figure~\ref{pics:twolayersconfig} the agents' location, traveled paths, and predicted trajectories, as well as the current Voronoi partitions and their centroids, for three instances of time during operation. \rahel{It can thus be seen that the locational optimization cost is decreasing}.    
\vspace{-0.9em}
\subsection{Results Two-Layers, Learning-Based Coverage MPC}
\label{subsec:experimenttwolayerlearning}
\begin{figure} [h!]
\centering
\input{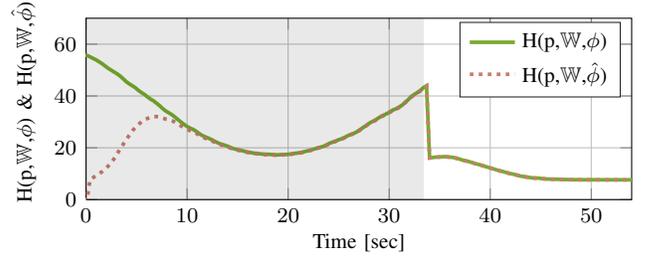}
\vspace{-0.7em}
\caption{\rahel{Locational optimization cost (green) as well as estimated locational optimization cost versus time (rose), applying Algorithm~\ref{alg:twolayermpcalglearningimp} in consideration of an initially unknown $\phi_{2}$. The grey background indicates time instances for which the agents are exploring.}}
\label{fig:twolayerlearningcoveragecost}
\vspace{-0.8em}
\end{figure}

\begin{figure}[h!]
\begin{minipage}[t]{0.15\textwidth}
\centering
\includegraphics[trim={5.2cm 8.1cm 4.6cm 7.7cm},clip, width = 0.8\textwidth]{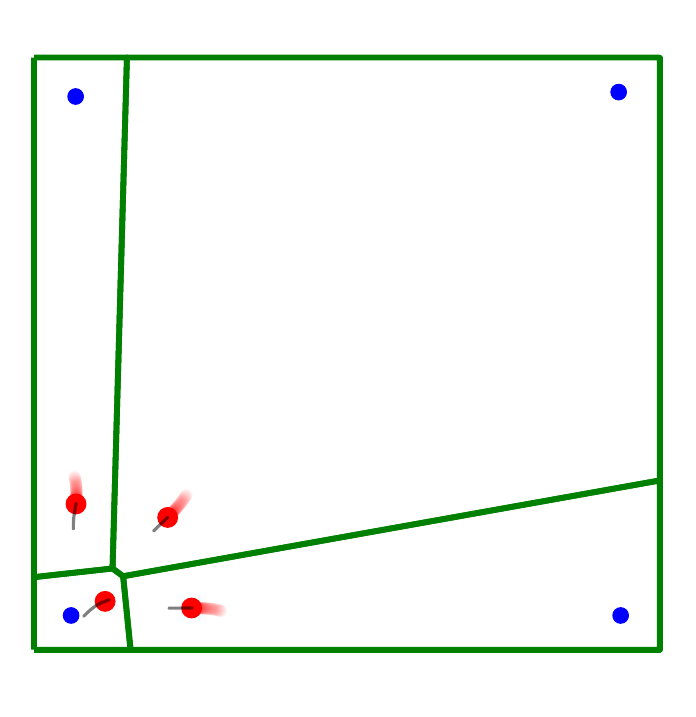}
\centering
\end{minipage}
\hfill
\begin{minipage}[t]{0.15\textwidth}
\centering
\includegraphics[trim={5.2cm 8.1cm 4.6cm 7.7cm},clip, width = 0.8\textwidth]{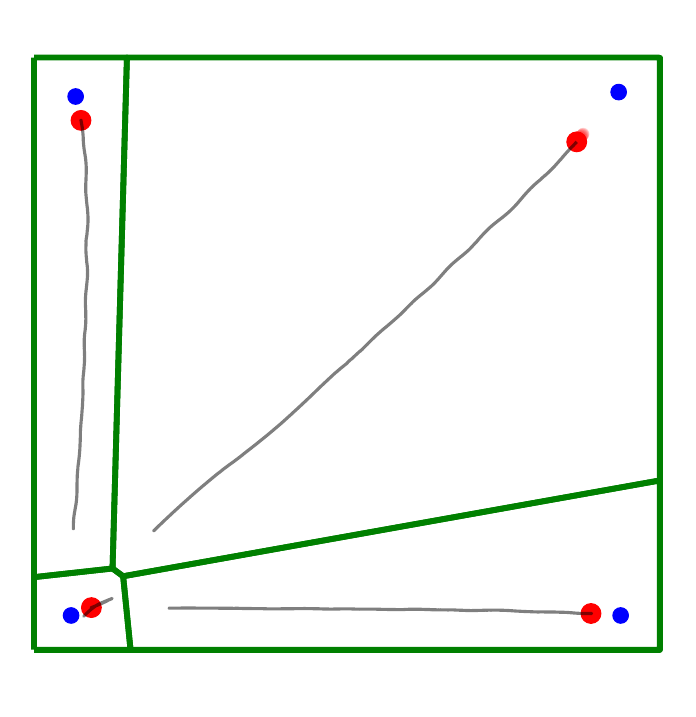}
\centering
\end{minipage}
\hfill
\begin{minipage}[t]{0.15\textwidth}
\centering
\includegraphics[trim={5.2cm 8.1cm 4.6cm 7.7cm},clip, width = 0.8\textwidth]{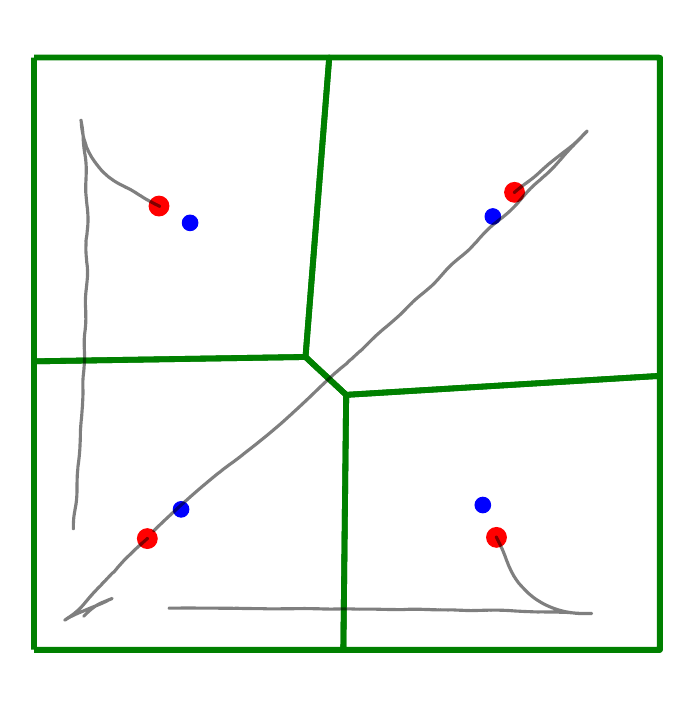}
\centering
\end{minipage}
\caption{\rahel{Configurations of cars at 2, 33, and 53 seconds applying  Algorithm~\ref{alg:twolayermpcalglearningimp} in the described set-up. The agents' location and their predicted trajectory are given in red, the Voronoi partitions in green, the references in blue, and the traveled paths are visualized in light grey. While the first two instances visualize the initial exploration movement, the third shows a covering behavior.}}
\label{pics:twolayerslearningconfig}
\vspace{-0.4em}
\end{figure}

Figure~\ref{fig:twolayerlearningcoveragecost} shows the locational optimization cost and the estimated cost when applying Algorithm~\ref{alg:twolayermpcalglearningimp} with the initially unknown density~$\phi_2$. Further, Figure~\ref{pics:twolayerslearningconfig} shows the obtained configurations at three instances in time.  

\begin{figure} [h!]
\centering
\input{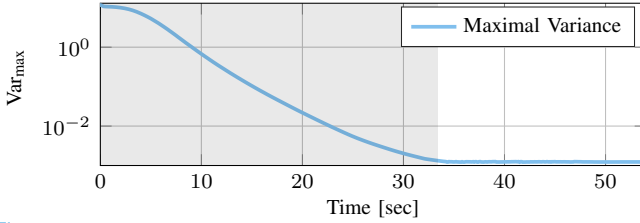}
\vspace{-0.7em}
\caption{\rahel{Maximal variance decrease over time applying Algorithm~\ref{alg:twolayermpcalglearningimp} in consideration of an initially unknown $\phi_{2}$. The grey background indicates time instances for which the agents are exploring.}}
\label{fig:twolayerslearningmaxvar}
\vspace{-0.7em}
\end{figure}

\begin{figure} [h!]
\centering
\input{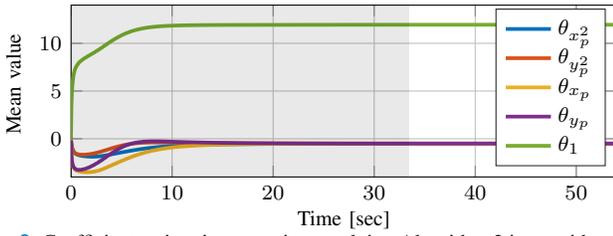}
\vspace{-0.7em}
\caption{\rahel{Coefficient estimation over time applying Algorithm~\ref{alg:twolayermpcalglearningimp} in consideration of an initially unknown $\phi_{2}$. The grey background indicates time instances for which the agents are exploring.}}
\label{fig:twolayerslearningcoef}
\vspace{-1.0em}
\end{figure}

\noindent Defining~$F(\text{Var}_{\max,t}) = \frac{\text{Var}_{\max,t}}{\text{Var}_{\max,0}}$, the first decision in Algorithm~\ref{alg:twolayermpcalglearningimp} is guaranteed to be exploration, and the agents drive towards the point of maximal variance within their initial Voronoi partition. Allowing for a better insight into the learning progress, the decreasing behaviour of the maximal variance over time is presented in Figure~\ref{fig:twolayerslearningmaxvar}. In Figure~\ref{fig:twolayerslearningcoef}, the development of estimated parameter vector $\theta$ is visualized.
In Figure~\ref{fig:twolayerlearningcoveragecost}, as well as in Figures~\ref{fig:twolayerslearningmaxvar} and~\ref{fig:twolayerslearningcoef}, the time instances for which the cars are exploring are indicated with a light grey background. It can be seen that, for the first \rahel{33} seconds, the agents keep exploring: correspondingly, any decrease or increase in cost is only accidental. However, by collecting measurements at each time step during exploration, after \rahel{15} seconds the mean estimate of $\theta$ is quite precise and further improvements are minor. This is also reflected in the small remaining maximal variance of the estimate. By the time the agents leave exploration mode, all mean estimates of the true coefficient $\theta$ show a maximal error of less than \rahel{0.06} and a maximal variance of approximately $10^{-3}$. Due to the remaining small uncertainty, by the time the experiment is interrupted the probability of another exploration movement for future instances is strictly positive, but very small.

\vspace{-0.4em}
\subsection{Results One-Layer Coverage MPC}
\label{subsec:experimentonelayer}
\begin{figure} [h!]
\centering
%
%
\definecolor{mycolor1}{rgb}{0.47000,0.67000,0.19000}%
\definecolor{mycolor2}{rgb}{0.89000,0.59000,0.53000}%
\begin{tikzpicture}

\begin{axis}[%
height=0.85in,
width=0.40\textwidth,
yshift=0.8cm,
at={(0.0in,0.0in)},
scale only axis,
xmin=0,
xmax=22,
xlabel style={font=\color{white!15!black}},
xlabel style={font=\footnotesize},
xlabel style={yshift=0.6ex,},
xlabel={Time [sec]},
ymin=0,
ymax=70,
ylabel style={font=\color{white!15!black}},
ylabel style={font=\footnotesize},
ylabel style={xshift=0.6ex,},
ylabel={H(p,$\mathbb{W}$) \& H($\bar{p}$,$\mathbb{W}$)},
axis background/.style={fill=white},
tick label style={font=\footnotesize},
xmajorgrids,
ymajorgrids,
legend style={legend cell align=left, align=left, draw=white!15!black},
legend style={font=\footnotesize}
]
\addplot [color=mycolor1, line width=1.4pt]
  table[row sep=crcr]{%
-1.14863276481628	65.562193201352\\
-0.9499192237854	65.5624347419556\\
-0.74800181388855	65.5609329674382\\
-0.548374652862549	65.2822277839491\\
-0.348526477813721	65.0303386888569\\
-0.146867752075195	64.3525586599515\\
0.0536508560180664	63.3171338541951\\
0.254810094833374	62.328330128019\\
0.452388525009155	61.3898731351064\\
0.653592109680176	60.3493058384483\\
0.855054140090942	59.2457307249114\\
1.05077505111694	58.3207244305405\\
1.25211906433105	57.2236254489755\\
1.45325183868408	56.0760044833987\\
1.65678715705872	54.8605719037616\\
1.85086393356323	53.773023367262\\
2.05186653137207	52.5128513380267\\
2.2526798248291	51.2926420275172\\
2.45258402824402	50.0099449309508\\
2.65606117248535	48.7309763530255\\
2.86030268669128	47.6195702193418\\
3.05085372924805	46.4754120151667\\
3.25225734710693	45.1484445480075\\
3.45727562904358	43.8475214178303\\
3.65331101417542	42.6404587216085\\
3.85134172439575	41.509864736026\\
4.05105924606323	40.1871730257659\\
4.25282216072083	38.9286886695794\\
4.45276713371277	37.6999430989049\\
4.65254688262939	36.4447367606383\\
4.85205054283142	35.2889340970934\\
5.05226540565491	34.1067977397443\\
5.25319838523865	32.8289335533659\\
5.45644354820251	31.9367821664029\\
5.65510845184326	30.6456142239336\\
5.85332989692688	29.4693239590976\\
6.05252838134766	28.3861047297461\\
6.25445294380188	27.0183760487706\\
6.46015214920044	25.905552687018\\
6.65498781204224	24.8764802416462\\
6.85395312309265	23.8645700332329\\
7.05144262313843	22.8743926276081\\
7.25302982330322	21.8226515302632\\
7.45259237289429	20.8498981183392\\
7.65178751945496	19.957950482174\\
7.85272026062012	19.0657912901285\\
8.05256104469299	18.1877506863703\\
8.25225877761841	17.4217944352581\\
8.45404386520386	16.5954410798695\\
8.65307760238647	15.7682165240718\\
8.85222029685974	14.9648181657544\\
9.05230665206909	14.1323156000588\\
9.25369954109192	13.3189219443762\\
9.45354962348938	12.5514388092422\\
9.6605167388916	11.8010218445597\\
9.85368680953979	11.1050340674596\\
10.0528147220612	10.8020312150005\\
10.2548942565918	10.1171839341643\\
10.4560463428497	9.33008651742342\\
10.6537096500397	8.70989280118209\\
10.852068901062	8.03524231914427\\
11.060142993927	7.44238125393548\\
11.2601866722107	6.94278358139857\\
11.453777551651	7.32208323535132\\
11.6524991989136	6.83365635289609\\
11.8529818058014	6.34175185751041\\
12.056131362915	5.77047468900845\\
12.2538182735443	5.28006582155066\\
12.4540915489197	5.51241340222508\\
12.6534292697906	5.01978566371832\\
12.852790594101	4.81567696009788\\
13.0552260875702	4.75377938461791\\
13.2522768974304	4.58126525518347\\
13.4556972980499	4.18152189290704\\
13.6580104827881	4.13796743892915\\
13.8601644039154	3.71228822124642\\
14.053829908371	3.49217254587481\\
14.2517249584198	3.23820424471994\\
14.4562628269196	3.04667752550112\\
14.6524398326874	2.84352885815419\\
14.8541240692139	2.69258297693489\\
15.0627791881561	2.61331699740651\\
15.2535579204559	2.48846082211692\\
15.4558203220367	2.34031304423724\\
15.6550364494324	2.23859256397869\\
15.853856086731	2.16405872275812\\
16.0536758899689	2.10244873744507\\
16.2539458274841	2.07627310813357\\
16.4536452293396	2.05658222181167\\
16.6541950702667	2.0303097331946\\
16.8526608943939	1.99458243081649\\
17.0542204380035	1.9588635763843\\
17.25421833992	1.92722415584726\\
17.4558839797974	1.89774068567154\\
17.6550118923187	1.87518103937933\\
17.8538489341736	1.8590554486696\\
18.0541191101074	1.84503837558478\\
18.25346326828	1.82541058036337\\
18.4529554843903	1.80984249510558\\
18.6553075313568	1.79847834759857\\
18.8579256534576	1.79420429428453\\
19.0531616210938	1.79390771014297\\
19.2545323371887	1.79735861391717\\
19.4557473659515	1.79764517956737\\
19.6673305034637	1.79333963301262\\
19.8524878025055	1.79010651031494\\
20.0534772872925	1.78926342507603\\
20.2548887729645	1.78924748815141\\
20.4541759490967	1.79134845389358\\
20.654465675354	1.79320624091296\\
20.8551850318909	1.79078057287302\\
21.0532202720642	1.78970154931529\\
21.2552809715271	1.78938267537597\\
21.4548418521881	1.78990505638254\\
21.6552832126617	1.79337986339714\\
21.8562107086182	1.79157638908723\\
22.0539293289185	1.78996910279734\\
22.2615072727203	1.78973453507758\\
};
\addlegendentry{H(p,W)}

\addplot [color=mycolor2, line width=1.4pt]
  table[row sep=crcr]{%
-1.14863276481628	4.82682191063969\\
-0.9499192237854	4.82682191063969\\
-0.74800181388855	5.18783651934293\\
-0.548374652862549	8.09393827843221\\
-0.348526477813721	12.2152705177284\\
-0.146867752075195	15.2022015626517\\
0.0536508560180664	15.6784610496897\\
0.254810094833374	15.3241592751222\\
0.452388525009155	14.9420886289247\\
0.653592109680176	14.5634875431525\\
0.855054140090942	14.1706006139336\\
1.05077505111694	13.7963053359219\\
1.25211906433105	13.3839974045554\\
1.45325183868408	12.9950163049752\\
1.65678715705872	12.5863879105637\\
1.85086393356323	12.1880321235988\\
2.05186653137207	11.7371531518192\\
2.2526798248291	11.3556276691377\\
2.45258402824402	10.9076012184155\\
2.65606117248535	10.5417216366012\\
2.86030268669128	10.1556514170574\\
3.05085372924805	9.79698410083848\\
3.25225734710693	9.42663582088015\\
3.45727562904358	9.05110002582203\\
3.65331101417542	8.71821829867112\\
3.85134172439575	8.40069574321629\\
4.05105924606323	8.07257895962971\\
4.25282216072083	7.73527326196105\\
4.45276713371277	7.44132066342783\\
4.65254688262939	7.1800521708258\\
4.85205054283142	6.91575574543485\\
5.05226540565491	6.65305281167643\\
5.25319838523865	6.42329877143479\\
5.45644354820251	6.20350286498015\\
5.65510845184326	6.01263056378562\\
5.85332989692688	5.83727697105999\\
6.05252838134766	5.67405977630017\\
6.25445294380188	4.86353653658054\\
6.46015214920044	4.41266911639833\\
6.65498781204224	4.09275741262297\\
6.85395312309265	3.98603371078364\\
7.05144262313843	3.89288785440247\\
7.25302982330322	3.8171757480794\\
7.45259237289429	3.74568689519841\\
7.65178751945496	3.69986506029771\\
7.85272026062012	3.65626231730288\\
8.05256104469299	3.62504639736417\\
8.25225877761841	3.60009623187219\\
8.45404386520386	3.58112633144317\\
8.65307760238647	3.56674747810836\\
8.85222029685974	3.55671910399462\\
9.05230665206909	3.54994951876785\\
9.25369954109192	3.54530090533223\\
9.45354962348938	3.54239101330009\\
9.6605167388916	3.54075027399415\\
9.85368680953979	3.53968124478077\\
10.0528147220612	2.82378958486142\\
10.2548942565918	2.57383343166985\\
10.4560463428497	2.45444175849939\\
10.6537096500397	2.45419794996822\\
10.852068901062	2.45386431791954\\
11.060142993927	2.45370043420302\\
11.2601866722107	2.45362748011544\\
11.453777551651	2.12377546297446\\
11.6524991989136	2.05272612448421\\
11.8529818058014	1.93783525919921\\
12.056131362915	1.93718634967001\\
12.2538182735443	1.93680043874723\\
12.4540915489197	1.82512050980072\\
12.6534292697906	1.79855932352391\\
12.852790594101	1.72494126595411\\
13.0552260875702	1.69429234498362\\
13.2522768974304	1.68688711185345\\
13.4556972980499	1.65832145309634\\
13.6580104827881	1.64788755926484\\
13.8601644039154	1.63932854401094\\
14.053829908371	1.64445108836856\\
14.2517249584198	1.6392472517355\\
14.4562628269196	1.63161252224022\\
14.6524398326874	1.63092709560399\\
14.8541240692139	1.63286171531763\\
15.0627791881561	1.62805126480581\\
15.2535579204559	1.63235923479764\\
15.4558203220367	1.63335088684578\\
15.6550364494324	1.63402580788887\\
15.853856086731	1.63365569063574\\
16.0536758899689	1.63365286341622\\
16.2539458274841	1.63365557838884\\
16.4536452293396	1.63365561466941\\
16.6541950702667	1.63365561825339\\
16.8526608943939	1.63365562787564\\
17.0542204380035	1.63365563689708\\
17.25421833992	1.63365563589644\\
17.4558839797974	1.63365563400017\\
17.6550118923187	1.63365560346543\\
17.8538489341736	1.63365553786641\\
18.0541191101074	1.63365540998605\\
18.25346326828	1.63365506864634\\
18.4529554843903	1.63365420995785\\
18.6553075313568	1.63365084811231\\
18.8579256534576	1.63365409048682\\
19.0531616210938	1.63365587691128\\
19.2545323371887	1.63365493218485\\
19.4557473659515	1.63365528769902\\
19.6673305034637	1.63365417037062\\
19.8524878025055	1.63364928183907\\
20.0534772872925	1.63365716967745\\
20.2548887729645	1.63365570900968\\
20.4541759490967	1.63365552768294\\
20.654465675354	1.63365537775702\\
20.8551850318909	1.63365384007193\\
21.0532202720642	1.63365645297589\\
21.2552809715271	1.63365565433818\\
21.4548418521881	1.63365478745126\\
21.6552832126617	1.63365544804109\\
21.8562107086182	1.63365444889696\\
22.0539293289185	1.63365674356721\\
22.2615072727203	1.63365565456281\\
};
\addlegendentry{H($\bar{p}$,$\mathbb{W}$)}

\end{axis}
\end{tikzpicture}%
\vspace{-0.7em}
\caption{\rahel{Locational optimization cost decrease (green) as well as the over all agents summed up target cost decrease (rose) over time, applying Algorithm~\ref{alg:onelayeralg} with respect to a known density $\phi_{1}$.}}
\label{fig:onelayercoveragecost}
\vspace{-0.5em}
\end{figure}
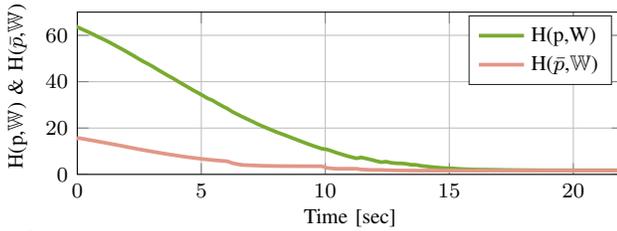

Applying Algorithm~\ref{alg:onelayeralg} in the described set-up using $\phi_{1}$, we obtain the locational optimization cost as well as the cumulative target cost decrease presented in Figure~\ref{fig:onelayercoveragecost}. 

\begin{figure}[h!]
\begin{minipage}[t]{0.15\textwidth}
\centering
\includegraphics[trim={5.2cm 8.1cm 4.6cm 7.7cm},clip, width = 0.8\textwidth]{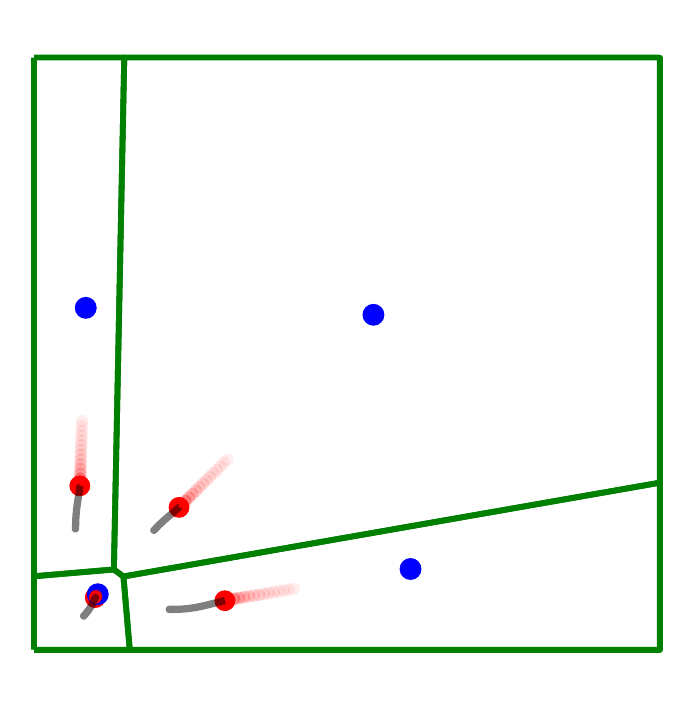}
\centering
\end{minipage}
\hfill
\begin{minipage}[t]{0.15\textwidth}
\centering
\includegraphics[trim={5.2cm 8.1cm 4.6cm 7.7cm},clip, width = 0.8\textwidth]{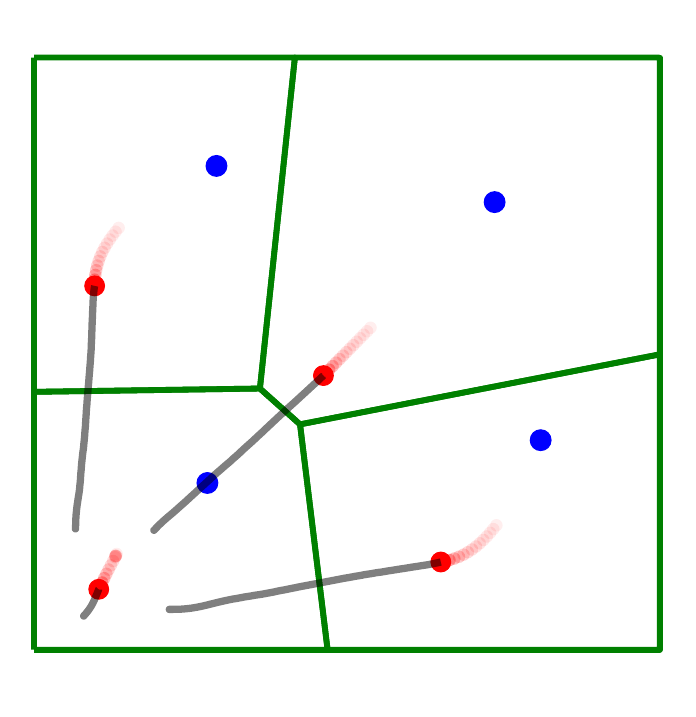}
\centering
\end{minipage}
\hfill
\begin{minipage}[t]{0.15\textwidth}
\centering
\includegraphics[trim={5.2cm 8.1cm 4.6cm 7.7cm},clip, width = 0.8\textwidth]{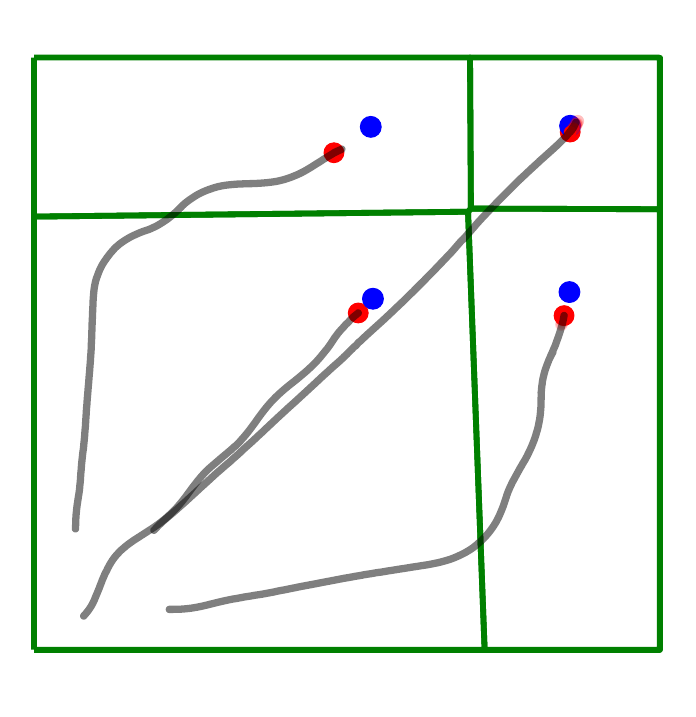}
\centering
\end{minipage}
\caption{\rahel{Configurations of cars at 2, 8, and 21 seconds applying Algorithm~\ref{alg:onelayeralg} in the described set-up. The agents' location and their predicted trajectory are given in red, the Voronoi partitions in green, the artificial setpoints in blue, and the traveled paths are visualized in light grey.}}
\label{pics:onelayerconfig}
\vspace{-0.7em}
\end{figure}

\noindent It shows a comparable behavior to the two-layers approach regarding the time required until convergence, as well as the final configuration. While at certain instances in time the locational optimization cost increases slightly, the summed up target cost is strictly decreasing. 
\subsection{Results One-Layer, Learning-Based Coverage MPC}
\label{subsec:experimentonelayerlearning}
In a final experiment, Algorithm~\ref{alg:onelayerlearningalg} is applied, with density $\phi_{2}$ initially unknown and $S$=2.5. 
\begin{figure} [h!]
%
%
\definecolor{mycolor1}{rgb}{0.47000,0.67000,0.19000}%
\definecolor{mycolor2}{rgb}{0.79000,0.49000,0.43000}%
\begin{tikzpicture}

\begin{axis}[%
height=0.85in,
width=0.40\textwidth,
yshift=0.8cm,
at={(0.0in,0.0in)},
scale only axis,
xmin=0,
xmax=21,
xlabel style={font=\color{white!15!black}},
xlabel style={font=\footnotesize},
xlabel style={yshift=0.6ex,},
xlabel={Time [sec]},
ymin=0,
ymax=70,
ylabel style={font=\color{white!15!black}},
ylabel style={font=\footnotesize},
ylabel style={xshift = 0.6ex,},
ylabel={H(p,$\mathbb{W}$,$\phi$) \& H(p,$\mathbb{W}$,$\hat{\phi}$)},
axis background/.style={fill=white},
tick label style={font=\footnotesize},
xmajorgrids,
ymajorgrids,
legend style={legend cell align=left, align=left, draw=white!15!black},
legend style={font=\footnotesize}
]
\addplot [color=mycolor1, line width=1.4pt]
  table[row sep=crcr]{%
-0.117774486541748	55.6416455837685\\
0.280490159988403	55.6453627549429\\
0.483649969100952	55.361083891311\\
0.682579755783081	54.8773618427026\\
0.883516311645508	54.4365677614694\\
1.08183526992798	53.8771787478607\\
1.28105568885803	52.7542535672557\\
1.4802074432373	51.6136157983385\\
1.6821653842926	50.4360011825191\\
1.8819944858551	49.4240281906488\\
2.08089518547058	48.4058542677896\\
2.28198099136353	47.3090191264453\\
2.48251509666443	46.3847905869479\\
2.68301582336426	45.6359750824814\\
2.8805935382843	45.029363172859\\
3.08038401603699	44.4180656231274\\
3.2818455696106	43.5742824500728\\
3.48093891143799	42.8493418005347\\
3.68154144287109	42.2362567113429\\
3.88178730010986	41.5072348794846\\
4.08241820335388	40.9321376153649\\
4.28357720375061	40.3427885174429\\
4.48154473304749	39.6726479886624\\
4.68079805374146	39.0314837426816\\
4.88374209403992	38.2690356066683\\
5.08170390129089	37.4957600645149\\
5.28213214874268	36.6608746608028\\
5.48064565658569	35.798442589528\\
5.6825487613678	34.693506745643\\
5.88136625289917	33.7996035091865\\
6.08316135406494	32.7448490956232\\
6.28040170669556	31.7546724163282\\
6.48473501205444	30.7255765475058\\
6.68359088897705	29.7565531932265\\
6.8818998336792	28.844052477089\\
7.08079028129578	28.0449804973609\\
7.28340697288513	27.2355326202412\\
7.4817681312561	26.4590145323513\\
7.68242835998535	25.7297394324212\\
7.88226366043091	25.0451935509965\\
8.08400750160217	21.3410933718077\\
8.28272938728333	20.7647311329084\\
8.48226618766785	20.1426341001193\\
8.68258595466614	19.52805074703\\
8.88428592681885	18.9668370496759\\
9.08221316337585	18.4405659707563\\
9.28181314468384	17.9469920160315\\
9.48220610618591	17.4299038113296\\
9.6818208694458	18.3242499781337\\
9.88172578811646	17.9735928518606\\
10.0821895599365	17.669638109797\\
10.2835392951965	17.3846642457329\\
10.4819307327271	17.1264844208831\\
10.6823532581329	16.6955435589931\\
10.8809492588043	16.324843202953\\
11.0806744098663	15.9516226898534\\
11.2827413082123	15.5441400341789\\
11.4807531833649	15.1649371911249\\
11.6827719211578	14.7359950204209\\
11.8808121681213	14.204652680973\\
12.0822365283966	13.7627185072749\\
12.2819607257843	13.2446165806699\\
12.482825756073	12.7426418872089\\
12.6820688247681	13.6182524074829\\
12.8831100463867	12.9932256099771\\
13.0828530788422	12.434824934557\\
13.2831931114197	12.848650264405\\
13.4814794063568	12.3879141763603\\
13.6822912693024	12.2708504510673\\
13.8833720684052	11.9899648522295\\
14.0813863277435	11.7371021230744\\
14.2816686630249	11.5166552083777\\
14.4826619625092	11.2839397286643\\
14.6806054115295	10.9476933473091\\
14.8825922012329	10.5617598837446\\
15.0800611972809	10.1971533405991\\
15.2834858894348	9.86447391244495\\
15.4836881160736	9.55526590479222\\
15.6811118125916	9.31695098847601\\
15.8810572624207	9.11008817662245\\
16.0815989971161	8.92641957598823\\
16.2841033935547	8.70922488775289\\
16.4819452762604	8.49097062595203\\
16.6800713539124	8.29315991785297\\
16.881413936615	8.13959434449154\\
17.0820443630219	8.03574777971558\\
17.2811107635498	8.03160866844738\\
17.4821639060974	8.02398643818547\\
17.6829349994659	8.01370806790568\\
17.8820822238922	7.96308694941008\\
18.082211971283	7.86668473000067\\
18.2817482948303	7.81194675754939\\
18.4820413589478	7.800862629438\\
18.6805729866028	7.80084693423395\\
18.8808255195618	7.80114947930034\\
19.0800879001617	7.80099185310746\\
19.2812404632568	7.8011984181981\\
19.4819047451019	7.8037018629369\\
19.6810040473938	7.80737793468351\\
19.8802058696747	7.80778929055742\\
20.081912279129	7.80842422173864\\
20.2811563014984	7.81327702564884\\
20.4812014102936	7.81614655889011\\
20.6810367107391	7.81611511608263\\
20.8817536830902	7.8164073447897\\
21.0817904472351	7.81649368399212\\
21.2805547714233	7.81879393713733\\
21.4817001819611	7.82025101883706\\
21.6815104484558	7.82105345996906\\
};
\addlegendentry{H(p,$\mathbb{W}$,$\phi$)}

\addplot [color=mycolor2, dotted, line width=1.4pt]
  table[row sep=crcr]{%
-0.117774486541748	0\\
0.280490159988403	5.01004814758321\\
0.483649969100952	6.58125594200134\\
0.682579755783081	7.43546088675415\\
0.883516311645508	8.30547631840922\\
1.08183526992798	9.79947514525108\\
1.28105568885803	11.4852891880853\\
1.4802074432373	13.5225527201427\\
1.6821653842926	15.779279662806\\
1.8819944858551	18.3177005873401\\
2.08089518547058	20.924079349164\\
2.28198099136353	23.5584393611047\\
2.48251509666443	26.2968107621151\\
2.68301582336426	29.0085142052109\\
2.8805935382843	31.6513171451911\\
3.08038401603699	34.0424879842239\\
3.2818455696106	35.9442474621927\\
3.48093891143799	37.620181898234\\
3.68154144287109	39.0402009464921\\
3.88178730010986	40.002763878236\\
4.08241820335388	40.7911141981991\\
4.28357720375061	41.2518007428867\\
4.48154473304749	41.3452060256601\\
4.68079805374146	41.2273358379846\\
4.88374209403992	40.7775567672026\\
5.08170390129089	40.1603621529442\\
5.28213214874268	39.3584878016996\\
5.48064565658569	38.4414985511979\\
5.6825487613678	37.2035576918124\\
5.88136625289917	36.1621549470681\\
6.08316135406494	34.9335389076995\\
6.28040170669556	33.7649651855327\\
6.48473501205444	32.5556951984043\\
6.68359088897705	31.4215310540775\\
6.8818998336792	30.3590771239661\\
7.08079028129578	29.4288444559117\\
7.28340697288513	28.5102477334512\\
7.4817681312561	27.6455868598057\\
7.68242835998535	26.8416545433883\\
7.88226366043091	26.092894716062\\
8.08400750160217	22.3217015314451\\
8.28272938728333	21.6891398417603\\
8.48226618766785	20.9980455679924\\
8.68258595466614	20.32173289701\\
8.88428592681885	19.7067657204088\\
9.08221316337585	19.1281227279209\\
9.28181314468384	18.5910780690894\\
9.48220610618591	18.036346768048\\
9.6818208694458	18.9019513344367\\
9.88172578811646	18.5243069614519\\
10.0821895599365	18.1952165330867\\
10.2835392951965	17.8873222638923\\
10.4819307327271	17.6090498681704\\
10.6823532581329	17.1570800762272\\
10.8809492588043	16.7628644918813\\
11.0806744098663	16.3650182976867\\
11.2827413082123	15.9323844290872\\
11.4807531833649	15.5324814002502\\
11.6827719211578	15.0835282181021\\
11.8808121681213	14.5330918118539\\
12.0822365283966	14.0700612358026\\
12.2819607257843	13.5309162371155\\
12.482825756073	13.0094650196548\\
12.6820688247681	13.8768285908668\\
12.8831100463867	13.2260533591586\\
13.0828530788422	12.6417166628174\\
13.2831931114197	13.0396586141099\\
13.4814794063568	12.5567486544518\\
13.6822912693024	12.4247238171294\\
13.8833720684052	12.1240601836556\\
14.0813863277435	11.8571121692981\\
14.2816686630249	11.6152620018859\\
14.4826619625092	11.3650247944024\\
14.6806054115295	11.0116585668762\\
14.8825922012329	10.6102443171718\\
15.0800611972809	10.2354556739546\\
15.2834858894348	9.89300860784614\\
15.4836881160736	9.57548757764806\\
15.6811118125916	9.3293640822962\\
15.8810572624207	9.11356538304075\\
16.0815989971161	8.92130998333853\\
16.2841033935547	8.69747925405955\\
16.4819452762604	8.47408157789424\\
16.6800713539124	8.2715294333268\\
16.881413936615	8.1134766655674\\
17.0820443630219	8.00581457808346\\
17.2811107635498	7.99852300447833\\
17.4821639060974	7.98763074788972\\
17.6829349994659	7.97435657218688\\
17.8820822238922	7.9205794945088\\
18.082211971283	7.82112121148361\\
18.2817482948303	7.76385912447984\\
18.4820413589478	7.7506163607135\\
18.6805729866028	7.74855502627953\\
18.8808255195618	7.74691551663029\\
19.0800879001617	7.74496149681459\\
19.2812404632568	7.74349603381707\\
19.4819047451019	7.74442190074292\\
19.6810040473938	7.74657260146394\\
19.8802058696747	7.74560350285448\\
20.081912279129	7.74494909081099\\
20.2811563014984	7.74848730863917\\
20.4812014102936	7.75021705136891\\
20.6810367107391	7.74909563439767\\
20.8817536830902	7.74835421852609\\
21.0817904472351	7.74746135002338\\
21.2805547714233	7.74876982322129\\
21.4817001819611	7.74929804564092\\
21.6815104484558	7.7492348773708\\
};
\addlegendentry{H(p,$\mathbb{W}$,$\hat{\phi}$)}

\end{axis}

\begin{axis}[%
width=0.0in,
height=0.0in,
at={(0in,0.0in)},
scale only axis,
xmin=0,
xmax=1,
ymin=0,
ymax=1,
axis line style={draw=none},
ticks=none,
axis x line*=bottom,
axis y line*=left
]
\end{axis}
\end{tikzpicture}%
\vspace{-0.7em}
\caption{\rahel{Locational optimization cost (green) as well as estimated locational optimization cost over time (rose), applying Algorithm~\ref{alg:onelayerlearningalg} for an initially unknown $\phi_{2}$.}}
\label{fig:onelayerlearningcoveragecost}
\vspace{-0.7em}
\end{figure}
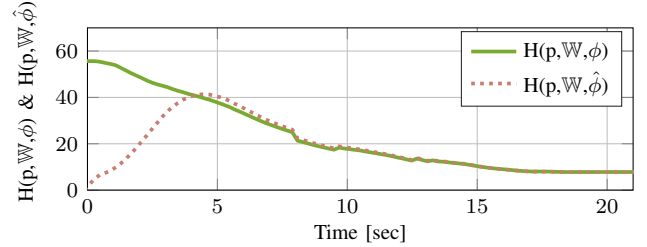

\begin{figure}[h!]
\begin{minipage}[t]{0.15\textwidth}
\centering
\includegraphics[trim={5.2cm 8.1cm 4.6cm 7.7cm},clip, width = 0.8\textwidth]{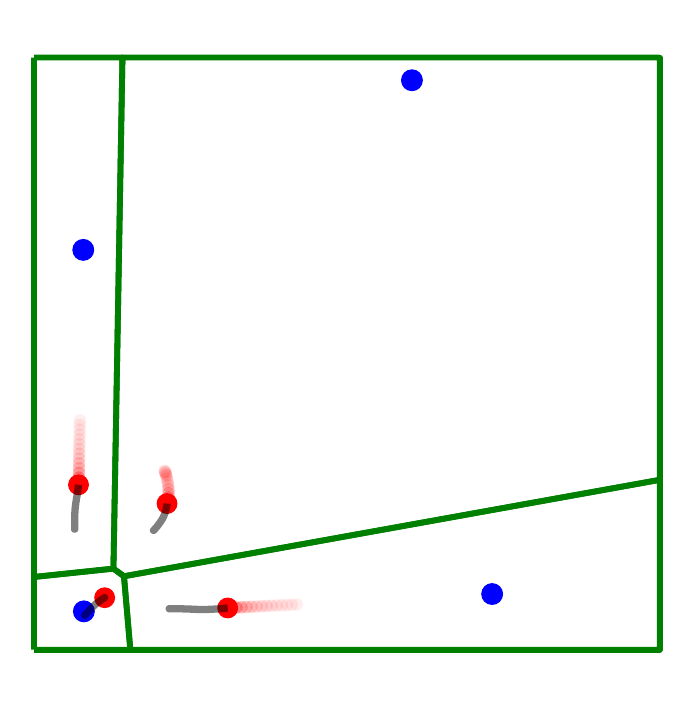}
\centering
\end{minipage}
\hfill
\begin{minipage}[t]{0.15\textwidth}
\centering
\includegraphics[trim={5.2cm 8.1cm 4.6cm 7.7cm},clip, width = 0.8\textwidth]{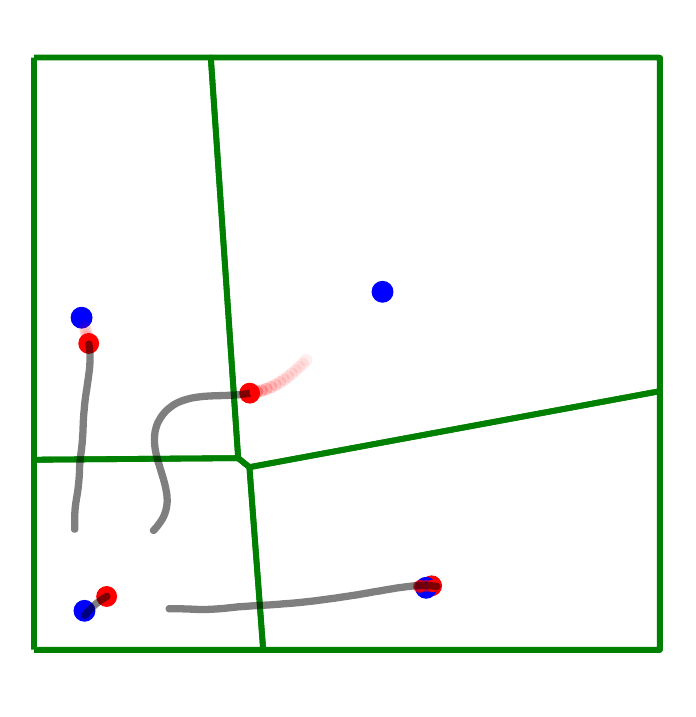}
\centering
\end{minipage}
\hfill
\begin{minipage}[t]{0.15\textwidth}
\centering
\includegraphics[trim={5.2cm 8.1cm 4.6cm 7.7cm},clip, width = 0.8\textwidth]{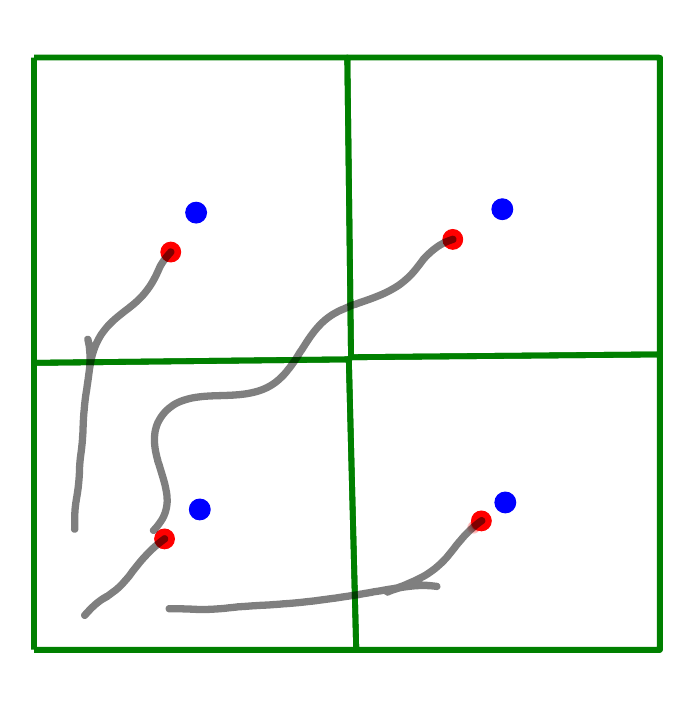}
\centering
\end{minipage}
\caption{\rahel{Configurations of cars at 2, 8, and 19 seconds applying Algorithm~\ref{alg:onelayerlearningalg} in the described set-up. The agents' location and their predicted trajectory are given in red, the Voronoi partitions in green, the artificial setpoints in blue, and the traveled paths are visualized in light grey.}}
\label{pics:onelayerlearningconfig}
\vspace{-0.7em}
\end{figure}

\begin{figure} [h]
\centering
%
%
\definecolor{mycolor1}{rgb}{0.50000,0.75000,0.93000}%
\begin{tikzpicture}

\begin{axis}[%
height=0.85in,
width=0.40\textwidth,
yshift=0.8cm,
at={(0.0in,0.0in)},
scale only axis,
xmin=0,
xmax=21,
xlabel style={font=\color{white!15!black}},
xlabel style={font=\footnotesize},
xlabel style={yshift=0.6ex,},
xlabel={Time [sec]},
ymode=log,
ymin=0.001,
ymax=16,
yminorticks=true,
ylabel style={font=\color{white!15!black}},
ylabel style={font=\footnotesize},
ylabel style={xshift=0.6ex,},
ylabel={$\textup{Var}_{\textup{max}}$},
axis background/.style={fill=white},
xmajorgrids,
ymajorgrids,
tick label style={font=\footnotesize},
legend style={legend cell align=left, align=left, draw=white!15!black},
legend style={font=\footnotesize}
]
\addplot [color=mycolor1, line width=1.4pt]
  table[row sep=crcr]{%
0.0824691772460939	11.2337052528901\\
0.4823326587677	10.5834347445309\\
0.683907461166382	10.4061974846819\\
0.883398008346558	10.3354384794997\\
1.08480448722839	10.2818153068785\\
1.28220982551575	10.2204790628016\\
1.48219556808472	10.1323660477821\\
1.68094701766968	10.0053900948755\\
1.88225836753845	9.82408095091294\\
2.08258218765259	9.59112377292213\\
2.2808596611023	9.29913952176296\\
2.48196144104004	8.94703675651548\\
2.68293137550354	8.53658175788563\\
2.88405508995056	8.07128169054354\\
3.08245701789856	7.55296232245986\\
3.28137726783752	6.99860414966163\\
3.48191614151001	6.42310359934903\\
3.68113107681274	5.84317107685419\\
3.88187260627747	5.2664518989589\\
4.08333582878113	4.70616917185519\\
4.28269715309143	4.17263326435623\\
4.48381371498108	3.6762895750628\\
4.68152732849121	3.22386750685917\\
4.88279004096985	2.81580338147141\\
5.08373804092407	2.45072128311792\\
5.28267259597778	2.12887830728859\\
5.48323984146118	1.84837001349638\\
5.68104691505432	1.60601493124275\\
5.88389511108398	1.39661087706699\\
6.08287544250488	1.21484168417437\\
6.28336806297302	1.0590220222861\\
6.48188943862915	0.923991761114672\\
6.68472023010254	0.806200499104494\\
6.8842405796051	0.705939462445048\\
7.08238549232483	0.625696316323029\\
7.28184170722961	0.562319007430952\\
7.48439021110535	0.509984810043683\\
7.68310136795044	0.466189474829822\\
7.88349456787109	0.429028948957453\\
8.08247299194336	0.396833193076347\\
8.2843870639801	0.333261330811213\\
8.4837438583374	0.312696687996926\\
8.68244643211365	0.295508510609875\\
8.8829309463501	0.280165109836974\\
9.08465547561645	0.267516057492097\\
9.2827887058258	0.257512727187911\\
9.48206562995911	0.248932539309797\\
9.68278117179871	0.241664533717542\\
9.88401408195496	0.241339579086565\\
10.0824365139008	0.23538743273719\\
10.2821797847748	0.229637126899569\\
10.4837989330292	0.223927427930646\\
10.6825706481934	0.218196215831969\\
10.8826281547546	0.212176326133362\\
11.0825271129608	0.205808639400006\\
11.2814244747162	0.199032069362536\\
11.4827241420746	0.191898015255592\\
11.6821047782898	0.184710950527391\\
11.8838531494141	0.177591485211253\\
12.0817265033722	0.170498813820227\\
12.2838096141815	0.163439087492615\\
12.484863948822	0.156294093750672\\
12.6830825328827	0.148975366014786\\
12.8830816268921	0.161841139402378\\
13.0848724365234	0.153071112254505\\
13.2828008651733	0.144064740665686\\
13.4831900119781	0.1341764995115\\
13.6818553924561	0.124956744614517\\
13.8834168434143	0.115050977117623\\
14.0844420909882	0.106360875883648\\
14.2820920467377	0.0991818053534859\\
14.4824532985687	0.0930265314485906\\
14.6831182956696	0.0872917197029991\\
14.881488275528	0.0825366207049508\\
15.0828029632568	0.0777433250492063\\
15.2814306735992	0.0730081374593279\\
15.4833886146545	0.0684125273087926\\
15.6849286079407	0.0640936098605778\\
15.8820456981659	0.0602642504062827\\
16.0815531730652	0.0569464397761256\\
16.2836560726166	0.0540106877732077\\
16.48409075737	0.0513814085461716\\
16.6819352626801	0.0490157823376638\\
16.8810507774353	0.0468855763544784\\
17.0838777542114	0.0449503714112135\\
17.2820245742798	0.0431881698980211\\
17.481378030777	0.0416058370373098\\
17.6838738441467	0.0401777079008201\\
17.8839389801025	0.0388819537210294\\
18.0822820186615	0.0376947549997477\\
18.2837690830231	0.0365897220303644\\
18.4817387580872	0.0355625031166764\\
18.6826481342316	0.0346153559752507\\
18.8818845272064	0.0337424943581691\\
19.0817939758301	0.032935628248512\\
19.2805633068085	0.0321868891538603\\
19.4819049358368	0.0314901884846144\\
19.6831392765045	0.0308406835927521\\
19.8814782619476	0.0302331541395085\\
20.0811514377594	0.0296633590094357\\
20.2840866565704	0.029127819526675\\
20.4836422920227	0.0286236977570073\\
20.6818365573883	0.0281480096076934\\
20.8817939281464	0.0276980069061547\\
21.0834691047668	0.0272715168931879\\
21.2821485519409	0.0268666820895049\\
21.4827281951904	0.0264821955919711\\
21.6823594093323	0.0261161636836615\\
21.8826069355011	0.0257671785670892\\
22.0820765018463	0.0254341736544502\\
22.2821170806885	0.0251160933380441\\
22.4816207408905	0.0248118713004267\\
22.6831359386444	0.0245203431247609\\
22.8840643882751	0.0242406837178302\\
};
\addlegendentry{Maximal Variance}

\end{axis}

\begin{axis}[%
width=0.854in,
height=0.438in,
at={(0in,0in)},
scale only axis,
xmin=0,
xmax=1,
ymin=0,
ymax=1,
axis line style={draw=none},
ticks=none,
axis x line*=bottom,
axis y line*=left
]
\end{axis}
\end{tikzpicture}%
\vspace{-0.7em}
\caption{\rahel{Maximal variance over time applying Algorithm~\ref{alg:onelayerlearningalg} for an initially unknown $\phi_{2}$.}}
\label{fig:onelayerlearningmaxvar}
\vspace{-0.7em}
\end{figure}
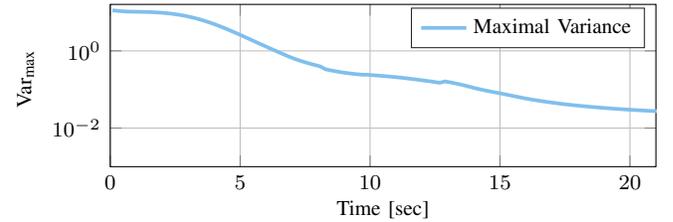

\begin{figure} [h!]
\centering
\input{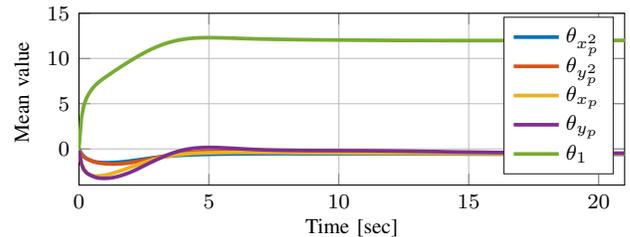}
\vspace{-0.7em}
\caption{\rahel{Coefficient estimation over time applying Algorithm~\ref{alg:onelayerlearningalg} for an initially unknown $\phi_{2}$.}}
\label{fig:onelayerlearningcoef}
\vspace{-0.7em}
\end{figure}
\noindent For the first half of the considered time-span, the variance dominates the target cost, and the setpoints are placed close to the locations of maximal variance within the partitions. The influence of the variance diminishes with its decreasing maximal value, and the goal continuously shifts to the coverage problem. It is important to note that the timescale of the graphics highly differs from the two-layers learning approach presented in Section~\ref{subsec:experimenttwolayerlearning}: for the conducted experiments, the one-layer learning approach takes \rahel{approximately half} of the time to steer the agents in a nearly optimal configuration. However, by the time the experiment was interrupted, all mean estimates showed a maximal error of less than \rahel{0.093} concerning the true coefficient value, and the maximal variance reads as approximately \rahel{0.02}. Both are slightly higher than in the two-layers learning-based approach, and the obtained estimates of the parameter vector $\theta$ are slightly less certain and precise due to a reduction in exploration movements. Both values would decrease if a large variance scaling factor $S$ was used.
\vspace{-0.7em}

\section{Discussion and conclusions}
\label{sec:conclusions}
This work presents a framework to optimally cover a predefined convex area using a non-homogeneous fleet of agents, whose movements are determined by their known nonlinear dynamics. Specifically, two methods based on tracking MPC without terminal constraints and accounting for collision avoidance constraints are introduced. While the first method relies on a two-layered structure and passes an online calculated reference to the individual MPC of each agent, the second method overcomes the hierarchical structure and directly integrates the calculation of the succeeding optimal configuration into the cost function of each agent's MPC. The developed methods are proposed for the case with a known environment, represented by a known density function $\phi$, and are individually extended to the scenario in which the density function needs to be actively learned by the agents. 
Recursive feasibility and convergence to an optimal configuration are formally proven for both methods in each scenario. Furthermore, hardware results are obtained for all the proposed methods using a miniature racing car platform. With the performance of the two methodologies being rather comparable in a known environment, one of the benefits of the one-layer approach is its simplicity. Furthermore, the one-layer learning approach allows for a more efficient operation in the considered experiments, since it does \textit{not} require an explicit exploration vs.~exploitation decision, but automatically adjusts it based on the remaining uncertainty. However, given the integration in its MPC cost function, the one-layer approach is computationally more expensive than its two-layers opponent. An interesting challenge for further research is a decentralized implementation for the learning and the Voronoi partition calculation~\cite{Bullo2012}, e.g., relying only on gossiping between the agents. 
\vspace{-0.7em}

\bibliographystyle{docstyle/IEEEtran} 
\bibliography{bibliography}

\appendix
\section{Proofs}
\label{sec:proofs}
In this section, the proofs of Theorems~\ref{theorem:twolayers}-\ref{theorem:onelayerlearning} are given. In all theorems, 
recursive feasibility is ensured by only updating the partitions if the condition presented in~\eqref{eq:feasibilitycond} is fulfilled. Furthermore, recall that $w$ indicates the timestep of the most recent partition update. 

\subsection{Proof of Theorem~\ref{theorem:twolayers}}
\label{subsec:twolayersproof}The proof of Theorem~\ref{theorem:twolayers} is divided into two steps. First, convergence of each agent to its reference $r$ is shown for a fixed Voronoi partition. This is followed by the proof that the fulfillment of the partition update requirement described in equations~\eqref{eq:updatereq1},~\eqref{eq:updatereq2} and~\eqref{eq:feasibilitycond} is satisfied after a finite time. Subsequently, convergence to a centroidal Voronoi configuration follows by  Proposition~\ref{proposition:cortesconvergence}.

\subsubsection{Convergence to Reference} 
The result will be proven by first introducing two intermediate results (Prop.~\ref{proposition:notminsteadynotcosetwolayers} and~\ref{proposition:upperlowerboudJtilde}). Prop.~\ref{proposition:notminsteadynotcosetwolayers} follows \cite[Proposition 1]{Soloperto2021} and shows that the optimal setpoint cannot be arbitrarily close to the current state-input pair, unless the latter belongs to the set $\rahel{\mathbb{T}_{\bar{p}_w^*,r,i}}$ defined in~\eqref{eq:Tid2layer}; Prop.~\ref{proposition:upperlowerboudJtilde} later provides a lower and upper bound for $\tilde{J}_{i}^{*}(x_{k}, \mathbb{W}_{\rahel{\bar{p}_w^*},i}) = J_{i}^{*}(x_{k}, \mathbb{W}_{\rahel{\bar{p}_w^*},i}) - l_{\mathbb{W}_{\rahel{\bar{p}_w^*}},r,i,\min}$.
\begin{proposition} Let Assumptions~\ref{assumption:dynamics},~\ref{assumption:expocostcontrollability},~\ref{assumption:boundedbyd} and~\ref{assumption:steadystatedecreasetwolayers} hold. Then there exist a constant $\bar{a}_{i} > 0$ such that, for any \rahel{$\bar{p}_{w,i}^*$} and corresponding reference $r$, and for any feasible state $x_{k} \in \mathbb{X}^{\mathrm{int}}: Cx_k \in \bar{\mathbb{W}}_{\rahel{\bar{p}_w^*},i}^{\mathrm{int}}$, the optimal one-step stage cost of problem~\eqref{eq:nonlineartrackingmpcwithcolavoidance} satisfies,
\begin{equation}
\begin{split}
    \ell_{i}^*(x_{k}, s_{k}^{*}) \geq \bar{a}_{i}d_{i}(s_{k}^{*})_{\rahel{\mathbb{T}_{\bar{p}_w^*,r,i}}}^{2}.
\end{split}
\label{eq:prop12layer}
\end{equation}
\label{proposition:notminsteadynotcosetwolayers}
\end{proposition}
\vspace{-1.9em}
\noindent \emph{Proof:} For time step $k\in \mathbb{N}$, assume for contradiction that $\ell_{i}^*(x_{k}, s_{k}^{*}) < \bar{a}_{i}d_{i}(s_{k}^{*})_{\rahel{\mathbb{T}_{\bar{p}_w^*,r,i}}}^{2}$. Following~\cite[Proposition 1]{Soloperto2021}, using compact constraints and $\bar{a}_{i}$ small enough, we can define a setpoint $s'_{k} \in \mathbb{S}_{\mathbb{W}_{\rahel{\bar{p}_w^*},i}}$ according to Assumption~\ref{assumption:steadystatedecreasetwolayers}, such that
\begin{align*}
    &V_{N,i}^{*}(x_{k},s'_{k},\mathbb{W}_{\rahel{\bar{p}_w^*},i}) - V_{N,i}^{*}(x_{k},s_{k}^{*},\mathbb{W}_{\rahel{\bar{p}_w^*},i}) + \ell_{T,i}(\bar{p}'_{k} - r_{i,k})\\
    & -\ell_{T,i}(\bar{p}^*_{k} - r_{i,k}) \overset{\eqref{eq:assumption2gammavmax},~\eqref{eq:assumption32twolayer}}{\leq} \gamma_{i} \ell_{i}^{*}(x_{k},s'_{k}) - \beta_{2,i}\epsilon^{\prime} d_{i}( s_{k}^{*} )_{\rahel{\mathbb{T}_{\bar{p}_w^*,r,i}}}^{2} \\
    &\overset{\eqref{eq:boundtwosteadystatestwolayer}}{\leq} \! \gamma_{i} \xi_{1,i} \ell_{i}^*(x_{k},s_{k}^{*}) \! +  \! \gamma_{i}\xi_{2,i} d_{i}(s'_{k} \! - \! s_{k}^{*})^{2} \! - \! \beta_{2,i}\epsilon^{\prime} d_{i}( s_{k}^{*} )_{\rahel{\mathbb{T}_{\bar{p}_w^*,r,i}}}^{2} \\ 
    &\overset{\eqref{eq:assumption31twolayer}}{<} (\gamma_{i} \xi_{1,i}\bar{a}_i + \gamma_{i} \xi_{2,i}\beta_{1,i}^{2}\epsilon^{\prime,2}- \beta_{2,i}\epsilon^{\prime}) d_{i}( s_{k}^{*} )_{\rahel{\mathbb{T}_{\bar{p}_w^*,r,i}}}^{2} \\ &:= - \bar{c}_id_{i}( s_{k}^{*} )_{\rahel{\mathbb{T}_{\bar{p}_w^*,r,i}}}^{2}. 
\label{eq:prop1}
\end{align*}
Given $\bar{a}_i$ small enough, an $\epsilon^{\prime} > 0$ can be found such that $\bar{c}_i > 0$. This implies that the new setpoint $s^{\prime}_k$ yields a smaller cost, which contradicts the optimality of $s_k^*$.\hfill{$\blacksquare$}

\begin{proposition}
Let Assumptions~\ref{assumption:dynamics},~\ref{assumption:expocostcontrollability},~\ref{assumption:boundedbyd},and~\ref{assumption:steadystatedecreasetwolayers} hold. Define  $a_{i} = \frac{1}{2} \min{(\bar{a}_{i},\alpha_{1,i})}$ and recall that $\bar{u}$ describes the minimizer of $\min_{u \in \mathbb{U}_{i}} \ell_{i}(x,u,s)$. Then, for any $\rahel{\bar{p}_w^*} \in \mathbb{A}^M$, the corresponding reference $r$ and setpoint $s_k^* \in \mathbb{S}_{\mathbb{W}_{\rahel{\bar{p}_w^*},i}}$, and for any feasible state $x_{k} \in \mathbb{X}^{\mathrm{int}}: Cx_k \in \bar{\mathbb{W}}_{\rahel{\bar{p}_w^*},i}^{\mathrm{int}}$, it holds
\begin{equation}
    \tilde{J}_{i}^{*}(x_{k}, \mathbb{W}_{\rahel{\bar{p}_w^*},i}) \! \geq \! a_{i}(d_{i}(s_{k}^{*})_{\rahel{\mathbb{T}_{\bar{p}_w^*,r,i}}}^{2} \! + \! d_{i}((x_{k},u_{x_k,s_k^*}^{*}) \! - \! s_{k}^{*})^2).
\label{eq:lowerbound2layer}
\end{equation}
Further, defining $\gamma_{i,V_{\max,i}} = \max{(\gamma_{i},\frac{V_{\max,i}}{\chi_{i}},2)}$ and denoting by $b_{i} \! = \! \max{(\gamma_{i,V_{\max,i}}\alpha_{2,i}, \beta_{T,i})}$, we have 
\begin{equation}
     \tilde{J}_{i}^{*}(x_{k}, \mathbb{W}_{\rahel{\bar{p}_w^*},i}) \! \leq \! b_{i} (d_{i}(s_{k}^{*})_{\rahel{\mathbb{T}_{\bar{p}_w^*,r,i}}}^{2} \! + \! d_{i}((x_{k},u_{x_k,s_k^*}^{*}) \! - \! s_{k}^{*})^2).
\label{eq:upperbound2layer}
\end{equation}
\label{proposition:upperlowerboudJtilde}
\end{proposition}
\vspace{-1.9em}
\noindent \emph{Proof:} As regards the lower bound, we have 
\begin{align}
     &\tilde{J}_{i}^{*}(x_{k}, \mathbb{W}_{\rahel{\bar{p}_w^*},i}) \geq  V_{N,i}^{*}(x_{k},s_{k}^{*},\mathbb{W}_{\rahel{\bar{p}_w^*},i}) \geq \ell_{i}(x_{k},u_{x_k,s_k^*}^{*},s_{k}^{*}) \notag \\ &\overset{\eqref{eq:boundstagecosttwolayer},\eqref{eq:prop12layer}}{\geq} \frac{\bar{a}_{i}}{2}d_{i}(s_{k}^{*})_{\rahel{\mathbb{T}_{\bar{p}_w^*,r,i}}}^{2} + \frac{\alpha_{1,i}}{2}d_{i}((x_{k},u_{x_k,s_k^*}^{*}) - s_{k}^{*})^2 \label{eq:lowerbound2layertmp} 
\end{align}
To show the existence of an upper bound, we leverage Assumptions~\ref{assumption:expocostcontrollability} and~\ref{assumption:steadystatedecreasetwolayers}. Therefore, we combine~\cite{Boccia2014} with the constraints $V_{N,i} \leq V_{\max,i}$,  to obtain $V_{N,i}^{*}(x,s) \leq \gamma_{i,V_{\max,i}} \ell_{i}^{*}(x,s)$, which yields: 
\begin{align}
     &\tilde{J}_{i}^{*}(x_{k}, \mathbb{W}_{\rahel{\bar{p}_w^*},i}) \notag \\ & = V_{N,i}^{*}(x_{k},s_{k}^{*},\mathbb{W}_{\rahel{\bar{p}_w^*},i}) + \ell_{T,i}(s_{k}^{*}-r) -   l_{\mathbb{W}_{\rahel{\bar{p}_w^*},i},r,\min} \label{eq:upperbound2layer1} \\ & \leq
     \gamma_{i,V_{\max,i}}\alpha_{2,i} d_{i}((x_{k},u_{x_k,s_k^*}^{*}) - s_{k}^{*}))^2 + \beta_{T,i}d_{i}(s_{k}^{*})_{\rahel{\mathbb{T}_{\bar{p}_w^*,r,i}}}^{2} \notag 
     \hspace{0.1em}\hfill\ensuremath{\blacksquare}
\end{align}
\vspace{-0.1em}

At this point, considering a fixed partition $\mathbb{W}_{\rahel{\bar{p}_w^*},i}$ and using the candidate $s_{k+1}=s_{k}^{*}$, combining Theorem~\ref{theorem:theo4boccia} with Proposition~\ref{proposition:notminsteadynotcosetwolayers} and Assumption~\ref{assumption:boundedbyd} yields \begin{align*}
    &\tilde{J}_{i}^{*}(x_{k+1},\mathbb{W}_{\rahel{\bar{p}_w^*},i}) - \tilde{J}_{i}^{*}(x_{k}, \mathbb{W}_{\rahel{\bar{p}_w^*},i})\\ 
    &\leq V^{*}_{N,i}(x_{k+1},s_{k}^{*},\mathbb{W}_{\rahel{\bar{p}_w^*},i}) + \ell_{T,i}(s_{k}^{*} - r_{i}) - l_{\mathbb{W}_{\rahel{\bar{p}_w^*}},r,i,\min} \\ 
    &- V^{*}_{N,i}(x_{k},s_{k}^{*},\mathbb{W}_{\rahel{\bar{p}_w^*},i}) - \ell_{T,i}(s_{k}^{*} - r_{i}) + l_{\mathbb{W}_{\rahel{\bar{p}_w^*}},r,i,\min} \\
    &\overset{\eqref{eq:theorem5twolayer}}{\leq} - \bar{\alpha}_{N,i}\ell_{i}^{*}(x_{k},s_{k}^{*})\\ &\overset{\eqref{eq:prop12layer}}{\leq} -\frac{\bar{\alpha}_{N,i}}{2}(\ell_{i}^{*}(x_{k},s_{k}^{*}) + \bar{a}_{i}d_{i}(s_{k}^{*})_{\rahel{\mathbb{T}_{\bar{p}_w^*,r,i}}}^{2}) \\ 
    &\overset{\eqref{eq:boundstagecosttwolayer}}{\leq} -\frac{\bar{\alpha}_{N,i}}{2}(\alpha_{1,i}d_{i}((x_{k},u_{x_{k},s_{k}^{*}}^{*})-s_{k}^{*})^{2} + \bar{a}_{i}d_{i}(s_{k}^{*})_{\rahel{\mathbb{T}_{\bar{p}_w^*,r,i}}}^{2})\\  & \leq - \tilde{\alpha}_{N,i}(d_{i}((x_{k},u_{x_{k},s_{k}^{*}}^{*})-s_{k}^{*})^{2} + d_{i}(s_{k}^{*})_{\rahel{\mathbb{T}_{\bar{p}_w^*,r,i}}}^{2}),
    \label{eq:2theorem2twolayer}
\end{align*}
where we defined $\tilde{\alpha}_{N,i}=\frac{\bar{\alpha}_{N,i}}{2}\text{min}\{\alpha_{1,i},\bar{\alpha}_i\}$. Using Proposition~\ref{proposition:upperlowerboudJtilde}, we have 
\begin{equation*}
    \begin{aligned}
    &\tilde{J}_{i}^{*}(x_{k+1},\mathbb{W}_{\rahel{\bar{p}_w^*},i}) - \tilde{J}_{i}^{*}(x_{k}, \mathbb{W}_{\rahel{\bar{p}_w^*},i}) \\
    &\leq - \tilde{\alpha}_{N,i}(d_{i}(s_{k}^{*})_{\rahel{\mathbb{T}_{\bar{p}_w^*,r,i}}}^{2} + d_{i}((x_{k},u_{x_{k},s_{k}^{*}}^{*})-s_{k}^{*})^{2})\\ 
    & \leq  -\frac{\tilde{\alpha}_{N,i}}{b_{i}}\tilde{J}_{i}^{*}(x_{k}, \mathbb{W}_{\rahel{\bar{p}_w^*},i})
    \end{aligned}
\end{equation*}

\noindent and obtain the exponential convergence relation for $\tilde{J}_{i}^{*}(x_{k+\tilde{k}},\mathbb{W}_{\rahel{\bar{p}_w^*},i})$
\begin{equation}
\begin{split}
    \tilde{J}_{i}^{*}(x_{k+\tilde{k}},\mathbb{W}_{\rahel{\bar{p}_w^*},i})
    \leq \Big(1 - \frac{\tilde{\alpha}_{N,i}}{b_{i}} \Big)^{\tilde{k}} \tilde{J}_{i}^{*}(x_{k}, \mathbb{W}_{\rahel{\bar{p}_w^*},i}),
\end{split}
\label{eq:4fintitetimecovergencetwolayer}
\end{equation}
which is guaranteed to converge, because $b_{i} >\tilde{\alpha}_{N,i}>0$.

\subsubsection{Finite Time Update Condition Fulfillment, Part 1}\label{subsection:finitetimeconditiontwolayersp1}

Given \eqref{eq:4fintitetimecovergencetwolayer}, and the fact that $\tilde{J}_{i}^{*}(x_{k}, \mathbb{W}_{\rahel{\bar{p}_w^*},i})$ admits a uniform bound due to compact constraints and continuity of the cost, it follows that,
for any $\bar{\epsilon} > 0$, there exists a finite time step $\tilde{k}'>0$ such that $\tilde{J}_{i}^{*}(x_{k+\tilde{k}'},\mathbb{W}_{\rahel{\bar{p}_w^*},i}) < \bar{\epsilon}$. 
Hence, by choosing $\bar{\epsilon} > 0$ small enough and considering the lower bound~\eqref{eq:lowerbound2layer}, the state at time step $\tilde{k}'$ is arbitrary close to the set $\rahel{\mathbb{T}_{\bar{p}_w^*,r,i}}$, and thus the update conditions~\eqref{eq:updatereq1},~\eqref{eq:updatereq2} hold (assuming we are not already in a centroidal Voronoi configuration).

\subsubsection{Finite Time Update Condition Fulfillment, Part 2}\label{subsection:finitetimeconditiontwolayersp2}We now need to ensure that the remaining Voronoi partition update condition~\eqref{eq:feasibilitycond} is fulfilled in finite time with the candidate state and input sequences proposed in~\eqref{eq:lemma1proposal}.  
Specifically, we need to find an upper bound for 
\begin{align}
    & V_{N,i}(x_{k+\tilde{k}'}, \hat{u}, s_{k}^{*}, \mathbb{W}_{\rahel{\bar{p}_w^*}}) = \sum_{l=0}^{N-2}\ell_{i}(x_{l\vert k+\tilde{k}'}^{*}, u_{l \vert k+\tilde{k}'}^{*},s_{k}^{*}) + \notag \\ &  \ell_{i}(x_{N-1\vert k+\tilde{k}'}^{*},\bar{u}_{k+\tilde{k}'}^{*},s_{k}^{*}) + \notag \\ & \ell_{i}(f(x_{N-1\vert k+\tilde{k}'}^{*},\bar{u}_{k+\tilde{k}'}^{*}),\bar{u}_{k+\tilde{k}'}^{*},s_{k}^{*}), 
    \label{eq:trackingcostcandidate}
\end{align}
and such a bound is required to be decreasing as $\tilde{k}'$ increases, along the lines of~\eqref{eq:4fintitetimecovergencetwolayer}. To this aim, we will make use of $\bar{\epsilon}$ as introduced above that can be seen as a decreasing function of $\tilde{k}'$.
The first term of~\eqref{eq:trackingcostcandidate} can be upper bounded by~\eqref{eq:4fintitetimecovergencetwolayer} as 
\begin{align}
    &\epsilon \geq V_{N,i}^{*}(x_{k+\tilde{k}'},s_{k}^{*},\mathbb{W}_{\rahel{\bar{p}_w^*}})  \geq \sum_{l=0}^{N-2}\ell_{i}(x_{l\vert k+\tilde{k}'}^{*}, u_{l \vert k+\tilde{k}'}^{*},s_{k}^{*}). \notag 
\label{eq:7finitetimeconvergence}
\end{align}
The upper bound for the last two terms is given by the following proposition.
\begin{proposition} Let Assumption~\ref{assumption:dynamics} and~\ref{assumption:boundedbyd} hold. 
Then, for any $x \in \mathbb{R}^{n_{i}}$ and any $s = (\bar{x},\bar{u}) \in \mathbb{S}_i$, it holds that
\begin{equation}
    \ell_{i}(x,\bar{u},s) + \ell_{i}(f(x,\bar{u}),\bar{u},s)\leq \gamma_{i}' \cdot \ell_{i}^{*}(x,s),
\label{eq:assumption22}
\end{equation}
where $\gamma_{i}'=\frac{2\alpha_{2,i}}{\alpha_{1,i}}\max_j\{1+\mathcal{L}_{i}^{\eta_j}\}$.
\label{proposition:upperboundsuccedingstagecost}
\end{proposition}

\noindent \emph{Proof:} 
In accordance to equation~\eqref{eq:boundstagecosttwolayer}, it holds that
\begin{align}
    &\ell_{i}(x,\bar{u},s) + \ell_{i}(f_{i}(x,\bar{u}),\bar{u},s)\leq \label{eq:fulfillassumption22}\\ 
    &2\alpha_{2,i} {\sum_{j=1}^{n_i}  \Big(\vert x_{j}-\bar{x}_j \vert^{\eta_j} + \vert f_{i,j}(x,\bar{u})-\bar{x}_j \vert^{\eta_j}\Big)}.\notag 
\end{align}
Because $\bar{x}$ is a steady state, we have that $f_{i,j}(x,\bar{u}) - \bar{x}_j = f_{i,j}(x,\bar{u}) - f_{i,j}(\bar{x},\bar{u})$. Thus, by Lipschitz continuity and reusing \eqref{eq:boundstagecosttwolayer}, we have that~\eqref{eq:fulfillassumption22} is upper bounded by 
\begin{align*}
    \hspace{3.3em}&2\alpha_{2,i}\sum_{j=1}^{n_i} (1 + \mathcal{L}_{i}^{\eta_j})\vert x_{j}-\bar{x}_j \vert^{\eta_j} \leq \gamma_i'\ell_i^*(x,s).
    \hspace{3.3em}\hfill\ensuremath{\blacksquare}
\end{align*}
\vspace{-1.0em}

\noindent Finally, noting also that
\begin{equation}
{\displaystyle \bar{\epsilon} \geq \ell_{i}(x_{N-1 \vert k+\tilde{k}'}^{*}, u_{N-1 \vert k+\tilde{k}'}^{*}, s_{k}^{*}) \geq \ell^{*}_{i}(x_{N-1 \vert k+\tilde{k}'}^{*},s_{k}^{*})},\notag 
\label{eq:8finitetimeconvergence}
\end{equation}
we obtain the desired bound
\begin{align}
     & V_{N,i}(x_{k+\tilde{k}'}, \hat{u}, s_{k}^{*}, \mathbb{W}_{\rahel{\bar{p}_w^*}}) \qquad \text{(see~\eqref{eq:trackingcostcandidate})} \notag \\
     &  \leq \sum_{l=0}^{N-2}\ell_{i}(x_{l\vert k+\tilde{k}'}^{*}, u_{l \vert k+\tilde{k}'}^{*},s_{k}^{*}) + \gamma_{i}' \ell^{*}_{i}(x_{N-1\vert k+\tilde{k}'}^{*},s_{k}^{*}) \notag \\ &
     \leq (1+ \gamma_{i}')\bar{\epsilon}.
\label{eq:9finitetimeconvergence}
\end{align}
Since it holds for any of the addenda, which in turn are lower-bounded by a function describing the distance between state and setpoint, they will become arbitrarily close. \rahel{In combination with Lemma~\ref{lemma:inclusioninnewinterior}, the update condition in~\eqref{eq:feasibilitycond} will then hold for a finite~$\tilde{k}'$.}

\subsection{Proof of Theorem~\ref{theorem:twolayerslearning}}
\label{subsec:twolayerslearningproof}
The result of Theorem~\ref{theorem:twolayerslearning} builds upon~\cite[Proposition 1]{Todescato2017}. It ensures that the learning-based solution converges to \rrahel{a centroidal} Voronoi partition if the estimate $\hat{\phi}_t$ converges in probability to the true density $\phi$. Then, convergence to a centroidal Voronoi configuration follows by Theorem~\ref{theorem:twolayers}. 

\begin{proposition}~\cite[Proposition 1]{Todescato2017} Assume $\hat{\phi}_{t}(p) \xrightarrow{\mathcal{P}} \phi(p)$, where $\mathcal{P}$ denotes convergence in probability in the space of continuous functions. Moreover, assume it is possible to start the Lloyd algorithm from the configuration reached by Algorithm~\ref{alg:twolayermpcalglearningimp} at time $\bar{t}$. Then the centroids $c^{L}_{\bar{t}}$ obtained from the Lloyd algorithm and the estimated ones $\hat{c}_{\bar{t}}$ coincide. Furthermore, for any $0<\bar{\delta} < 1$, $\varepsilon > 0$ an integer $N$ can be picked, such that there exists a $\bar{t}$ sufficiently large such that
\begin{equation}
     \mathcal{P}[ \Vert \hat{c}_{\bar{t} + t} - c^{L}_{\bar{t} + t} \Vert < \varepsilon ] > 1 - \bar{\delta}, \ \ t = 0, \hdots, N,
\label{eq:learningprop1}
\end{equation}
where $\mathcal{P}$ is associated to the Gaussian distribution of the estimated coefficients.
\label{proposition:convergenceofcentroidstwolayerslearning}
\end{proposition}
\noindent By construction, the locations of the agents never coincide, so it is possible to start Lloyd's algorithm from an arbitrary instant in time. Then, it remains to show that $\hat{\phi}_{t}(p) \xrightarrow{\mathcal{P}} \phi(p)$ holds by applying Algorithm~\ref{alg:twolayermpcalglearningimp}.

\begin{lemma}
\rahel{Let Assumptions~\ref{assumption:dynamics},~\ref{assumption:expocostcontrollability},~\ref{assumption:boundedbyd},~\ref{assumption:bayesianconvergence},and~\ref{assumption:steadystatedecreasetwolayers} hold.}
Consider a horizon length $N \geq N^{*}$, constants $\rho>0, r_{i,\max} >0$, and $\varepsilon>0$ sufficiently small. Then, for any $p \in \mathbb{A}$, $\hat{\phi}_t(p) \xrightarrow{\mathcal{P}} \phi(p)$.
\label{lemma:learning}
\end{lemma}
\noindent \emph{Proof:} We prove this by showing that, given any partition configuration $\mathbb{W}_p$, the maximum variance value inside the $i-$th partition, $\text{Var}_{i,\max,t} = \max_{p \in \mathbb{W}_{p,i}} \text{Var}_t(p)$, converges to zero for each $i$. This implies that $\text{Var}_t(p)$ at any point $p$ inside the partition
converges to zero with increasing $t$: this is equivalent to proving $\mathscr{L}^2$-convergence of $\phi_t(p)$, from which convergence in probability follows.\\
For a time instant $t$, define $p^{\text{maxvar}}_{t,i} = \arg\max_{p \in \mathbb{W}_{p,i}} \text{Var}_t(p)$, and consider the position of the $i-$th agent, $p_{t,i}$, such that $\|p^{\text{maxvar}}_{t,i} - p_{t,i}\| \leq (\rho + r_{\max} + \varepsilon)$. Next, we show $\text{Var}(p_{t,i}) \geq \tilde{c}\text{Var}(p^{\text{maxvar}}_{t,i})$ for some $\tilde{c}>0$. To this aim, we have
\begin{align*}
    \text{Var}(p^{\text{maxvar}}_{t,i}) &\leq 2\Phi(p_{t,i})\Sigma_{t}\Phi(p_{t,i})^{\top}  \\
    + &2(\Phi(p_{t,i}) - \Phi(p^{\text{maxvar}}_{t,i}))\Sigma_t(\Phi(p_{t,i}) - \Phi(p^{\text{maxvar}}_{t,i}))^{\top}\\
    &\leq 2 \text{Var}(p_{t,i}) + 2\mathcal{L}_{\Phi}^2(\rho + r_{\max} + \varepsilon)^2\|\Sigma_t\|,
\end{align*}
with some Lipschitz constant $\mathcal{L}_\Phi>0$ from Assumption~\ref{assumption:bayesianconvergence}. Moreover, by the linear independence of the features in $\Phi$ stated in the same Assumption, we have $\text{Var}(p^{\text{maxvar}}_{t,i}) \geq c_{\Phi}\|\Sigma_t\|$. Applying this inequality yields
\begin{equation*}
   \text{Var}(p_{t,i}) \geq \underbrace{(1 - 2\mathcal{L}_{\Phi}^2(\rho+r_{\max} + \varepsilon)^2/c_{\Phi})/2}_{=:\tilde{c}} \text{Var}(p^{\text{maxvar}}_{t,i}), 
\end{equation*}
with $\tilde{c}>0$ for $\gamma>0, r_{i,\max} >0, \varepsilon>0$ sufficiently small.\\
The proof is concluded by showing that $\text{Var}(p_{t,i})$ converges to zero. Assume for contradiction that \mbox{$\lim\sup_{t \rightarrow +\infty} \Phi(p_{t,i})\Sigma_t\Phi(p_{t,i})^{\top} \! = \! c^{\prime} > 0$.} We can isolate a subsequence $\{t_{\tau}\}_{\tau}$ such that $\lim_{\tau \rightarrow +\infty} \Phi(p_{t_{\tau},i})\Sigma_{t_{\tau}}\Phi(p_{t_{\tau},i})^{\top} \! = \! c^{\prime}$. However, from recursive application of equation~\eqref{eq:bayessigmaupdate} we have that \mbox{$\lim_{\tau\rightarrow\infty}\Sigma_{t_\tau}^{-1} \! \succeq \! \sum_{\tau=1}^{\infty}\Phi(p_{t_{\tau},i})^{\top}\Phi(p_{t_{\tau},i})$}. 
This implies that $\Phi(p_{t_\tau,i})\Sigma_{t_\tau}\Phi(p_{t_\tau,i})^\top$ tends to 0, contradicting the starting hypothesis. 
\hfill$\blacksquare$

\subsection{Proof of Theorem~\ref{theorem:onelayer}}
\label{subsubsec:onelayerknownenvironmenttheory}
We first present two preliminary results that will be needed in the main proof. In particular, Proposition~\ref{proposition:notminsteadynotcoseonelayer} generalizes Proposition~\ref{proposition:notminsteadynotcosetwolayers} to non-convex target costs, and Proposition~\ref{proposition:voronoiupdatecostdecrease} ensures that an update of the partitions in accordance to the current steady state does not increase the cost~\eqref{eq:cortescost}.

\begin{proposition} Let Assumptions~\ref{assumption:dynamics},~\ref{assumption:expocostcontrollability},~\ref{assumption:boundedbyd},~\ref{assumption:ballwithminconvex} and~\ref{assumption:steadystatedecreaseonelayer} hold. Then there exists a constant $\bar{a}_{i} > 0$ such that, for any $\bar{p}_{w,i}^* \in \mathbb{S}_{\mathbb{A},i}^{\mathrm{p}}$ any feasible state $x_{k} \in \mathbb{X}^{\mathrm{int}}:Cx_k \in \bar{\mathbb{W}}_{\bar{p}_{w}^*,i}^{\mathrm{int}}$, the optimal solution of the MPC problem~\eqref{eq:fullnonlineartrackingmpconelayer} satisfies
\begin{equation}
    \ell_{i}^*(x_{k}, s_{k}^{*}) \geq \bar{a}_{i}\kappa(\Vert \bar{p}_{k}^{*} \Vert)_{\rahel{\mathbb{T}_{\bar{p}_{k}^{*},\bar{p}_w^*,i}}}.
\label{eq:prop1}
\end{equation}
\label{proposition:notminsteadynotcoseonelayer}
\end{proposition}
\vspace{-1.2em}
\noindent \emph{Proof:} Follows the same steps as the proof of Proposition 2, with Assumption~\ref{assumption:steadystatedecreasetwolayers} replaced by Assumptions~\ref{assumption:ballwithminconvex} and~\ref{assumption:steadystatedecreaseonelayer}.\hfill{$\blacksquare$} 

\begin{proposition} 
For any $\bar{p},\bar{p}_k\in\mathbb{A}^M$, it holds
\begin{equation}
    H(\bar{p}_{k},\mathbb{W}_{\bar{p}_{k}}) \leq  H(\bar{p}_{k},\mathbb{W}_{\bar{p}}).
\label{eq:condassumption6}
\end{equation}
\label{proposition:voronoiupdatecostdecrease}
\end{proposition}
\vspace{-1.5em}
\noindent \emph{Proof:} 
Given setpoint positions $\bar{p}$, each partition $\mathbb{W}_{\bar{p},i}$ admits itself a partition $\{\mathbb{W}_{\bar{p} \rightarrow \bar{p}_{k},i \rightarrow j}\}_{j=1}^M$, explicitly depending on $\bar{p}_{k}$, where each $\mathbb{W}_{\bar{p} \rightarrow \bar{p}_{k},i \rightarrow j}$ describes the set of points which are currently assigned to the steady-state of agent $i$, but would be assigned to the steady-state of agent $j$ in case of a Voronoi partition update in accordance to $\bar{p}_{k}$. 
Therefore, with $g$ non-decreasing, it holds:
\begin{align*}
    &H(\bar{p}_{k},\mathbb{W}_{\bar{p}})  = \sum_{i=1}^{M}\int_{\mathbb{W}_{\bar{p},i}} g(\Vert q - \bar{p}_{i} \Vert)\phi(q)dq \\ 
    &= \sum_{i=1}^{M}\sum_{j=1}^{M}\int_{\mathbb{W}_{\bar{p}\rightarrow\bar{p}_{k},i\rightarrow j}} g(\Vert q - \bar{p}_{i} \Vert)\phi(q)dq \\ 
    &\overset{\eqref{eq:voronoidef}}{\geq} \sum_{i=1}^{M}\sum_{j=1}^{M}\int_{\mathbb{W}_{\bar{p}\rightarrow\bar{p}_{k},i\rightarrow j}} g(\Vert q - \bar{p}_{j} \Vert)\phi(q)dq \\
    &= \sum_{i=1}^{M}\int_{\mathbb{W}_{\bar{p}_{k},i}} g(\Vert q - \bar{p}_{i} \Vert)\phi(q)dq = H(\bar{p}_{k},\mathbb{W}_{\bar{p}_{k}}).
    \label{eq:assumption6}
    \hspace{3.0em}\hfill\ensuremath{\blacksquare}
\end{align*}

\noindent
The remainder of the proof amounts to showing the following facts: 1) convergence to a fixed setpoint, taking into account a possible update of the Voronoi tessellation; 2) convergence to a centroidal Voronoi partition; and 3) finite-time partition update. The following results thereby rely on a Lyapunov like decrease of the storage function for the MPC in~\eqref{eq:fullnonlineartrackingmpconelayer}.\\
The storage function we are considering is the summation of the optimal MPC cost over all agents, that is
\begin{equation}
\begin{aligned}
&J^{*}(x_{k}, \mathbb{W}_{\rahel{\bar{p}^{*}_{w}}}) = V_{N}^{*}(x_{k}, s_{k}^{*},\mathbb{W}_{\bar{p}^{*}_{w}}) + \lambda  H(\bar{p}_{k},\mathbb{W}_{\bar{p}^{*}_{w}}). 
\end{aligned}
\label{eq:lyapunovonelayer}
\end{equation}
When the subscript $i$ is dropped, then we refer to quantities of the overall system. The storage function~\eqref{eq:lyapunovonelayer} has an upper and lower bound. The first exists thanks to the compactness of $\mathbb{A},\mathbb{X},\mathbb{U}$ and the continuity of the cost. Given that the locational optimization cost is always greater or equal to zero, the latter follows by
\begin{align}
     &J^{*}(x_{k},\mathbb{W}_{\rahel{\bar{p}_w^*}}) \geq V_{N}^{*}(x_{k},s_{k}^{*},\mathbb{W}_{\rahel{\bar{p}_w^*}}) \geq \ell^{*}(x_{k},s_{k}^{*}) \notag \\ &\overset{\eqref{eq:boundstagecosttwolayer}}{\geq} \alpha_{1}d((x_{k},u_{x_{k},s_{k}^{*}}^{*})-s_{k}^{*})^{2} \geq 0.
\label{eq:lowerbound}
\end{align}

\subsubsection{Convergence to Setpoint} We now prove that the optimal cost at time $k+1$ on a candidate update partition $\mathbb{W}_{\bar{p}_{k}^{*}}$ decreases with respect to its value corresponding to the Voronoi partition induced by the setpoint positions $\bar{p}$ at time $w$. In other words, we show that $J^{*}(x_{k+1},\mathbb{W}_{\bar{p}_{k}^{*}}) < J^{*}(x_{k},\mathbb{W}_{\rahel{\bar{p}_w^*}})$. By definition of the involved storage function, this in turn implies that $x_k$ converges to the current steady-state $s_k^*$.\\
Recalling the definitions of $\gamma_{i}'$ and $\gamma_{i,V_{\max,i}}$ given in Proposition~\ref{proposition:upperboundsuccedingstagecost} and the proof of Proposition~\ref{proposition:upperlowerboudJtilde}, respectively, let $\gamma_{\max} := \max_{i=1,...,M}\max{\{\gamma_{i}',\gamma_{i,V_{\max,i}}\}}$ and $\gamma' = \max_{i=1,...,M}{\{\gamma_{i}'\}}$. Moreover, we will make use of the following result taken from the proof of Theorem~\ref{theorem:theo4boccia}:
\begin{equation}
\begin{split}
    &\left( \frac{\gamma_{i,V_{\max,i}}}{\gamma_{i,V_{\max,i}}-1} \right)^{N-1} \ell_{i}^{*}(\hat{x}_{N-1\vert k},s_{k}^*) \\ & \leq V_{N,i}(x_{k},s_k^*,\mathbb{W}_{\rahel{\bar{p}_w^*},i}) \leq \gamma_{i,V_{\max,i}} \ell^{*}_{i}(x_{k},s_k^*).
\end{split}
\label{eq:prooftheom4}
\end{equation}

Next, we get the following storage function decrease: 
\begin{align}
    &J^{*}(x_{k+1},\mathbb{W}'_{\bar{p}_{k}^{*}}) \leq J^{*}(x_{k+1}, s^{*}_{k},\mathbb{W}_{\bar{p}_{k}^{*}}) \label{eq:intermediateJ1} \\ & = 
    V_{N}^{*}(x_{k+1},s_{k}^{*},\mathbb{W}_{\bar{p}_{k}^{*}}) + \lambda H(\bar{p}_{k}^{*},\mathbb{W}'_{\bar{p}_{k}^{*}}) 
    \notag \\ &\overset{\eqref{eq:lemma1proposal},\eqref{eq:feasibilitycond}}{\leq} V_{N}(\hat{x}_{\cdot \vert k}, \hat{u}_{\cdot \vert k},s_{k}^{*}) + \lambda H(\bar{p}_{k}^{*},\mathbb{W}_{\bar{p}_{k}^{*}})  
    \notag \\ &\overset{\eqref{eq:condassumption6},\eqref{eq:feasibilitycond}}{\leq} V_{N}(\hat{x}_{\cdot \vert k}, \hat{u}_{\cdot \vert k},s_{k}^{*}) + \lambda H(\bar{p}_{k}^{*},\mathbb{W}_{\rahel{\bar{p}_w^*}}) 
    \notag \\
    &\overset{\eqref{eq:lemma1proposal}}{=} J^{*}(x_{k},\mathbb{W}_{\rahel{\bar{p}_w^*}}) - \ell(x_{k},u_{0\vert k}^{*},s_{k}^{*}) - \ell(x_{N-1\vert k}^{*},u_{N-1\vert k}^{*},s_{k}^{*}) \notag \\ & + \ell(\hat{x}_{N-1 \vert k},\bar{u}^{*}_{k},s_{k}^{*}) + \ell(f(\hat{x}_{N-1 \vert k},\bar{u}^{*}_{k}),\bar{u}^{*}_{k},s_{k}^{*}) \notag \\
    & \leq J^{*}(x_{k},\mathbb{W}_{\rahel{\bar{p}_w^*}}) - \ell^{*}(x_{k},s_{k}^{*}) - \ell^{*}(x_{N-1\vert k}^{*},s_{k}^{*}) \notag \\ & + \ell(\hat{x}_{N-1\vert k},\bar{u}^{*}_{k},s_{k}^{*}) + \ell(f(\hat{x}_{N-1 \vert k},\bar{u}^{*}_{k}),\bar{u}^{*}_{k},s_{k}^{*}) \notag \\
    &\overset{\eqref{eq:assumption22}}{\leq} J^{*}(x_{k},\mathbb{W}_{\rahel{\bar{p}_w^*}}) - \ell^{*}(x_{k},s_{k}^{*}) + (\gamma' -1)\ell^{*}(x_{N-1\vert k}^{*},s_{k}^{*}) \notag \\
    &\overset{\eqref{eq:prooftheom4}}{\leq} J^{*}(x_{k},\mathbb{W}_{\rahel{\bar{p}_w^*}}) - \ell^{*}(x_{k},s_{k}^{*})  \notag \\ & +(\gamma_{\max} - 1)\left( \frac{\gamma_{\max}-1}{\gamma_{\max}} \right)^{N-1} \gamma_{\max} \ell^{*}(x_{k},s_{k}^{*}). \notag
\end{align}
Thus, we obtain 
\begin{equation}
 J^{*}(x_{k+1},\mathbb{W}'_{\bar{p}_{k}^{*}}) \leq J^{*}(x_{k},\mathbb{W}_{\rahel{\bar{p}_w^*}}) - \bar{\alpha}_{N}\ell^{*}(x_{k},s_{k}^{*}),
 \label{eq:1theorem2}
\end{equation}
where $\bar{\alpha}_{N} = 1-\left( \dfrac{(\gamma_{\max}-1)}{\gamma_{\max}} \right)^{N} \gamma_{\max}^{2} > 0$ by choosing $N>N^{*}$ large enough. Thus, inequality~\eqref{eq:1theorem2}, in combination with lower and upper bounds on $J^{*}$ and Assumption~\ref{assumption:boundedbyd}, ensures that $\lim_{k\rightarrow \infty} \|  x_{k} - s_{k}^{*} \|  = 0$.

\subsubsection{Convergence to Centroidal Voronoi partition} Combining Proposition~\ref{proposition:notminsteadynotcoseonelayer} with the result seen in~\eqref{eq:1theorem2}, it follows that $J^{*}$ decreases as long as $x_{k} \neq s_{k}^{*}$ or $\bar{p}_{i,k}^{*} \notin \rahel{\mathbb{T}_{\bar{p}_{i,k}^{*},\bar{p}_w^*,i}}$ for all $i \in \{1,\hdots,M \}$. In fact,
\begin{align}
    &J^{*}(x_{k+1}, \mathbb{W}_{\rahel{\bar{p}_w^*}}) - J^{*}(x_{k}, \mathbb{W}_{\rahel{\bar{p}_w^*}})
    \leq - \bar{\alpha}_{N}\ell^{*}(x_{k},s_{k}^{*}) \notag \\
    &\leq -\frac{\bar{\alpha}_{N}}{2}(\ell^{*}(x_{k},s_{k}^{*}) + \bar{a}\kappa(\Vert \bar{p}_{k}^{*} \Vert)_{\rahel{\mathbb{T}_{\bar{p}_{k}^{*},\bar{p}_w^*}}}^{2}),
\label{eq:2theorem2}
\end{align}
with $\bar{a}\kappa(\Vert \bar{p}_{k}^{*} \Vert)_{\rahel{\mathbb{T}_{\bar{p}_{k}^{*},\bar{p}_w^*}}}^{2} = \max_{i} \bar{a}_i \sum_{i=1}^{M}\kappa(\Vert \bar{p}_{i,k}^{*} \Vert)_{\rahel{\mathbb{T}_{\bar{p}_{k}^{*},\bar{p}_w^*,i}}}^{2}$.
Accordingly, it can be concluded that $\lim_{k \to \infty}p_{i,k} \in \rahel{\mathbb{T}_{\bar{p}_{k}^{*},\bar{p}_w^*,i}}$. 

\subsubsection{Finite Time Partition Update} Finally, it has to be shown that equation~\eqref{eq:feasibilitycond} is fulfilled in finite time and the Voronoi partition can be updated. For this purpose, we leverage the result in~\eqref{eq:1theorem2} with respect to a fixed set of partitions $\mathbb{W}_{\rahel{\bar{p}_w^*}}$ and candidate $s_{k+1} = s_k^*$, and combine them with Assumption~\ref{assumption:expocostcontrollability}. Then, we obtain 
\begin{equation}
\begin{split}
     V_{N}^{*}(x_{k+k^{\prime}},s_{k}^{*},\mathbb{W}_{\rahel{\bar{p}_w^*}}) 
     \leq \Big(1 - \frac{\bar{\alpha}_{N}}{\gamma_{\max}} \Big)^{k^{\prime}} V_{N}^{*}(x_{k},s_{k}^{*},\mathbb{W}_{\rahel{\bar{p}_w^*}}),
\end{split}
\label{eq:4fintitetimecovergence}
\end{equation}
that converges because by definition $\gamma_{\max}> \bar{\alpha}_{N} > 0$. At this point, the same reasoning carried out in Appendix~\ref{subsection:finitetimeconditiontwolayersp2} can be used to show that condition~\eqref{eq:feasibilitycond} is met in finite time.

\noindent In conclusion, the algorithm does converge to a configuration in which each agent's position defines a local minimum of $H(p, \mathbb{W})$ and the Voronoi partition does not update anymore: that is, a centroidal Voronoi configuration is obtained. 

\subsection{Proof of Theorem~\ref{theorem:onelayerlearning}}
\label{subsec:onelayerunknownenvironmenttheory}
For this proof we first show convergence to a steady state describing a local minimum of the locational optimization function considering the converged estimate. In a second step, an upper bound for the remaining variance at each reachable position is defined.
\subsubsection{Convergence} Similar to the proof of Theorem~\ref{theorem:onelayer}, we define the storage function as
\begin{align}
&J^{*}(x_{k}, \mathbb{W}_{\bar{p}^{*}_{w}},\hat{\phi}_{k}) \notag \\ & =  V_{N}^{*}(x_{k}, s_{k}^{*},\mathbb{W}_{\bar{p}^{*}_{w}}) + \lambda (H(\bar{p}_{k}^{*},\mathbb{W}_{\bar{p}^{*}_{w}}, \hat{\phi}_{k}) - S\text{Var}_{k}(\bar{p}_{k}^{*}))\notag \\
& = \sum_{i=1}^{M} \bigg( \sum_{l=0}^{N-1}\ell_{i}(x_{i,l\vert k}^{*}, u_{i,l \vert k}^{*}, s_{i,k}^{*}) \label{eq:lyapunovonelayerlearning} \\  &+ \lambda (H_{i}(\bar{p}_{i,k}^{*},\mathbb{W}_{\bar{p}^{*}_{w},i},\hat{\phi}_{k}) - S\text{Var}_{M,k}(\bar{p}_{i,k}^{*}) \bigg), \notag
\end{align}
Following the reasoning in equations~\eqref{eq:intermediateJ1} and~\eqref{eq:1theorem2} for the adapted storage function, we end up with
\begin{align}
&J^{*}(x_{k+1}, \mathbb{W}'_{\bar{p}^{*}_{k}},\hat{\phi}_{k+1}) \leq J^{*}(x_{k}, \mathbb{W}_{\bar{p}_w^*},\hat{\phi}_{k}) - \underbrace{\bar{\alpha}_{N}\ell^{*}(x_{k},s_{k}^{*})}_{\geq 0} 
\notag \\ &- \lambda \sum_{i=1}^{M} \int_{\mathbb{W}_{\bar{p}_{w}^*},i} g(\Vert q - \bar{p}^{*}_{i,k} \Vert)(\hat{\phi}(q)_{k}-\hat{\phi}(q)_{k+1})dq \notag \\ &- \underbrace{\lambda S (\text{Var}_{M,k}(\bar{p}^{*}_{k}) - \text{Var}_{M,k+1}(\bar{p}^{*}_{k}))}_{\geq 0}. \label{eq:lyth5}
\end{align}
Using the Bernstein-von Mises Theorem~\cite[Chapter 10.2]{vanderVaart1998} we have that $\hat{\phi}_{k}(q_k)$ converges in probability to the Gaussian distribution with mean equal to the maximum likelihood estimate, and with variance equal to $1/k$ times the asymptotic variance of the maximum likelihood estimate (i.e., the inverse of the Fisher information matrix). Using~\eqref{eq:bayesupdate}, this convergence implies that $(\hat{\theta}_{k}-\hat{\theta}_{k+1}) \xrightarrow{\mathcal{P}} 0$. Thus, the storage function decrease in~\eqref{eq:lyth5} implies $\lim_{k\rightarrow\infty}\|x_k-s_k^*\|=0$ in probability. Applying Proposition~\ref{proposition:notminsteadynotcoseonelayer} ensures convergence of the steady state to a local minimum of the target cost with respect to the current density~$\phi_k$, similarly to Equation~\eqref{eq:2theorem2}.

\subsubsection{Remaining Variance} 
Given the converged state, the following proposition provides a bound on the variance for all feasible position $p\in\mathbb{F}:=\lim_{k\rightarrow\infty}\mathbb{F}_k$.  

\begin{proposition} 
Suppose the closed loop converges to a setpoint $\overline{s}_i^*$ with state $\bar{x}_{i}^{*}$ and position $\bar{p}_{i}^{*}$. 
Then, for each agent $i \in \{1,\hdots,M\}$, there exists a uniform constant $\Delta H_i\geq 0$ such that, for any $p\in \mathbb{F}_i$ with corresponding $s_{p}\in\mathbb{S}_{\rahel{\bar{p}_w^*},i}$, the variance is bounded by
\begin{align}
    &\lim_{k \rightarrow \infty}\mathrm{Var}_{k}(p) \leq  \frac{\Delta H_i}{S}. 
\label{eq:upperboundvaronelayerlearningsimplified}
\end{align}
\label{proposition:onelayerlearning}
\end{proposition}
\vspace{-1.5em}
\noindent \emph{Proof:} 
Convergence ensures $\lim_{k\rightarrow\infty} V_N(\overline{x}^*,\overline{s}^*)=0$, $\text{Var}_\infty(\overline{p}^*)=0$ and hence the minimum cost in the MPC problem is given by $\lambda H(\overline{p}^*,\mathbb{W}_{\overline{p}^*,i},\hat{\phi}_\infty)$. 
The cost of any feasible position $p\in\mathbb{F}_i$ with corresponding setpoint $s_p\in\mathbb{S}_i$ is larger than or equal to this minimum, i.e.,
\begin{align}
    &\lambda H(\overline{p}^*,\mathbb{W}_{\overline{p}^*,i},\hat{\phi}_\infty) \label{eq:contraditiononelayerlearning2}\\
    &\leq V_{N}^{*}(\bar{x}_{i}^{*}, s_{p},\mathbb{W}_{\bar{p},i}) +\lambda (H(\overline{p},\mathbb{W}_{\overline{p}^*,i},\hat{\phi}_\infty)-S\text{Var}_\infty({p})).
    \nonumber
\end{align}
Note that $V_N(\overline{x}_i^*,p)\leq V_{\max,i}$; moreover, the definition of the coverage cost $H$ in~\eqref{eq:cortescost}   
with $g$ continuous and $\mathbb{W}_{\overline{p}^*,i}\subseteq\mathbb{A}$ compact ensures that there exists a uniform constant $\Delta H_i\geq 0$ such that
\begin{align*}
    &\dfrac{1}{\lambda}V_{N}^{*}(\bar{x}_{i}^{*}, s_{p},\mathbb{W}_{\bar{p},i})  + H(\overline{p},\mathbb{W}_{\overline{p}^*,i},\hat{\phi}_\infty)-H(\overline{p}^*,\mathbb{W}_{\overline{p}^*,i},\hat{\phi}_\infty)\\
    & \leq \Delta H_i.
\end{align*}
Inequality~\eqref{eq:upperboundvaronelayerlearningsimplified} follows by applying this bound to inequality~\eqref{eq:contraditiononelayerlearning2}. 
$\blacksquare$\\

\noindent Therefore, for every $\tilde{\epsilon} > 0$, a scaling $S>0$ can be found such that
$\lim_{k\rightarrow\infty}\text{Var}_{M,k}(p) \leq \frac{\Delta H}{S} \leq \tilde{\epsilon}$, $\ \forall p \in \mathbb{F}$, where $\Delta H = \max_{\Delta H_i}\{\Delta H_1,\hdots,\Delta H_M\}$. 
\vspace{-1.7em}
\begin{IEEEbiography}[{\includegraphics[width=1in,height=1.25in,clip,keepaspectratio]{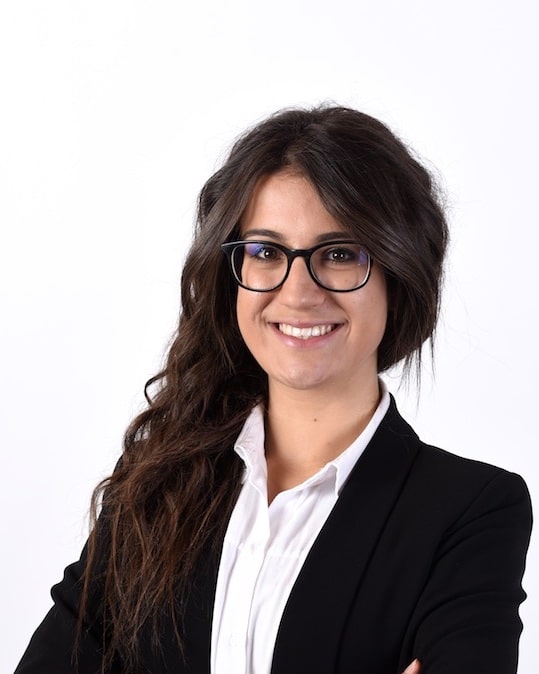}}]{Rahel Rickenbach} is a PhD candidate at ETH Zürich. She received her bachelors degree in mechanical engineering in 2019 and her masters degree in robotics, systems and control in 2021, both from ETH Zürich. In 2022 she was awarded the SGA-Förderpreis for her master thesis. Her main research interests lie in the areas of multi agent and coverage control, as well as inverse optimal control. 
\end{IEEEbiography}

\begin{IEEEbiography}[{\includegraphics[width=1in,height=1.25in,clip,keepaspectratio]{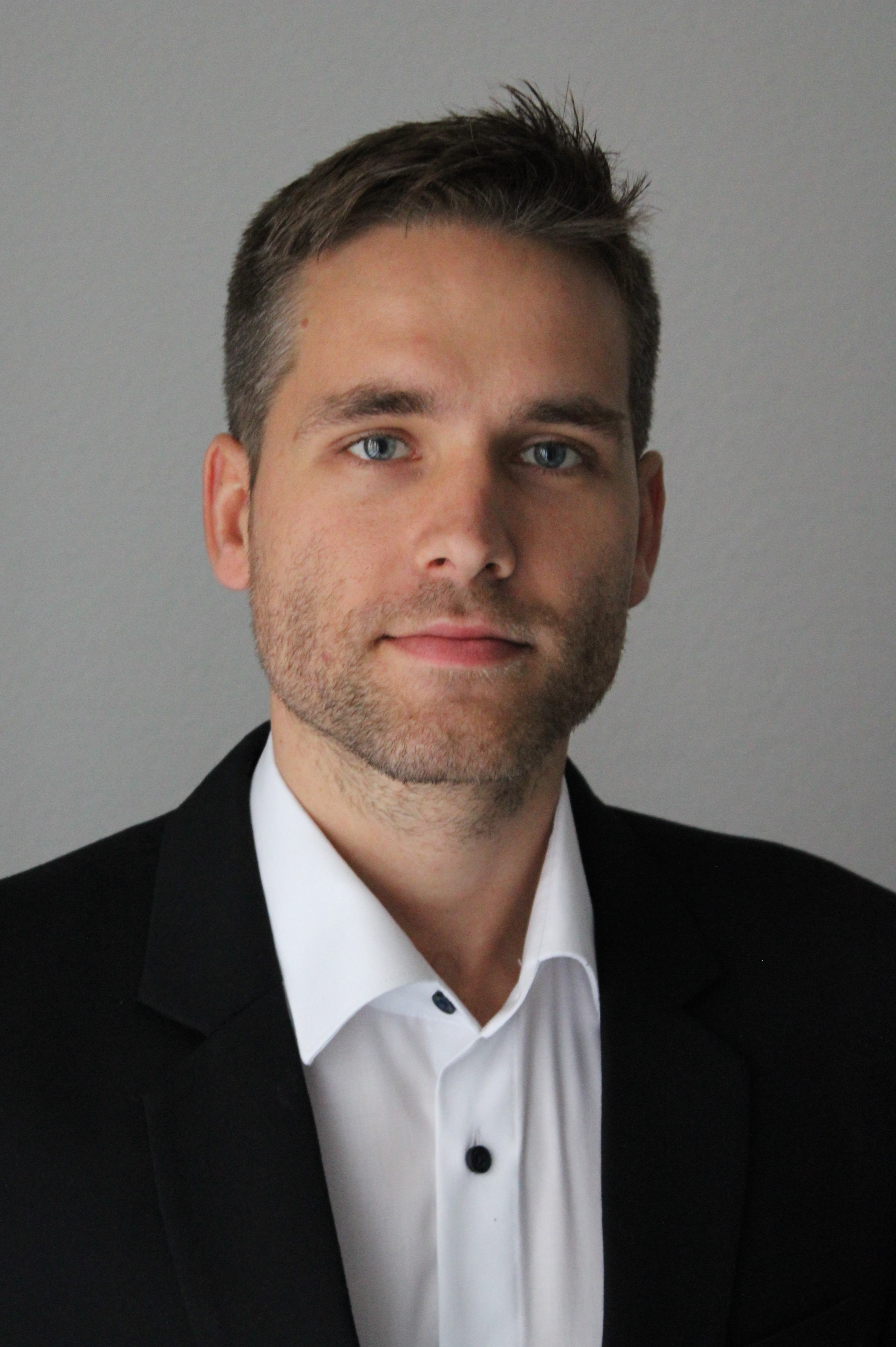}}]{Johannes K\"ohler} received his Master degree in Engineering Cybernetics from the University of Stuttgart, Germany, in 2017. 
In 2021, he obtained a Ph.D. in mechanical engineering, also from the University of Stuttgart,
Germany, for which he received the 2021 European Systems \& Control PhD award.
He is currently a postdoctoral researcher at the Institute for Dynamic Systems and Control (IDSC) at ETH Zürich.
His research interests are in the area of model predictive control and control and estimation for nonlinear uncertain systems. 
\end{IEEEbiography}

\begin{IEEEbiography}[{\includegraphics[width=1in,height=1.25in,clip,keepaspectratio]{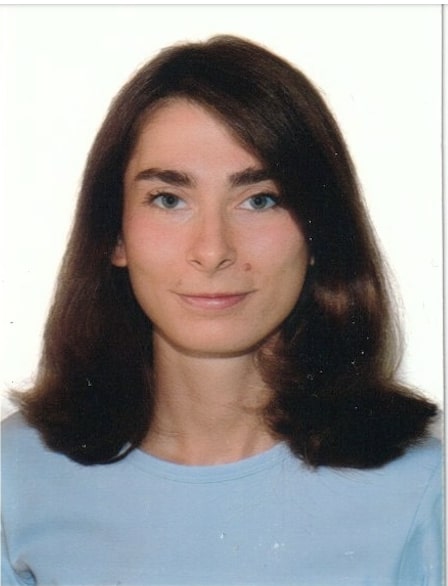}}]{Anna Scampicchio} was born in 1993. She received in 2015 the Bachelor degree in Information Engineering and in 2017 the Masters degree in Automation Engineering, both cum laude, from the University of Padova. In 2017 she was awarded with the Roberto Rocca scholarship for her career during the Masters Degree. She held a visiting position at the Department of Applied Mathematics of University of Washington, Seattle, in 2019. In 2021 she received the Ph.D. in Information Engineering from the University of Padova, and now she is a
postdoctoral researcher at the Institute for Dynamic Systems and Control, ETH Z{\"u}rich. Her research interests lie at the interplay among system identification, machine learning and control design.
\end{IEEEbiography}

\begin{IEEEbiography}[{\includegraphics[width=1in,height=1.25in,clip,keepaspectratio]{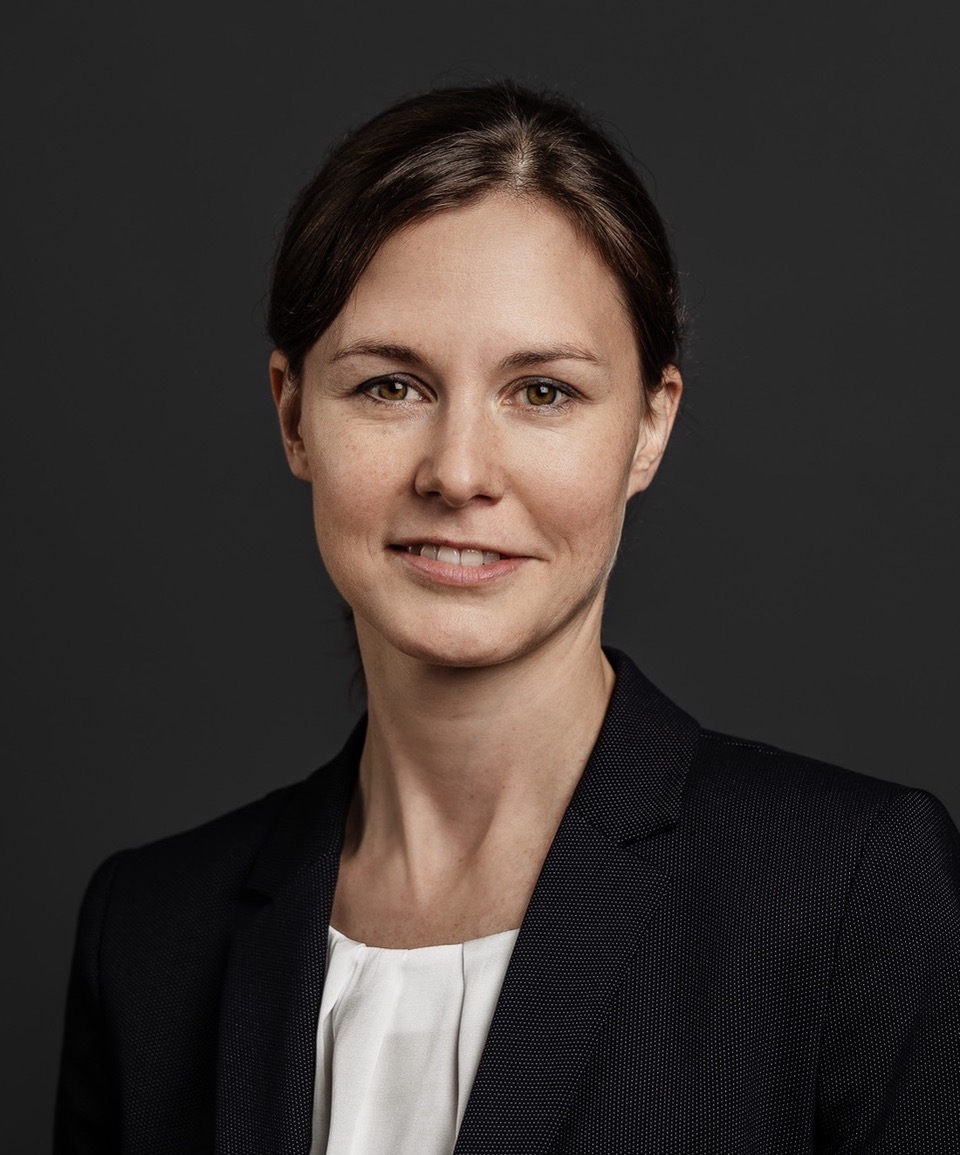}}]{Melanie N. Zeilinger} 
is an Associate Professor at ETH Zurich, Switzerland. She received the Diploma degree in engineering cybernetics from the University of Stuttgart, Germany, in 2006, and the Ph.D. degree with honors in electrical engineering from ETH Zurich, Switzerland, in 2011. From 2011 to 2012 she was a Postdoctoral Fellow with the Ecole Polytechnique Federale de Lausanne (EPFL), Switzerland. She was a Marie Curie Fellow and Postdoctoral Researcher with the Max Planck Institute for Intelligent Systems, Tübingen, Germany until 2015 and with the Department of Electrical Engineering and Computer Sciences at the University of California at Berkeley, CA, USA, from 2012 to 2014. From 2018 to 2019 she was a professor at the University of Freiburg, Germany. Her current research interests include safe learning-based control, as well as distributed control and optimization, with applications to robotics and human-in-the-loop control.
\end{IEEEbiography}
 \vspace{-30em}

\begin{IEEEbiography}[{\includegraphics[width=1in,height=1.25in,clip,keepaspectratio]{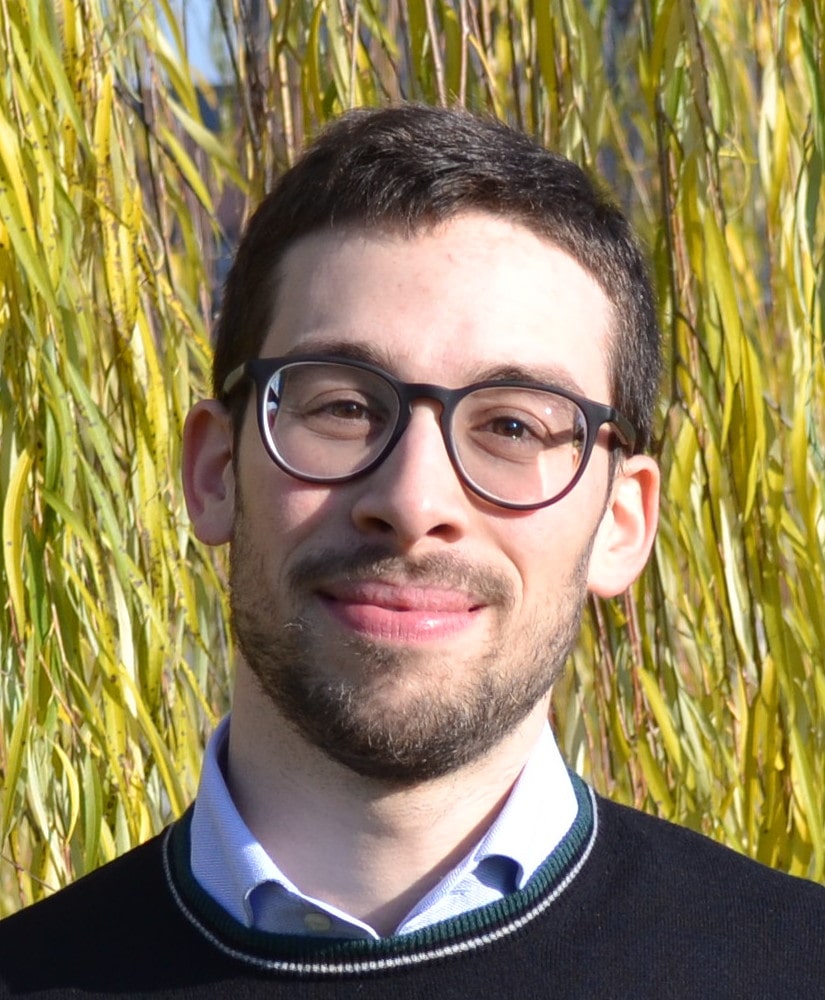}}]{Andrea Carron} is a Senior Lecturer at ETH Zürich. He received the bachelor’s, master’s, and Ph.D. degrees in control engineering from the University of Padova, Padova, Italy. During his master and Ph.D. studies, he spent three stays abroad as a Visiting Researcher: the first at the University of California at Riverside, Riverside, CA, USA, the second at the Max Planck Institute, Tubingen, Germany, and the third at the University of California at Santa Barbara, Santa Barbara, CA, USA. From 2016 to 2019, he was a Post-Doctoral Fellow with the Intelligent Control Systems Group, ETH Zürich, Zürich, Switzerland. His is research interests include safe-learning, learning-based control, multiagent systems, and robotics.
\end{IEEEbiography}.
\newpage

\onecolumn 

\section*{Look-up notation table in alphabetic order}

In the following, we present and contextualize all the symbols used throughout the paper. The list is divided in four subsections: general sets, operators and functions; indices; set cardinalities and vector dimensions; and variables and sets. Within each subsection, symbols are ordered in a dictionary-like lexicographic order, where capital letters are presented separately from lower-case ones, and grouped by font type.

\subsection*{General sets, operators and functions}
\begin{flalign*}
\notn{\|\cdot \|}{Euclidean norm}
\notn{\ominus}{Pontryagin set difference}
\notn{f(y)_{\mathbb{D}}}{$\min_{\tilde{{y}} \in \mathbb{D}}f({y} - \tilde{{y}})$, for continuous $f(\cdot)$ and compact $\mathbb{D}$}
\notn{\text{Var}_{t}(p_i)}{variance at arbitrary position $p_i$ at time $t$, \eqref{eq:variancephiblr}}
\notn{\text{Var}_{M,t}(p)}{summed up $\text{Var}_{t}(p_i)$ over all agents}
\notn{\mathcal{B}(\cdot)}{Bernoulli random variable with success probability $\cdot$}
\notn{\mathcal{N}(\mu_0,\Sigma_0)}{Gaussian distribution with mean $\mu_0$ and covariance $\Sigma_0$}
\notn{\mathcal{P}[\cdot]}{Probability operator}
\notn{\mathbb{B}_{o}^{b}}{Ball with radius $o$: $\{x \in \mathbb{R}^b\vert \,\Vert x \Vert\leq o \}$}
\notn{\mathbb{N}}{Set of natural numbers}
\notn{\mathbb{R}_+}{Set of non-negative real numbers}
\notn{\mathbb{V}^{\mathrm{int}},\; \mathbb{V}\subset\mathbb{R}^b}{$\mathbb{V} \ominus \mathbb{B}_{\epsilon}^{b}$ for problem-dependent, fixed $\epsilon >0$ \hspace{14.7em}}
\end{flalign*}

\subsection*{Indices}
\begin{flalign*}
\notn{h}{time index at which a density measurement is collected}
    \notn{i}{agent index}
    \notn{j}{dummy index}
    \notn{k}{time step}
    \notn{l}{prediction step}
    \notn{t}{time index for data collection \eqref{eq:setofmeasurements}}
    \notn{w}{time index for partition update}
    \notn{w_+}{next partition update time index \hspace{19.5em}}
\end{flalign*}

\subsection*{Set cardinalities and vector dimensions}
\begin{flalign*}
\notn{D}{coverage domain dimension (usually $D=2$ or $D=3$)}
\notn{M}{number of agents}
\notn{N}{horizon length of MPC problem}
\notn{N^*}{minimum horizon length as in Theorem \ref{theorem:theo4boccia}}
\notn{}{}
\notn{m_i}{dimension of input vector $u_{i,k}$}
\notn{n_i}{dimension of state vector $x_{i,k}$} 
\notn{}{}
\notn{\upsilon}{number of features describing density function $\phi$ \hspace{12.5em}}
\end{flalign*}

\subsection*{Variables and sets}
\begin{flalign*}
\notn{C_i}{$D\times n_i$ matrix for $i-$th agent state-position map \eqref{eq:nonlineardynamics}}
\notn{F}{arbitrary, strictly monotonically increasing function such that \mbox{$F(\epsilon)=0 \Leftrightarrow \epsilon=0$}}
\notn{H(p,\mathbb{O})}{locational optimization cost in \eqref{eq:cortescost}}
\notn{H(p,\mathbb{O},\hat{\phi})}{locational optimization cost in \eqref{eq:cortescost} with respect to estimated density $\hat{\phi}$}
\notn{I_t}{set of measurements \eqref{eq:setofmeasurements}}
\notn{J_i}{MPC cost for the $i-$th agent \eqref{eq:generalnonlineartrackingmpccost}}
\notn{J^{*}(x_{k}, \mathbb{W}_{\bar{p}_w^*})}{summation of $J_{i}^{*}(x_{k}, \mathbb{W}_{\bar{p}_w^*,i})$ over all agents}
\notn{J^{*}(x_{k}, \mathbb{W}_{\bar{p}_w^*},\hat{\phi_k})}{$J^{*}(x_{k}, \mathbb{W}_{\bar{p}_w^*,i})$ with respect to estimated density at time step $k$}
\notn{J_{i}^{*}(x_{k}, \mathbb{W}_{\bar{p}_w^*,i})}{optimal $J_i$ with respect to initial state $x_{k}$ and partition $\mathbb{W}_{\bar{p}_w^*,i}$}
\notn{\tilde{J}_{i}^{*}(x_{k}, \mathbb{W}_{\bar{p}_w^*,i})}{Two-layers approach Lyapunov function, equal to $J_{i}^{*}(x_{k}, \mathbb{W}_{\bar{p}_w^*,i})- l_{\mathbb{W}_{\bar{p}_w^*},r,i,\min}$}
\notn{\mathcal{L}_i}{Lipshitz constant of the $i-$th agent dynamics}
\notn{\mathcal{L}_{\Phi}}{Lipshitz constant from Assumption \ref{assumption:bayesianconvergence}}
\notn{Q_j}{scaling matrix for state vector in stage cost, Section \ref{subsec:expermpccost}}
\notn{Q_j^{\prime}}{scaling matrix for state vector in target cost, Section \ref{subsec:expermpccost}}
\notn{R_z}{scaling matrix for input vector in stage cost, Section \ref{subsec:expermpccost}}
\notn{R_z^{\prime}}{scaling matrix for input vector in target cost, Section \ref{subsec:expermpccost}}
\notn{S}{scaling factor for variance term in one-layer MPC cost, \eqref{eq:fullnonlineartrackingmpccostonelayerlearning}}
\notn{T}{continuous mapping entering Proposition \ref{proposition:cortesconvergence}}
\notn{T_s}{sampling time}
\notn{\text{Var}_{\max}}{maximal variance value computed over the set $\mathbb{A}$}
\notn{\text{Var}_{i,\max,t}}{maximal variance value computed over the $i$-th partition at time step $t$}
\notn{V_{N,i}}{tracking cost for the $i-$th agent}
\notn{V_{N,i}^{*}(x_{i,k},s_{i,k}, \mathbb{P}_{i})}{optimal $V_{N,i}$ with arbitrary initial condition, setpoint and position constraint}
\notn{V_{N}^{*}(x_{k}, s_{k}^{*},\mathbb{W}_{\bar{p}^{*}_{w}})}{summation of $V_{N,i}^{*}(x_{i,k},s_{i,k}, \mathbb{W}_{\bar{p}^{*}_{w,i}})$ over all agents}
\notn{V_{\max,i}}{user chosen bound on tracking cost \eqref{eq:generalvarbound}}
\notn{}{}
\notn{a_{i}}{positive constant entering \eqref{eq:lowerbound2layer}, Proposition \ref{proposition:upperlowerboudJtilde} and \eqref{eq:prop1}, Proposition \ref{proposition:notminsteadynotcoseonelayer}}
\notn{\bar{a}}{maximal value of $\bar{a}_{i}$ over all agents}
\notn{\bar{a}_{i}}{positive constant in lower bound of optimal stage cost \eqref{eq:prop12layer}, Proposition \ref{proposition:notminsteadynotcosetwolayers}}
\notn{b_{i}}{positive constant entering \eqref{eq:upperbound2layer}, Proposition \ref{proposition:upperlowerboudJtilde}}
\notn{\tilde{c}}{constant in proof of Lemma \ref{lemma:learning}}
\notn{c^{\prime}}{constant in proof of Lemma \ref{lemma:learning}}
\notn{\bar{c}_{i}}{positive constant in proof of Proposition \ref{proposition:notminsteadynotcosetwolayers}}
\notn{c(\mathbb{W}_{p}, \phi)}{centroids of the Voronoi tessellation $\mathbb{W}_{p}$}
\notn{c_i(\mathbb{W}_{p,i}, \phi)}{$i-$th centroid \eqref{eq:voronoicenterdef}}
\notn{\hat{c}_{\bar{t}}}{estimated centroids at time step $\bar{t}$}
\notn{c^{L}_{\bar{t}}}{centroids at time step $\bar{t}$ obtained with Lloyd algorithm}
\notn{c_{\Phi}}{constant entering bound in Assumption \ref{assumption:bayesianconvergence}}
\notn{d(\cdot)}{summed up distance function over all agents}
\notn{d_i(\cdot)}{distance function defining the adopted stage cost, \eqref{eq:distance}}
\notn{e_{w}}{vector of errors $e_{i,w}$}
\notn{e_{i,w}}{$i-$th agent position error $\Vert \bar{p}_{i,w}^* \! - \! c_i(\mathbb{W}_{\bar{p}_{w}^*,i}, \phi) \Vert$}
\notn{e_{v,i,k}}{distance from point of maximum variance, $\Vert v_i(\mathbb{W}_{\bar{p}_w^*,i}, \text{Var}_t) - p_{i,k} \Vert$}
\notn{f_i(\cdot,\cdot)}{state transition map for $i-$th agent \eqref{eq:nonlineardynamics}}
\notn{g(\cdot)}{general sensing capability (Remark \ref{rmk:g})}
\notn{g_{1,i},g_{2,i}}{position and state-dependent dynamics vector of the $i$-th car}
\notn{g_{3,i},g_{4,i}}{Lie brackets of $g_{1,i}$ and $g_{2,i}$}
\notn{k'}{time steps into future in \eqref{eq:4fintitetimecovergence}}
\notn{\tilde{k}'}{Finite time step at which $\tilde{J}_{i}^{*}(x_{k+\tilde{k}'},\mathbb{W}_{\bar{p}_w^*,i}) < \bar{\epsilon}$}
\notn{l_{\mathbb{W}_{\bar{p}},r,i,\min}}{minimal target cost entering \eqref{eq:Tid2layer}, Assumption \ref{assumption:steadystatedecreasetwolayers}}
\notn{l_{\bar{p}',\mathbb{W}_{\bar{p}},i,\min}}{minimal target cost entering \eqref{eq:Tid}, Assumption \ref{assumption:steadystatedecreaseonelayer}}
\notn{\ell_i}{stage cost for the $i-$th agent \eqref{eq:generalnonlineartrackingmpccost}}
\notn{\ell^*_i(x_k,s_k)}{one-step optimal stage cost for $i-$th agent}
\notn{\ell^*(x_k,s_k)}{summed up $\ell^*_i(x_k,s_k)$ over all agents}
\notn{\ell_{T,i}}{target cost for the $i-$th agent \eqref{eq:generalnonlineartrackingmpccost}}
\notn{m_{i,h}}{density measure of $i-th$ agent at time $h$ \eqref{eq:measmod}}
\notn{\bar{m}_{t+1}}{vector of measurements $[m_{1,t+1} \: \cdots \: m_{M,t+1}]^{\top}$}
\notn{p}{Collection of positions $[p_1,\cdots,p_M]^{\top}$}
\notn{p_i}{position of the $i-$th agent}
    \notn{p_{i,k}}{position of the $i-$th agent at time $k$}
    \notn{\bar{p}_i}{position setpoint for the $i-$th agent}
    \notn{\bar{p}'}{position setpoint in Assumptions \ref{assumption:steadystatedecreasetwolayers} and \ref{assumption:steadystatedecreaseonelayer}}
    \notn{\bar{p}_{i,k}}{position setpoint for the $i-$th agent at time $k$}
    \notn{\bar{p}_{i,w}^*}{position setpoint on which the Voronoi partition is built}
    \notn{p^{\text{maxvar}}_{t,i}}{position with maximal variance in the $i$-th partition at time step $t$}
    \notn{r_i}{general external reference for the $i-$th agent}
     \notn{r_{i,k}}{external reference for the $i-$th agent at time $k$}
     \notn{r_{\max}}{maximum of all $r_{i,\max}$}
     \notn{r_{i,\max}}{radius of the ball covering the $i-$th agent, \eqref{eq:colavoidancerequirement}}
    \notn{s_i}{artificial setpoint for $i-$th agent, $s_i = (\bar{x}_{i},\bar{u}_{i})$}
    \notn{s_{i,k}}{artificial setpoint for the $i-$th agent at time $k$}
    \notn{s_{i,k}^*}{optimal artificial setpoint for the $i-$th agent at time $k$}
    \notn{\hat{s}}{candidate setpoint, Assumption \ref{assumption:steadystatedecreasetwolayers}}
    \notn{\bar{t}}{investigated time step in Proposition \ref{proposition:convergenceofcentroidstwolayerslearning}}
    \notn{u_{i,k}}{input of the $i-th$ agent at time instant $k$}
    \notn{u_{i,l\vert k}}{$l-$steps-ahead predicted input of agent $i$ at time $k$}
    \notn{u_{i,\cdot\vert k}}{predicted inputs $x_{i,l\vert k}$ with $l=0,...,N-1$}
    \notn{u_{i,\cdot\vert k}^*}{optimal input sequence for $i-$th agent}
    \notn{\bar{u}_i}{input setpoint for $i-$th agent}
    \notn{u_{d,i}}{steering input of the $i$-th car}
    \notn{u_{r,i}}{input reference of the $i$-th car}
    \notn{u_{v,i}}{velocity input of the $i$-th car}
    \notn{v(\mathbb{W}_{\bar{p}_w^*}, \text{Var})}{vector collecting all $v_i(\mathbb{W}_{\bar{p}_w^*,i}, \text{Var})$}
    \notn{v_i(\mathbb{W}_{\bar{p}_w^*,i}, \text{Var})}{point of maximum variance within the $i-$th partition}
    \notn{x_{i,k}}{state of the $i-th$ agent at time instant $k$}
    \notn{x_{i,l\vert k}}{$l-$steps-ahead predicted state of agent $i$ at time $k$}
    \notn{x_{i,\cdot\vert k}}{predicted states $x_{i,l\vert k}$ with $l=0,...,N-1$}
    \notn{x_{i,\cdot\vert k}^*}{optimal state sequence for $i-$th agent}
    \notn{\bar{x}_i}{state setpoint for $i-$th agent}
    \notn{x_{p,i}}{$x$-coordinate position of the $i$-th car}
    \notn{x_{r,i}}{state reference of the $i$-th car}
    \notn{y_{p,i}}{$y$-coordinate position of the $i$-th car}
    \notn{}{}
    \notn{\mathbb{A}}{convex area in $\mathbb{R}^D$}
    \notn{\mathbb{F}_{k}}{union of $\mathbb{F}_{i,k}$}
    \notn{\mathbb{F}_{i,k}}{set of positions that are feasible w.r.t.~\eqref{eq:fullnonlineartrackingmpconelayer}}
    \notn{\mathbb{O}}{arbitrary partition of $\mathbb{A}$, $\mathbb{O} = \{\mathbb{O}_{1}, \hdots, \mathbb{O}_{M}\}$}
    \notn{\mathbb{O}_i}{polytope associated to the $i-$th agent}
    \notn{\mathbb{P}_i}{general position constraint for the $i-$th agent}
    \notn{\mathbb{S}_{\mathbb{P},i}}{set of artificial setpoints \eqref{eq:steadystatesetwithcollavoidance}}
\notn{\mathbb{S}_{\mathbb{P},i}^{\mathrm{p}}}{set of positions satisfying the artificial setpoints set conditions}
\notn{\mathbb{S}_{\bar{p}^{\prime},\mathbb{P},i}^{\mathrm{p}}}{$\mathbb{B}_{r}^{D}(\bar{p}') \cap \mathbb{S}_{\mathbb{P},i}^{\mathrm{p}}$}
\notn{\mathbb{T}_{\bar{p},r,i}}{union of all $\bar{p} \in \mathbb{S}_{\mathbb{P},i}^{\mathrm{p}}$ resulting in a minimal target cost value, \eqref{eq:Tid2layer}}
\notn{\mathbb{T}_{\bar{p}^{\prime},\bar{p},r,i}}{union of all $\bar{p} \in \mathbb{S}_{\bar{p^{\prime}}\mathbb{P},i}^{\mathrm{p}}$ resulting in a minimal target cost value, \eqref{eq:Tid}}
    \notn{\mathbb{X}_i}{state constraints for $i-$th agent}
    \notn{\mathbb{U}_i}{input constraints for $i-$th agent}
    \notn{\mathbb{W}_{p}}{Voronoi tessellation built according to $p$}
    \notn{\mathbb{W}_{p,i}}{set of the Voronoi tessellation associated to the $i-$th agent \eqref{eq:voronoidef}}
    \notn{\bar{\mathbb{W}}_{\bar{p}_w^*},i}{$\mathbb{W}_{\bar{p}_w^*,i} \ominus \mathbb{B}_{r_{\max}}^D$}
    \notn{\mathbb{W}_{\bar{p} \rightarrow \bar{p}_{k},i \rightarrow j}}{set of points changing partition ($i \rightarrow j$) for a Voronoi update considering $\bar{p}_{k}$}
    \notn{}{}
    \notn{\Delta H}{uniform constant for asymptotic variance bound in Theorem \ref{theorem:onelayerlearning}}
    \notn{\Delta H_i}{individual $\Delta H$ for each agent with $\Delta H = \textstyle{\max_{\Delta H_i}}\{\Delta H_1,\hdots,\Delta H_M\}$}
    \notn{\Phi}{vector collecting density features, Assumption \ref{assumption:bayesianconvergence}}
    \notn{\bar{\Phi}_{t+1}}{matrix of features $[\Phi(p_{1,t+1})^{\top} \: \cdots \: \Phi(p_{M,t+1})^{\top}]^{\top}$}
    \notn{\Sigma}{positive semi-definite matrix, possibly associated to a covariance}
    \notn{}{}
    \notn{\alpha_{1}}{maximal value of $\alpha_{1,i}$ over all agents}
    \notn{\alpha_{1,i},\alpha_{2,i}}{constants entering bounds for stage cost \eqref{eq:boundstagecosttwolayer}, Assumption \ref{assumption:boundedbyd}}
    \notn{\bar{\alpha}_{N}}{maximal value of $\bar{\alpha}_{N,i}$ over all agents}
    \notn{\bar{\alpha}_{N,i}}{constant entering bound \eqref{eq:theorem5twolayer}, Theorem \ref{theorem:theo4boccia}}
    \notn{\tilde{\alpha}_{N,i}}{positive constant used to show exponential convergence in \eqref{eq:4fintitetimecovergencetwolayer}}
    \notn{\beta_{1,i},\beta_{2,i}}{constants entering bounds \eqref{eq:assumption3twolayer}, Assumption \ref{assumption:steadystatedecreasetwolayers} and \eqref{eq:assumption3}, Assumption \ref{assumption:steadystatedecreaseonelayer}}
    \notn{\beta_{T,i}}{constant entering bound \eqref{eq:upperboundtargetcost}, Assumption \ref{assumption:steadystatedecreasetwolayers} and \eqref{eq:assumption33}, Assumption \ref{assumption:steadystatedecreaseonelayer}}
    \notn{\gamma_i}{constant for tracking cost bound \eqref{eq:assumption2gammavmax}, Assumption \ref{assumption:expocostcontrollability}}
    \notn{\gamma'}{maximal value of $\gamma_{i}'$ over all agents}
    \notn{\gamma_{i}'}{constant entering \eqref{eq:assumption22}, Proposition \ref{proposition:upperboundsuccedingstagecost}}
    \notn{\gamma_{V_{\max}}}{maximal value of $\gamma_{i}'$ and $\gamma_{i,V_{\max,i}}$ over all agents}
    \notn{\gamma_{i,V_{\max,i}}}{constant entering \eqref{eq:upperbound2layer1}, in proof of Proposition \ref{proposition:upperlowerboudJtilde}}
    \notn{\delta_i}{wheel orientation with respect to neutral position of the $i$-th car}
    \notn{\delta_{r,i}}{reference wheel orientation with respect to neutral position of the $i$-th car}
    \notn{\bar{\delta}}{constant between 0 and 1 entering \eqref{eq:learningprop1} in Proposition \ref{proposition:convergenceofcentroidstwolayerslearning}}
    \notn{\epsilon}{small but fixed positive constant to define the interior of a set}
    \notn{\epsilon^{\prime}}{constant between 0 and 1 entering bounds in Assumptions \ref{assumption:steadystatedecreasetwolayers} and \ref{assumption:steadystatedecreaseonelayer}}
    \notn{\varepsilon}{positive constant in \eqref{eq:learningprop1}, Proposition \ref{proposition:convergenceofcentroidstwolayerslearning}}
    \notn{\bar{\epsilon}}{positive constant to show fulfillment of \eqref{eq:updatereq1},~\eqref{eq:updatereq2} in Section \ref{subsection:finitetimeconditiontwolayersp1}}
    \notn{\tilde{\epsilon}}{positive constant in proof of Proposition \ref{proposition:onelayerlearning}}
    \notn{\zeta}{dummy variable entering the definition of $d_i$}
    \notn{\eta_{i,j}}{dummy exponent entering the definition of $d_i$}
    \notn{\theta}{vector of linear combinators describing the density, Assumption \ref{assumption:bayesianconvergence}}
    \notn{\kappa}{$\mathcal{K}$-function}
    \notn{\lambda}{scaling factor in one-layer MPC cost, \eqref{eq:fullnonlineartrackingmpccostonelayer}}
    \notn{\mu}{mean vector}
    \notn{\nu_{i,h}}{additive noise entering density measurement model \eqref{eq:measmod}}
    \notn{\xi_{1,i},\xi_{2,i}}{constants entering bound \eqref{eq:boundtwosteadystatestwolayer}, Assumption \ref{assumption:expocostcontrollability}}
    \notn{\rho}{bound for position error $e_{v,i,k}$}
    \notn{\rho_{1,...,4}, \rho_{u}}{stage cost exponents used in the experiments}
    \notn{\sigma^2}{noise variance entering density measurements \eqref{eq:measmod}}
    \notn{\phi}{density function, $\phi: \mathbb{A} \rightarrow \mathbb{R}_+$}
    \notn{\phi_1}{density function used in experiments for known environments}
    \notn{\phi_2}{density function used in experiments for unknown environments}
    \notn{\hat{\phi}_{t}}{estimated density at time index $t$}
    \notn{\hat{\phi}_{t}(p)}{estimated density at position $p$ and time index $t$ \eqref{eq:thetaupdate}}
    \notn{\hat{\phi}_{\infty}}{converged estimate for $\phi$}
    \notn{\chi_i}{bound on optimal stage cost, Assumption \ref{assumption:expocostcontrollability}}
    \notn{\psi_i}{orientation with respect to x-axis of the $i$-th car}
    \notn{\psi_{r,i}}{reference orientation with respect to x-axis of the $i$-th car}
\end{flalign*}

\end{document}